\documentclass[twocolumn,english,superscriptaddress]{revtex4-2}
\usepackage[T1]{fontenc}
\usepackage[latin9]{inputenc}
\usepackage{xcolor}
\usepackage{units}
\usepackage{amsmath}
\usepackage{amssymb}
\usepackage{graphicx}

\makeatletter

\usepackage{bbm}
\usepackage[offset=1.5em]{simpler-wick}
\usepackage{physics}
\usepackage[colorlinks,citecolor=blue,
urlcolor=blue,hypertexnames=true,linkcolor=blue]
{hyperref} 

\def\vac{{\vert\text{vac}\rangle}}

\def\rvac{\vert0\rangle}
\def\lvac{\langle0\vert}

\makeatother

\usepackage{babel}
\begin{document}
\title{Interferometric-Spectroscopy With Quantum-Light; Revealing Out-of-Time-Ordering
Correlators}
\author{Shahaf Asban}
\email{sasban@uci.edu}

\affiliation{Department of Chemistry and Physics \& Astronomy, University of California,
Irvine, California 92697-2025, USA}
\author{Konstantin E. Dorfman }
\email{dorfmank@lps.ecnu.edu.cn}

\affiliation{State Key Laboratory of Precision Spectroscopy, East China Normal
University, Shanghai 200062, China}
\author{Shaul Mukamel}
\email{smukamel@uci.edu}

\affiliation{Department of Chemistry and Physics \& Astronomy, University of California,
Irvine, California 92697-2025, USA}
\begin{abstract}
We survey the inclusion of interferometric elements in nonlinear spectroscopy
performed with quantum light. Controlled interference of electromagnetic
fields coupled to matter can induce constructive or destructive contributions
of microscopic coupling sequences (histories) of matter. Since quantum
fields do not commute, quantum light signals are sensitive to the
order of light-matter coupling sequence. Matter correlation functions
are thus imprinted by different field factors, which depend on that
order. We identify the associated quantum information obtained by
controlling the weights of different contributing pathways, and offer
several experimental schemes for recovering it. Nonlinear quantum
response functions include out-of-time-ordering matter correlators
(OTOC) which reveal how perturbations spread throughout a quantum
system (information scrambling). Their effect becomes most notable
when using ultrafast pulse sequences with respect to the path difference
induced by the interferometer. OTOC appear in quantum-informatics
studies in other fields, including black holes, high energy, and condensed
matter physics. 
\end{abstract}
\maketitle

\section{Introduction}

The quantum nature of light can affect and be utilized to steer optical
signals in many ways \citep{muk20}. First, unique properties such
as photon entanglement show nonclassical bandwidth characteristics,
offering new ways to study many-body correlations. Multi-photon collective
resonances \citep{dor14} excited by entangled light sources give
access to matter information not available with classical sources.
Second, low-intensity quantum light sources are useful for various
sensing applications. An entangled pair can be generated such that
each photon has a very different frequency regime. This provides a
convenient way to probe matter information in less accessible frequency
ranges (e.g. IR, XUV), while measuring visible photons \citep{kal16}.
Another property of quantum light is the larger parameter space which
enables sensing applications such as phase imaging \citep{hum13},
quantum sensing networks , and spectrally resolved optical phase profiles
\citep{gio11}. Third, quantum light allows to extend nonlinear spectroscopic
signals down to the few-photon level where the quantum nature of the
field must be taken into account. Observed effects include the strong
light-matter coupling in cavities \citep{her14}, the enhancement
of the medium's nonlinearity \citep{sil01} and linear pump-signal
scaling of the two-photon absorption processes \citep{var17}. The
parameter-set of the photon wave-function offers novel control knobs
that supplement classical parameters, such as frequencies and time
delays \citep{Dorfman_2016}. Quantum light opens a possibility to
shape and control excitation pathways of matter in a way not possible
by shaped classical pulses, and results in steering exciton relaxation
in molecular systems \citep{sch13}. Fourth, quantum light sources
enhance phase measurements beyond the shot-noise limit and have been
recently shown experimentally to enhance the performance of imaging
schemes \citep{Asban_2019}. The spatial resolution may be enhanced
in quantum imaging applications, quantum-optical coherence, as well
as in quantum lithographic applications. Quantum imaging with entangled
light can achieve enhanced resolution, and quantum metrology can overcome
the shot noise limit \citep{bri10}. 

In this perspective, we survey emerging novel spectroscopic techniques
made possible by interferometric setups. Each setup includes three
components: an incoming quantum light source (preparation), field-matter
coupling, and detection. Interference of optical fields has a rich
history of experimentally unraveling illusive physical phenomena.
Due to its linear dispersion, path differences of light are associated
with time delays, rendering controlled interference setups (interferometers)
valuable sensitive phase evaluation devices. Quantum probes are more
complex and potentially carry additional information \citep{nielsen_chuang_2010}.
This can be used to outperform purely classical schemes in precision
measurements, due to higher Fisher information and corresponding lower
$\text{Cramér-Rao}$ bound \citep{Helstrom_1976}. Setups based on
Mach-Zehnder \citet{rar90}, Hong-Ou-Mandel \citep{hom87}, and Franson
\citep{ray13} interferometers with quantum light are sensitive to
the change in photon statistics of quantum light upon coupling to
matter, and can be revealed by coincidence detection with multiple
detectors \citep{kal08,ray13,Lavoie_2020}. Quantum-enhanced interferometers
-- such as the ones used for the observation of gravitational waves
\citep{Caves_1981,LIGO_2019} -- indeed demonstrate unprecedented
phase estimation precision with high loss tolerance at lower photon
flux \citep{Hudelist_2014,Li_2014,Anderson_2017,Manceau_2017,Shaked_2018},
and in wide-field imaging \citep{Frascella_2019}. Generally, interferometers
shuffle the time ordering of the input fields, creating a superposition
of possible histories related to different paths. When the input field
is a composition of well separated pulses, this effect is expressed
in the output of time- resolved signals. This superimposed re-ordering
can be described via linear transformations, and further classified
into symmetry groups (Sec. $\text{\ref{Sec: building-blocks}}$),
suggesting a systematic classification of experimental setups. Here,
we consider coupling a quantum material system of interest, to auxiliary
electromagnetic fields under such conditions. The probe may propagate
through known interference at any stage -- prior, during or after
the coupling with matter -- and finally detected. Our approach \citep{Mukamel_1995}
is closely related to the space-time tomographic mapping of superdensity
operators \citep{Cotler_2018}. We connect quantum information contributions
to matter quantities.

Matter does not affect the state of coherent light, thus, all light-matter
coupling histories (Liouville pathways) contribute with the same weight
of the field. Their sum defines the classical response function. In
contrast, other states of light (e.g., fixed number of photons) may
carry different amplitude for each possible pathway. Each light-matter
interaction sequence is then associated with a unique generalized
response, which constitutes the classical (nonlinear) response \citep{Harbola_2008,Kryvohuz_2012_I,Kryvohuz_2012_II}.
From this point of view, the excess (quantum) information carried
by the probe, allows one to open the measurement black box and closely
observe the triggering sequence. Interferometric transformations of
such states of light, correspond to altering between superimposed
pathways judiciously (in lossless transformations). The interference
of classical probes results in an output modulated by the classical
response. The algebraic-geometric view of interferometry, implies
that invariant observables transform as scalars (e.g., total photon
number) and thus detected as constant flux. Others, (e.g., single
polarization after basis transformation) are sensitive to rotations
and thus show oscillations in measurements \citep{Scully_1982,Kim_2000}.
The latter may carry useful information and should be studied in more
detail.

The response to classical light is given by correlation functions
of the dipole operator $V\left(t_{n}\right)V\left(t_{n-1}\right)...V\left(t_{1}\right)$
with a specific prescription of time ordering, we label them time
ordered correlators (TOC). Multiphoton interferometric signals can
give rise to generalized response functions, composed of light-matter
coupling sequences in irregular time-ordering. These are broadly denoted
out-of-time-ordering correlators (OTOC). This terminology will be
precisely defined in Sec. $\text{\ref{Sec.: OTOC }}$. OTOC are attracting
considerable attention in other fields, connected to quantum information
dynamics of interacting many-body (closed) systems \citep{Larkin_1969,Kitaev_2014,Shenker_2014,Roberts_2015,Maldacena_2016,Aleiner_2016,yao_2016,Chen_2016,yoshida_2017,Kukuljan_2017,Swingle_2017,Pappalardi_2018,YungerHalpern_2019,Roberts_2019,Gonzales_2019,Landsman_2019,Yan_2020,Yan_2020b}.
They provide useful signatures of quantum information scrambling,
motivated by the quantum analogue of the ``butterfly effect''. In
chaotic quantum systems, they grow exponentially fast (in time) prior
to the Ehrenfest time (timescale in which quantum effects dominate)
\citep{Larkin_1969,Kitaev_2014,Aleiner_2016,Patel_2017,Mukamel_1996}.
Otherwise, it follows a powerlaw at most \citep{Chen_2016,Swingle_2017,Kukuljan_2017}.
Computation of multipoint space-time correlations in such setups can
be carried out using the density operator formalism in Liouville space,
introduced in \citep{Mukamel_1995} and more recently in \citep{Cotler_2018}.
Alternatively, it can be done using the wave-function (Hilbert space)
approach \citep{Aleiner_2016}. The latter circumvents perplexing
paradoxes one would inevitably encounter in \emph{time symmetric}
formulations to quantum mechanics \citep{Aharonov_2010,Mukamel_comments_2011}.

In Sec. $\text{\ref{Sec: building-blocks}}$ we describe the building
blocks of linear and nonlinear interferometry. In Sec. $\text{\ref{Sec: Main }}$
we present a general expression for the observable in Liouville space.
In $\text{\ref{Sec.: OTOC }}$ we discuss the contributions of OTOC
obtained by post-coupling interferometry (detection). We then introduce
novel pathway selection protocols such as exchange-phase cycling in
Sec. $\text{\ref{Phase-cycling}}$, and time domain sorting in Sec.
$\text{\ref{Sec.: Time-domain-QED}}$ -- both enabled by state-preparation
in interferometric setups. We discuss an approach for harnessing Einstein-Podolsky-Rosen
(EPR) correlations for enhanced joint time-frequency resolutions in
Sec. $\text{\ref{Joint resolutions}}$. Finally we summarize our results
in Sec. $\text{\ref{Sec: Summary}}$.

\section{Building blocks of nterferometric signals \label{Sec: building-blocks}}

Interferometry can be classified into two main types: passive-linear
or active-nonlinear wave-mixing. Introducing a group-theoretic description
of the interferometric elements, reveals clear notions regarding available
information in terms of conserved currents. Matter degrees of freedom
may introduce broken symmetries, altering otherwise-invariant quantities
in terms of photon flux. We consider optical modes described by boson
annihilation (creation) operators $a_{i}\,\left(a_{i}^{\dagger}\right)$,
satisfying $\left[a_{i},a_{j}^{\dagger}\right]=\delta_{ij}$ and $\left[a_{i},a_{j}\right]=0$.
In order to discuss their transformations under interferometric setups,
we adopt the vector notation $\boldsymbol{a}\equiv\left(a_{1},a_{2}\right)^{T}$.

\begin{figure}
\noindent \begin{centering}
\includegraphics[scale=0.37]{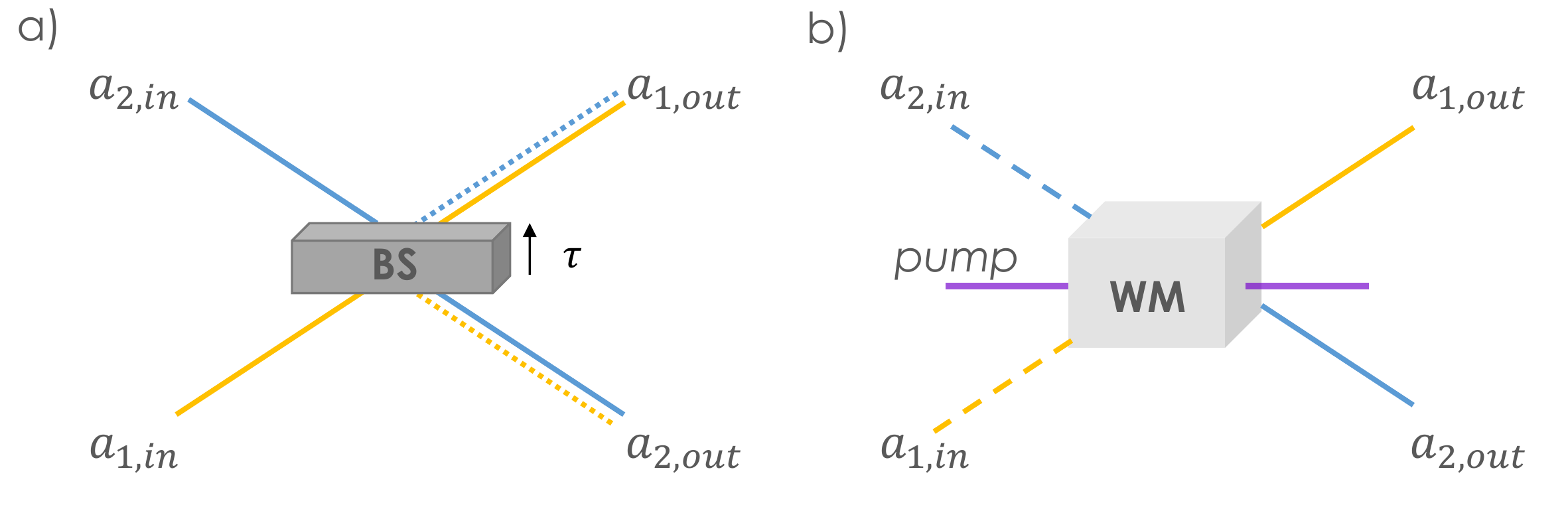}
\par\end{centering}
\caption{Interferometric elements. (a) Passive interferometric element, used
to combine two incoming modes to superposed outgoing modes with the
input output relation given by Eq. $\text{\ref{Passive transf.}}$,
where $\phi=\omega\tau$. (b) An active interferometric element, involves
a nonlinear parametric process such as three-wave mixing, and four-wave
mixing. Dashed lines represent modes that are initially in the vacuum
state. In four-wave mixing, one of the inputs is populated, e.g.,
attenuated pump as in Eq. $\text{\ref{Active Trnsf.}}$. \label{Fig 1}}
\end{figure}

\subsection{Linear-passive interferometric elements }

A generic linear interferometric setup is realized by arrays of beam-splitters
(BS) as depicted in Fig. $\text{\ref{Fig 1}a}$, mirrors and phase
elements. The input-output relations corresponding to the two-port
device (BS) is described by the unitary transformation $\boldsymbol{a}_{\text{out}}=\hat{{\cal R}}_{\phi}\boldsymbol{a}_{\text{in }}$and

\begin{equation}
\hat{{\cal R}}_{\phi}=\left(\begin{array}{cc}
T & iRe^{i\phi}\\
iRe^{-i\phi} & T
\end{array}\right)\label{Passive transf.}
\end{equation}

\noindent Here $T$ and $R$ are the reflection and transmission coefficients
such that $T^{2}+R^{2}=1$, and $\phi$ is a relative phase employed
e.g., by shifted BS or mirrors (assuming lossless BS). Such elements
are employed in many interferometric schemes -- historically highlighting
different physical realizations(i.e., Mach-Zehnder, Franson, Michelson
Sagnac, relying on combinations of optical modes). The symmetry group
of such transformations becomes more apparent by introducing the Hermitian
operators \citep{Yurke_1986}

\begin{subequations}
\begin{align}
J_{x} & =\frac{1}{2}\left(a_{1}^{\dagger}a_{2}+a_{2}^{\dagger}a_{1}\right),\\
J_{y} & =-\frac{i}{2}\left(a_{1}^{\dagger}a_{2}-a_{2}^{\dagger}a_{1}\right),\\
J_{z} & =\frac{1}{2}\left(a_{1}^{\dagger}a_{1}-a_{2}^{\dagger}a_{2}\right),
\end{align}
\end{subequations}

\noindent satisfying the commutation relations of the Lie algebra
of $SU\left(2\right)$, $\left[J_{i},J_{j}\right]=i\epsilon_{ijk}J_{k}$,
where $\epsilon_{ijk}$ is the antisymmetric tensor and and $i,j,k\in x,y,z$
(the number operator $N=a_{1}^{\dagger}a_{1}+a_{2}^{\dagger}a_{2}$
is proportional to the identity). Clearly, two-mode passive rotation
corresponds to an $SU\left(2\right)$ transformation with the invariant
(Casimir) $J^{2}=\frac{N}{2}\left(\frac{N}{2}+1\right)$. Coupling
to matter degrees of freedom in the interaction picture, is represented
by a relative shift in the unitary evolution. This stems from the
fact that for each path, the joint light-matter system is evolved
for different duration, giving rise to the out-of-time-ordering matter
correlators (OTOC) and discussed in Sec. $\text{\ref{Sec.: OTOC }}$.
The evolution operator is given by $\hat{U}\left(t,t'\right)\equiv\exp\left\{ -\frac{i}{\hbar}H\left(t-t'\right)\right\} $,
where $H=H_{\phi}+H_{\mu}+H_{\mu\phi}$. In the absence of matter,
the path difference merely yields a linear phase $\phi=\omega\tau$
corresponding to time translations of the combined modes. 

\subsection{Nonlinear-active interferometric elements }

Active interferometric elements constitute nonlinear combinations
of fields, e.g., n-wave mixing processes. Three-wave mixing generates
entangled photon pairs through parametric down conversion. A pump
photon is down converted to a pair of spontaneously generated entangled
photons. Four-wave mixing (FWM) induces further quadrature squeezing
\citet{Reid_1985}, which attracted considerable attention from the
early days of quantum enhanced metrology, aiming to improve the detection
of gravitation wave \citet{Caves_1981}.

Nonlinear interferometric techniques present several merits. They
utilize remarkable bandwidth extension with narrowband probes \citet{Shaked_2018},
improved contrast in phase measurements with sub-shotnoise scaling
-- while maintaining these enhanced features with impressive loss
tolerance \citet{Hudelist_2014,Li_2014,Manceau_2017,Anderson_2017,Du_2018,Frascella_2019}.
It can be employed in the detection process to characterize time-domain
light-matter pathways as further discussed in Sec. $\text{\ref{Sec.: Time-domain-QED}}$.

To characterize a two-photon FWM operation in terms of a transformation,
we introduce the operators

\begin{subequations}
\begin{align}
K_{x} & =\frac{1}{2}\left(a_{1}^{\dagger}a_{2}^{\dagger}+a_{1}a_{2}\right),\\
K_{y} & =-\frac{i}{2}\left(a_{1}^{\dagger}a_{2}^{\dagger}-a_{1}a_{2}\right),\\
K_{z} & =\frac{1}{2}\left(a_{1}^{\dagger}a_{1}+a_{2}a_{2}^{\dagger}\right),
\end{align}
\end{subequations}

\noindent satisfying the of the Lorentz group $SU\left(1,1\right)$;
$\left[K_{x},K_{y}\right]=-iK_{z}$, $\left[K_{y},K_{z}\right]=iK_{x}$
and $\left[K_{z},K_{x}\right]=iK_{y}$, and the Casimir (invariant)
$K^{2}=J_{z}\left(J_{z}+1\right)$. To demonstrate their effect, we
consider a realization of this transformation in which one of the
inputs in Fig. $\text{\ref{Fig 1}}$b is populated by an attenuated
pump, with a relative delay $\delta$ with respect to the activating
pump. The scattering matrix is then given by

\begin{equation}
\left(\begin{array}{c}
a_{1}\\
a_{2}^{\dagger}
\end{array}\right)_{\text{out}}=\left(\begin{array}{cc}
\cosh\beta & e^{-i\delta}\sinh\beta\\
e^{i\delta}\sinh\beta & \cosh\beta
\end{array}\right)\left(\begin{array}{c}
a_{1}\\
a_{2}^{\dagger}
\end{array}\right)_{\text{in}},\label{Active Trnsf.}
\end{equation}

\noindent where $\beta$ is related to the reflectivity of the FWM
\citet{Yurke_1986}. The $J$ operators transform under the passive
elements using $SU\left(2\right)$ rotations which are (almost) equivalent
to manifold-preserving rotations in 3D. In contrast, the active elements
impose Lorentz boosts on the $K$ operators which corresponds to quadrature
squeezing (manifold-shearing). In addition to the benefits derived
from narrowband pump operation (above), high-order mixing generate
particularly useful set for sorting through individual spontaneous
processes (Sec. $\text{\ref{Sec.: Time-domain-QED}}$).

\section{Interferometric quantum spectroscopy -- an open frontier \label{Sec: Main }}

\noindent Quantum fields are represented by operator quantities, in
contrast to classical fields, which are c-numbers. Pertubative expansion
of field-matter interactions yields optical signals expressed as multi-point
correlation functions. The relative order of the dipole operators
impacts the detected time ordering of the field-matter interactions
and their expectation values. For non commuting fields, each arrangement
of matter correlation function corresponds to a different field correlation
function. Thus, various detection schemes, can provide different information
regarding the many-body dynamics.

Interferometric setups typically mix several modes and thus the mapping
between physical interaction occurrences and their detection time
is not straightforward. Below, we survey several approaches to manipulate
and distinguish between time ordered events, and show how OTOC show
up in measurements.

\begin{figure}
\begin{centering}
\includegraphics[scale=0.45]{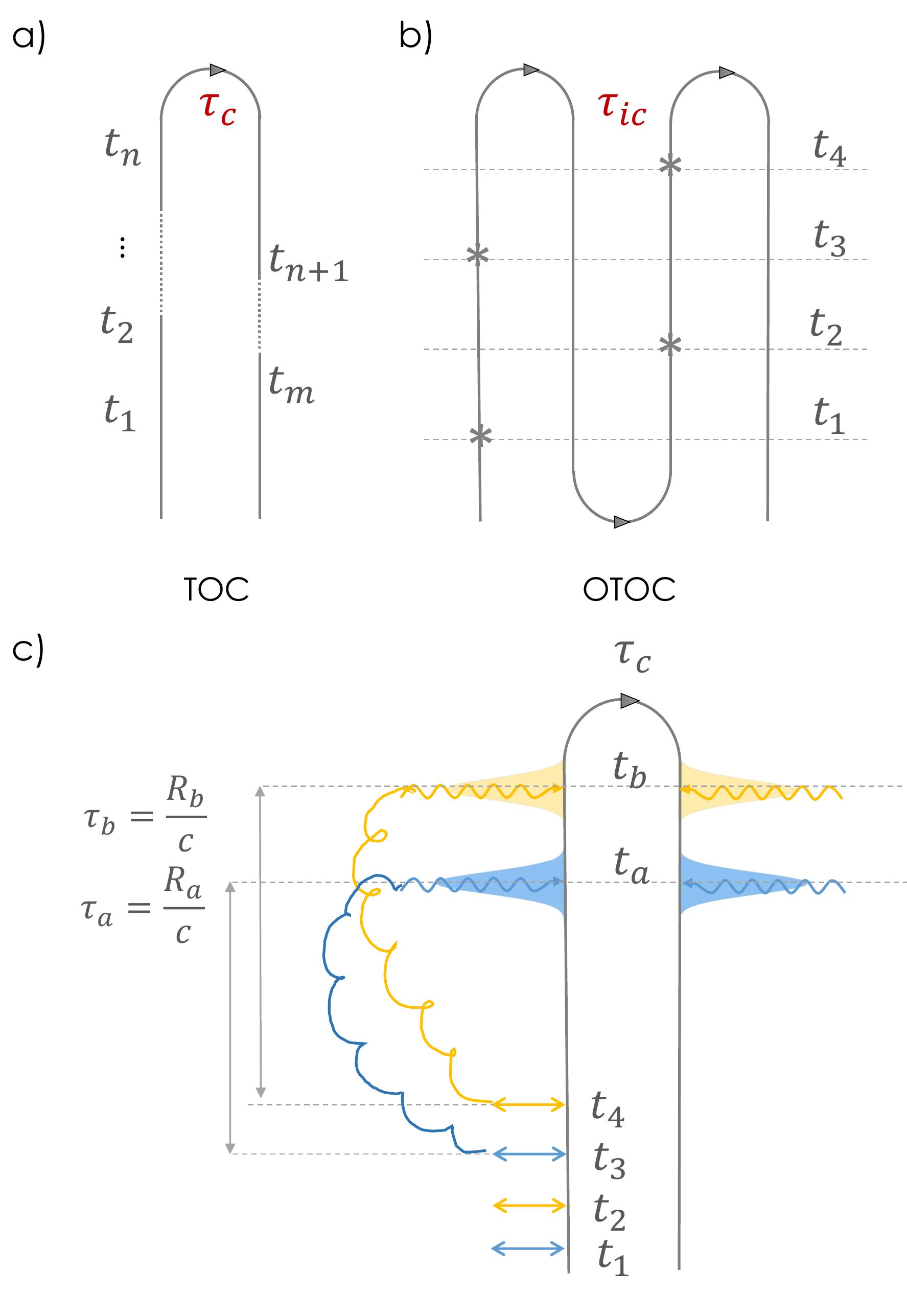}
\par\end{centering}
\caption{\textbf{Time-ordered Vs. out-of-time-ordering matter correlators}.
(a) A fully time-ordered matter correlator (TOC) computed along a
typical closed time contour $\tau_{c}$. Such contribution compose
the nonlinear response of matter upon coupling to classical light.
(b) An OTOC diagram. Different optical paths in the interferometer
re-arrange the radiative trajectories in the detection plane. The
resulting OTOC are computed along an irregular wiggling time contour
$\tau_{ic}$. \label{Fig 2}(c) A fully time-ordered loop diagrammatic
representation of a possible process that contributes to the $4^{\text{th}}$
order optical signal. Two photons interact with a sample, then detected
in coincidence following free propagation duration of $\tau_{a}$
and $\tau_{b}$. The correlators are computed along the closed time
contour $\tau_{c}$ from the distant past to the present back to the
past. }
\end{figure}

\subsection{Out-of-time-ordering correlators -- order of arrival Vs. order of
interaction \label{Sec.: OTOC }}

When a quantum system is coupled to a classical field the, response
is given by correlation functions of the form $\left\langle V\left(t_{m}\right)\cdots V\left(t_{n+1}\right)V\left(t_{n}\right)\cdots V\left(t_{2}\right)V\left(t_{1}\right)\right\rangle $
represented by the loop diagram in Fig. $\text{\ref{Fig 2}}$a. Time
proceeds forward in the left branch from $t_{1}$ to $t_{n}$, then
proceeds backward on the right branch from $t_{n+1}$ to $t_{m}$.
We denote these time ordered correlators (TOC). Interferometric measurements
with quantum light are given by more complex objects where time proceeds
forward and backward multiple times, as shown in Fig. $\text{\ref{Fig 2}}$b.
These are denoted OTOC. 

As an example we consider the loop diagram shown in Fig. $\text{\ref{Fig 2}}$c.
Matter correlation function in this example can be read off the diagram
as $\left\langle {\cal T}_{\text{c}}V\left(t_{4}\right)V\left(t_{3}\right)V\left(t_{2}\right)V\left(t_{1}\right)\right\rangle $,
where ${\cal T}_{\text{c}}$ is a time ordering operator corresponding
to the closed time contour $\tau_{c}$. Light-matter interaction events
are ordered along the loop, and propagated forward in time from the
distant past on the left branch, and backwards in time to the past
on the right branch. In this example, $t_{4}>t_{3}>t_{2}>t_{1}$ such
that the propagation is always forward in time as depicted in Fig.
$\text{\ref{Fig 2}}$c. An OTOC appears when the time-flow may be
inverted backwards in some intervals of the correlation function,
e.g., $\left\langle {\cal T}V\left(t_{4}\right)V\left(t_{2}\right)V\left(t_{3}\right)V\left(t_{1}\right)\right\rangle $,
which yields reverse evolution between the second and third operators
and depicted in Fig. $\text{\ref{Fig 2}}$b. Two-photon input in an
interferometer introduces several modified time-orderings, resulting
in irregular time-flow at the output. The coupling of light with matter
parametrizes the matter correlation functions along the wiggling contour
such as the one introduced in Ref. \citep{Aleiner_2016} for computations
of OTOC in closed systems. This can be interpreted as interference
of past and future contributions of matter multipoint correlation
functions. However, this terminology can be avoided. We next review
the interferometric transformation in terms of the detected signal.

\begin{figure}[t]
\begin{centering}
\includegraphics[scale=0.22]{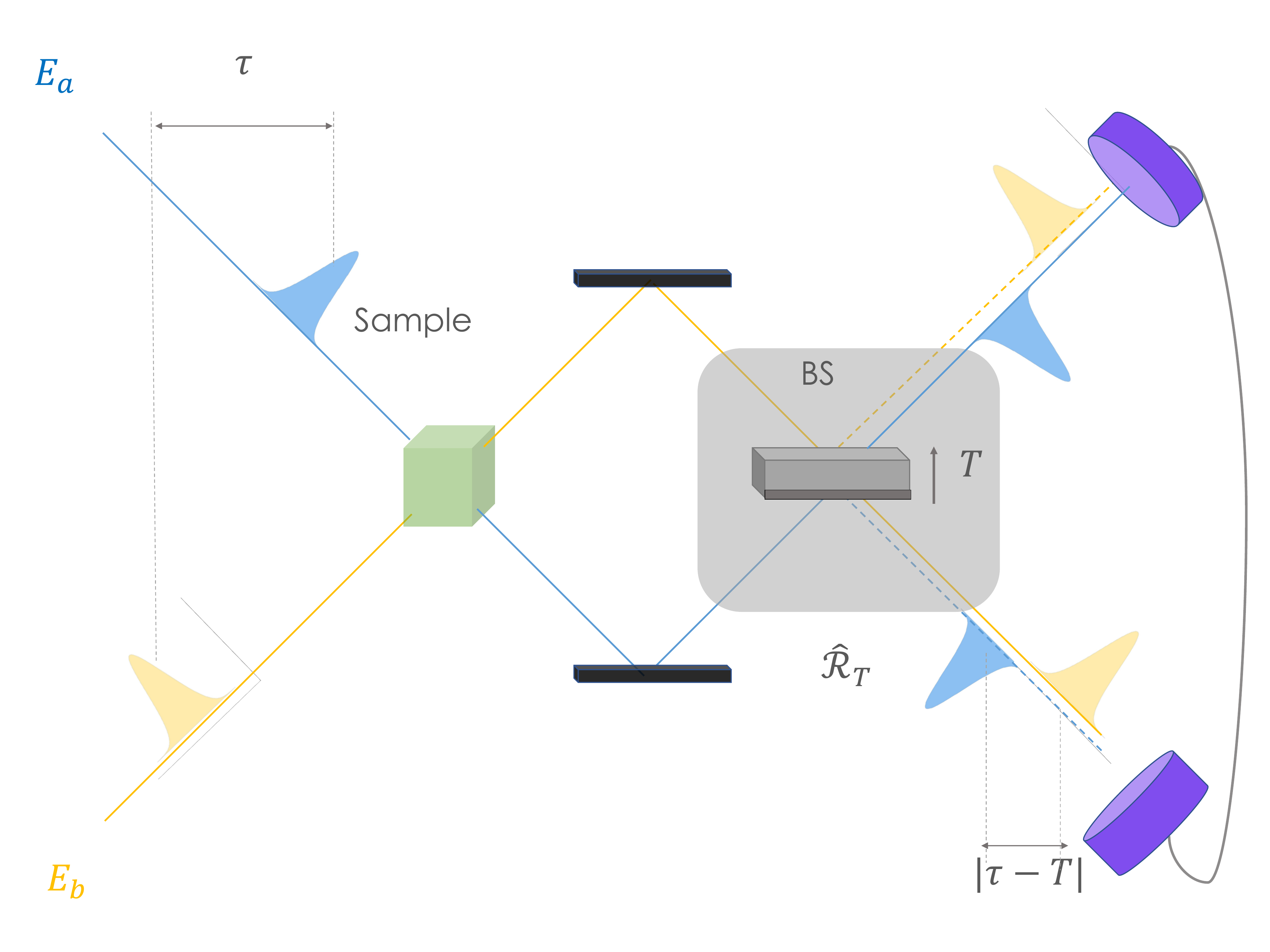}
\par\end{centering}
\caption{\label{Fig 3}\textbf{ OTOC detection} \textbf{via} \textbf{a} \textbf{Hong-Ou-Mandel}
\textbf{interferometer}. Two quantum probes arranged in space-time
wave-packets with relative delay time $\tau$. Then the photons interact
with matter (sample), and transformed using BS introducing several
possible propagation trajectories of light-matter interaction in a
Hong-Ou-Mandel interferometer configuration. The photons are time-resolved
and measured in coincidence, forming matter correlation function on
a wiggling time contour.}
\end{figure}

When the electromagnetic field propagation direction is known, the
Jordan-Schwinger map (JSM) is described by Stokes operators which
follow the Lie algebra of $SU\left(2\right)$ symmetry group \citep{Jauche_1976,Mota_2016,Mota_2004_2,Yurke_1986}.
Thus, (passive) interferometric setups can be described using a sequence
of $SU\left(2\right)$ rotations (see Sec. $\text{\ref{Sec: building-blocks}}$).
The Hong-Ou-Mandel (HOM) interferometer depicted in Fig. $\text{\ref{Fig 3}}$
\citep{Hong_1987}, is the simplest setup that gives rise to interference
between future and past matter events. It combines two optical modes
on a movable BS, which are then detected in coincidence using two
detectors. The shift of the BS with respect to the center introduces
controlled path differences between four distinct trajectories. Observables
in this setup are composed purely of field operators that evolve according
to the free electromagnetic Hamiltonian $H_{\phi}$. Measurements
are described using annihilation of modes in the far-field basis (post-rotation)
as described by Glauber \citep{Glauber_1963}.

The light-matter coupling $H_{\mu\phi}$ is generally composed of
operators from the joint Hilbert space. Light and matter degrees of
freedom thus become entangled upon energy exchange. We consider an
optical signal, generated from a general multipoint correlation function
of field operators, given by (see Sec. S2 of the SM)

{\small{}{} 
\begin{equation}
{\cal C}\left(t_{1},...,t_{n}\right)=\left\langle {\cal T}{\cal O}_{t_{1},...,t_{n}}\left(\boldsymbol{{\cal E}},\boldsymbol{{\cal E}}^{\dagger}\right)e^{-\frac{i}{\hbar}\underset{t_{0}}{\overset{t}{\int}}du{\cal H}_{\text{\ensuremath{\mu\phi}},-}\left(u\right)}\right\rangle .\label{Main Signal}
\end{equation}
}{\small\par}

\noindent Here ${\cal C}\left(t_{1},...,t_{n}\right)$ is an $n$-point
correlation function where $t$ is taken to be the latest time, ${\cal T}$
is the time-ordering superoperator, ${\cal O}_{t_{1},...,t_{n}}$
is a superoperator composition of the electric field operators in
Liouville space and $\left\langle \cdots\right\rangle \equiv\text{tr}\left\{ \cdots\rho\left(t_{0}\right)\right\} $
is the trace with respect to the initial state of the density operator.
${\cal H}_{\mu\phi,-}\left(t\right)$ is the interaction superoperator
operating on a Hilbert space operator $A$ as a commutator ${\cal H}_{\mu\phi,-}A\equiv{\cal H}_{\mu\phi}A-A{\cal H}_{\mu\phi}$
\citet{Mukamel_1995}. $\boldsymbol{{\cal E}}\left(\boldsymbol{{\cal E}}^{\dagger}\right)$
is the positive (negative) frequency electric field components $\boldsymbol{E}=\boldsymbol{{\cal E}}+\boldsymbol{{\cal E}}^{\dagger}$.
In the following illustration we are interested in expectation values
of intensity correlations. To that end it is convenient to introduce
the right-left superoperator notation, whereby left (right) superoperators
${\cal E}_{L}\left({\cal E}_{R}\right)$, act on the density according
to ${\cal E}_{L}\rho\equiv{\cal E}\rho$ $\left({\cal E}_{R}\rho\equiv\rho{\cal E}\right)$.
One can calculate the observables according to \emph{order of arrival}
at the detector, imposing the coupling description in this basis,
$\boldsymbol{{\cal E}}_{\text{detection}}=\hat{{\cal R}}\boldsymbol{{\cal E}}_{\text{interaction}}$
. Alternatively, the observable can be computed in the \emph{order
of interaction} with matter, in which ${\cal O}_{t_{1},...,t_{n}}\left(\boldsymbol{{\cal E}},\boldsymbol{{\cal E}}^{\dagger}\right)$
will be expressed in basis of the interaction operators (rotation
backwards, see Sec. S2 of the SM). In this description, the high dimensional
space -- accommodating both the electromagnetic field and the matter
field of $\dim H_{\mu}\oplus H_{\phi}$ -- is unraveled by fully
time-ordered correlation functions in the interaction picture. Since
only part of the system (i.e. the electromagnetic field) is detected,
path difference of the auxiliary degrees of freedom correspond to
effective time-flow wiggling in the measured matter correlation function.
When the observable ${\cal O}\left(\boldsymbol{{\cal E}},\boldsymbol{{\cal E}}^{\dagger}\right)$
is invariant under rotations (scalar e.g., $\boldsymbol{{\cal E}}^{\dagger}\cdot\boldsymbol{{\cal E}}$,
or any function of the Casimir of the Stokes operators), no interference
will be registered in the absence of matter. When the observable is
basis dependent, interference between future and past matter pathways
may show up in coincidence counting experiments. Notably, this is
also where one would look for superior quantum performance with respect
to phase-shift measurements \citep{Yurke_1986}.


\subsubsection*{Example of an OTOC contribution }

Consider the setup sketched in Fig. $\text{\ref{Fig 3}}$. Two modes
$\left\{ E_{a},E_{b}\right\} $ are prepared with a relative delay
time $\tau$, then interact with a sample. The modes are rotated by
a BS and measured in coincidence respectively at spacetime coordinates
$\left\{ \boldsymbol{r}_{a}t_{a},\boldsymbol{r}_{b}t_{b}\right\} $.
The recorded signal can be computed by Eq. $\text{\ref{Main Signal}}$,
by simultaneous annihilation of two photons from both sides of the
density operator

\begin{align}
{\cal O}_{t_{a},t_{b}} & \left(\boldsymbol{{\cal E}},\boldsymbol{{\cal E}}^{\dagger}\right)=\label{time coincidence observable}\\
 & {\cal E}_{R,a}^{\dagger}\left(\boldsymbol{r}_{a}t_{a}\right){\cal E}_{R,b}^{\dagger}\left(\boldsymbol{r}_{b}t_{b}\right){\cal E}_{L,b}\left(\boldsymbol{r}_{b}t_{b}\right){\cal E}_{L,a}\left(\boldsymbol{r}_{a}t_{a}\right).\nonumber 
\end{align}

\noindent Eq. $\text{\ref{time coincidence observable}}$ projects
the two-photon subspace of the density operator as a function of time
at two designated detection positions $\left\{ \boldsymbol{r}_{a},\boldsymbol{r}_{b}\right\} $,
annihilating two photons from the left and right. Crucially, both
modes $\left(a,b\right)$ are measured at the detection plane, and
require change of basis with respect to the interaction plane, using
the coupling Hamiltonian ${\cal H}_{\mu\phi}\left(t\right)=\boldsymbol{V}\left(\boldsymbol{r},t\right)\cdot\boldsymbol{E}\left(\boldsymbol{r},t\right)$
in the interaction-regime basis, and $\boldsymbol{V}=\boldsymbol{\mu}+\boldsymbol{\mu}^{\dagger}$
is the dipole operator (see Sec. S1 and S2 of the SM). A nonvanishing
signal is recorded only if two photons are detected. The lowest nonvanishing
order contributing to the signal (apart from noninteracting background)
includes four events, the two modes are annihilated then created as
a result of the interaction with the sample. Multiple processes contribute
to the overall signal, however, we are interested to demonstrate a
contribution which results in OTOC.\textbf{ }Such contributions appear
for example by considering the process displayed in Fig. $\text{\ref{Fig 2}}$a,
followed by interferometry from the sample to the detectors which
reorders the correlation function as shown in Fig. $\text{\ref{Fig 2}}$b. 

We consider the two-photon initial state of light $\vert\Psi_{0}\rangle=\int d\omega_{a}d\omega_{b}\Phi\left(\omega_{a},\omega_{b}\right)a^{\dagger}\left(\omega_{a}\right)b^{\dagger}\left(\omega_{b}\right)\vac$,
describing creation of the two modes from the vacuum with amplitude
$\Phi\left(\omega_{a},\omega_{b}\right)=\phi_{a}\left(\omega_{a}\right)\phi_{b}\left(\omega_{b}\right),$and
compute the process amplitude (see Sec. S2 of the SM). When the temporal
distribution of the wavepackets $\epsilon$ is narrow in comparison
to their relative delay $\tau$, and the matter response time, one
obtains sufficient temporal resolution to study the OTOC explicitly.
We show this by approximating the temporal envelope by delta distributions
$\phi_{a}\left(t\right)\rightarrow\delta_{\epsilon}\left(t-\tau\right)$,
$\phi_{b}\left(t\right)\rightarrow\delta_{\epsilon}\left(t\right)$
and obtain

\begin{equation}
{\cal C}_{ab}\left(\tau\right)\propto\left\langle V_{a}\left(\tau\right)V_{b}\left(0\right)V_{a}\left(\tau\right)V_{b}\left(0\right)\right\rangle ,\label{OTOC demonstration}
\end{equation}

\noindent for $\tau=2T$. Eq. $\text{\ref{OTOC demonstration}}$ clearly
reflects the wiggling (forward-backward-forward) time-flow of matter
correlation functions carried by an auxiliary probe through an interferometer
\footnote{S. Asban and S. Mukamel. Out-of-Time-Ordering Matter Correlators in
Quantum Interferometric Spectroscopy, -- \emph{to be published \label{fn:OTOC paper}}}. We stress that such processes contribute even with no temporal resolution,
the delta distributions are invoked purely for illustration purposes.
To illustrate this, we have computed the OTOC contribution obtained
by applying an entangled pair generated by a spontaneous parametric
down-converter pumped using a narrowband beam  in Sec. II.B of the
SM. Matter information is imprinted in the EM field according to the
incidence time. The interaction may occur at various times (which
are integrated upon in the interaction picture). The interference
of the detected beams depends on equally distributed times dictated
by the BS relative displacement, thanks to the linear group velocity
of the field. This can be viewed as interfering past and future from
the matter point of view (in the absence of losses).

Using this setup, one can measure rates of quantum information scrambling
in molecules. Consider for example a two-color measurement whereby
two wavepackets, each resonant with different bond (vibrational) or
localized electronic state (core electrons) as depicted in Fig. $\text{\ref{Fig 3}}$.
The signal will now carry quantum information regarding ``cross talk''
of these possibly far-apart channels as well as decoherence processes.
Such measurements become particularly interesting for complex bio-molecules,
since it can reveal the time and length-scales in which quantum dynamics
are important.

The field rotation effect on the time-flow of matter correlation functions,
bears a resemblance to the Keldysh rotation. The latter is used to
benefit from the linear relations between different kinds of matter
Green's functions \citep{kamenev2011field}. Here, the left and right
components of the field operators play a role corresponding to forward
and backward evolution, by modifying the bra and the ket. Intuitively,
the polarization degree of freedom transforms according to Pauli matrices,
similar to the augmented Keldysh contour reported in \citep{Aleiner_2016}.
The correspondence between augmented Keldysh contours and interference
of auxiliary fields merits further study in a unified framework.



\begin{figure}[th]
\begin{centering}
\includegraphics[bb=0bp 0bp 510bp 468bp,clip,scale=0.49]{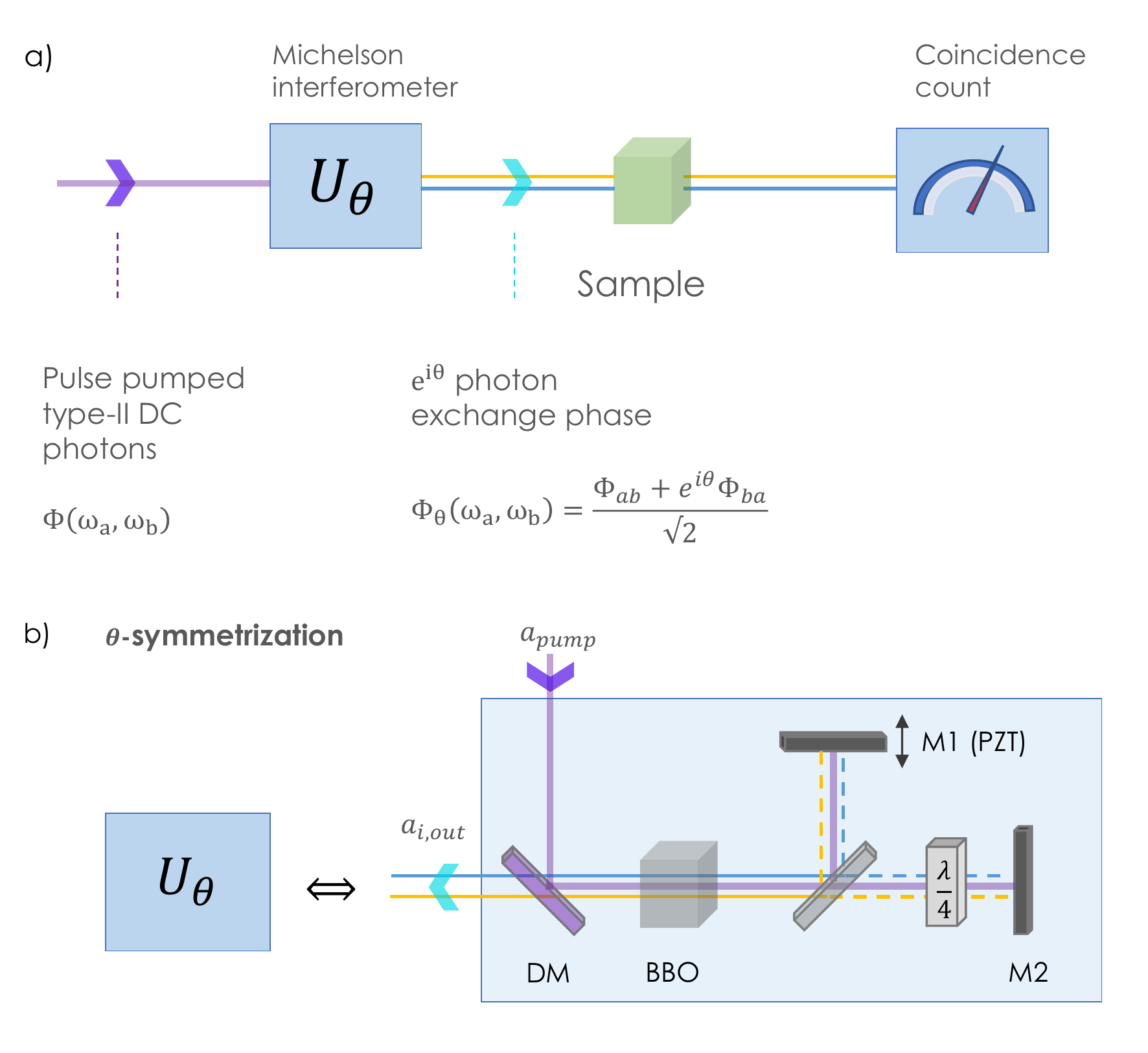}
\par\end{centering}
\caption{\label{Fig 4} Exchange-phase-cycling scheme. (a) Two-photon exchange
phase-cycling setup. The interferometric preparation process attributes
$\theta$ phase difference with respect to photon exchange, denoted
as $\theta$--symmetrization prior to the coupling with the sample.
Following the light-matter interaction, the photons are detected in
coincidence, scanning the two-photon subspace of the density operator.
(b) Implementation of $\theta$-symmetrization via a modified Michelson
interferometer (see text).}
\end{figure}

\begin{figure*}[t]
\begin{centering}
\includegraphics[scale=0.41]{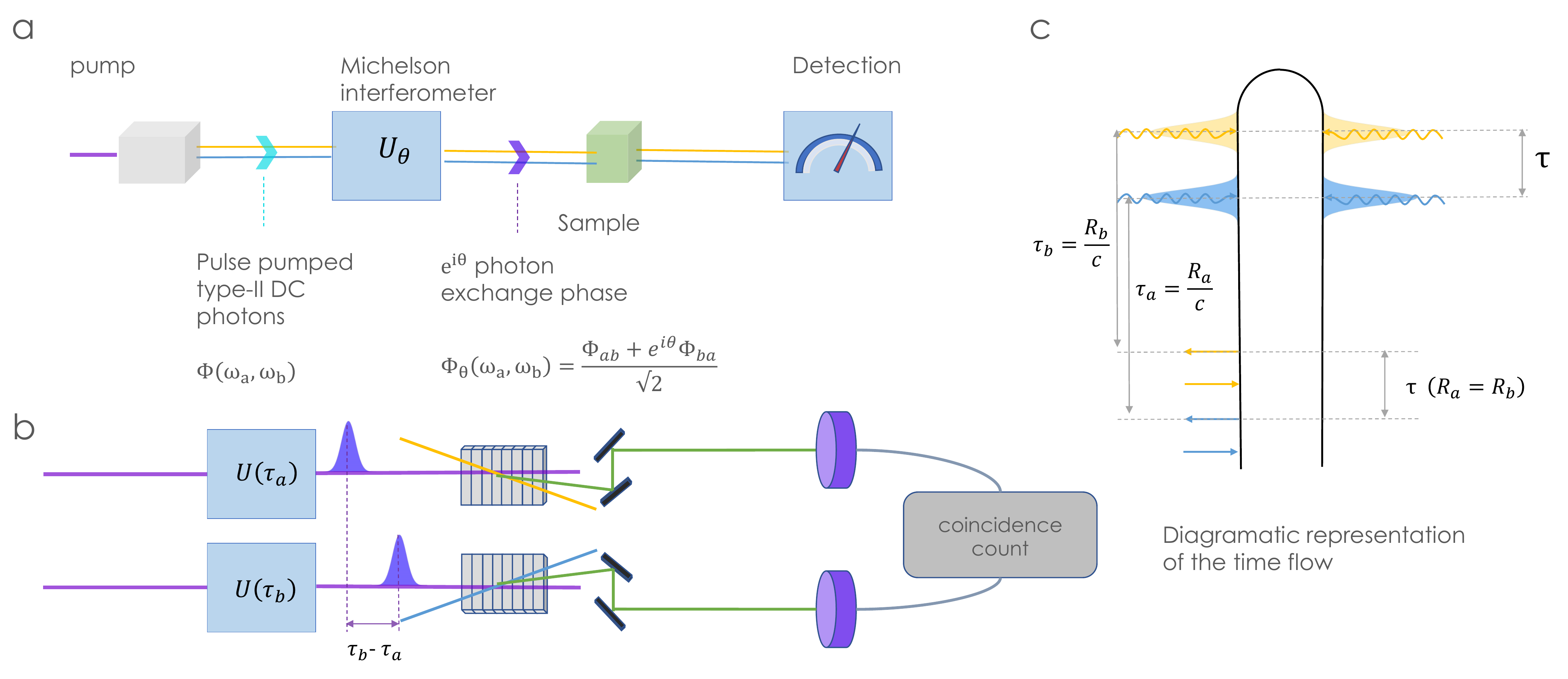}
\par\end{centering}
\caption{Time-domain switching platform. (a) A $\theta$--symmetrized entangled
pair is generated and directed and coupled to a sample. (b) The incident
photons (yellow and blue) are separated using a polarization beam
splitter (not shown), then scanned in time domain using an up conversion
with the same (generating) a delayed pump beam ($U\left(T_{a}\right)$
and $U\left(T_{b}\right)$). The up-converted photons are detected
in coincidence. (c) Time-domain diagrammatic description. Emitted
light from matter de-excitation undergoes free (retarded) propagation
to the detector. Ultrafast time-domain detection scheme with controlled
time delay $\left(\tau=\tau_{b}-\tau_{a}\right)$, used as a switch
between pathways. \label{Fig 5}}
\end{figure*}



\subsection{Exchange-phase-cycling protocols; quantum statistics and pathway
selectivity \label{Phase-cycling}}

Quantum statistics is known to play a central role in shaping the
interference patterns observed in coincidence detection of bosons
\citep{Mandel_1985}, fermions \citep{Bocquillon_2013} and fractional
charges, e.g., quantum-hall quasiparticles \citep{Chamon_1997}. Indistinguishable
photon wavefunctions, such as (ideal) entangled photon pairs, are
symmetric with respect to exchange. This is reflected in the generic
form of wave-function using a symmetrized pair amplitude $\vert\Psi_{\left\{ 2\right\} }\rangle=\frac{1}{\sqrt{2}}\int d\omega_{a}d\omega_{b}\left[\Phi\left(\omega_{a},\omega_{b}\right)+\Phi\left(\omega_{b},\omega_{a}\right)\right]a^{\dagger}\left(\omega_{a}\right)b^{\dagger}\left(\omega_{b}\right)\vac,$
where $a\left(\omega\right)$ $\left(a^{\dagger}\left(\omega\right)\right)$
and $b\left(\omega\right)$ $\left(b^{\dagger}\left(\omega\right)\right)$
are annihilation (creation) photon operators applied on the vacuum
$\vac$. In practice, entangled photon pairs are distinguishable owing
to variations in the quantum channel responsible for their generation.
Orthogonally polarized photon wave packets of entangled pair produced
in a type-II parametric down conversion using an ultrashort pump pulse,
may be rather distinguishable. Each polarization has a different bandwidth
due to the dispersion characteristics of the birefringent crystal
\citep{Branning_1999}. The time-frequency signature invokes some
degree of distinguishability resulting in reduced interference contrast
due to a nonvanishing exchange phase. Interestingly, the exchange
phase of such entangled pair can be set in a controlled manner using
the Michelson interferometer setup, producing a $\theta$--symmetrized
amplitude \citep{Branning_1999},

\begin{equation}
\vert\Psi_{\left\{ 2\right\} }\rangle=\int d\omega_{a}d\omega_{b}\Phi_{\theta}\left(\omega_{a},\omega_{b}\right)a^{\dagger}\left(\omega_{a}\right)b^{\dagger}\left(\omega_{b}\right)\vac,
\end{equation}

\noindent where $\Phi_{\theta}\left(\omega_{a},\omega_{b}\right)=\left[\Phi\left(\omega_{a},\omega_{b}\right)+e^{i\theta}\Phi\left(\omega_{a},\omega_{b}\right)\right]/\sqrt{2}$.
The exchange phase can thus be used to manipulate photonic pathways
\footnote{S.Asban and S. Mukamel, Exchange Phase Cycling -- pathway selection
in Hong-Ou-Mandel Interferometric Spectroscopy. \emph{In preparation}}. The pathway in which photon $a$ is coupled to the matter at time
$t_{1}$ preceding its entangled counterpart $\left(\omega_{a};t_{1}\right)\rightarrow\left(\omega_{b};t_{2}\right)$,
and the opposite trajectory $\left(\omega_{b};t_{1}\right)\rightarrow\left(\omega_{a};t_{2}\right)$,
carry a valuable phase difference. Repeating the measurement with
different values of $\theta$ renders a set of signals from which
single light-matter interaction pathways can be isolated (pathway
selectivity). Generally, preparation interferometric procedures can
extend this notion to an $N$ photons amplitudes. N photons exchange
phase cycling procedure introduces ${N \choose 2}$ independent phases
using the amplitude

\noindent 
\begin{equation}
\Phi_{\boldsymbol{\Theta}}\left(\omega_{1},...,\omega_{N}\right)={\cal N}^{-1}\sum_{\left\{ i,j\right\} }e^{i\theta_{ij}}\hat{{\cal P}}_{ij}\Phi\left(\omega_{1},...,\omega_{N}\right),
\end{equation}

\noindent summing over all pair permutations $\left\{ i,j\right\} $
with the normalization ${\cal N}=\sqrt{{N \choose 2}+1}$. $\hat{{\cal P}}_{ij}$
is the exchange operator between the $i$ and $j$ photons $\hat{{\cal P}}_{ij}\Phi_{\boldsymbol{\Theta}}\left(\omega_{1},...,\omega_{i},...,\omega_{j},...,\omega_{N}\right)=\Phi_{\boldsymbol{\Theta}}\left(\omega_{1},...,\omega_{j},...,\omega_{i},...,\omega_{N}\right)$.
Generating a set of signals with independent exchange phases, one
may independently access different pathways.


\subsection{Time-domain QED -- ultrafast pathway switching \label{Sec.: Time-domain-QED}}

Remarkable progress in time domain detection techniques, allows the
observation of quantum electrodynamic (QED) processes such as electric-field
vacuum fluctuations in subcycle scale \citep{Riek_2015,Riek_2017}
and bunching at the femtosecond timescale \citep{Boitier_2009}. This
offers novel experimental possibilities, such as unraveling light-matter
(spontaneous) pathways and sorting between relaxation mechanisms.

A Liouville pathway represents a distinct time ordering of events.
Thus, coincidence measurements with a controlled delay, may prove
useful for discriminating light-matter absorption-emission sequences.
It is possible to recover the temporal profile of each photon of an
entangled pair using setups such as the one reported in \citep{Kuzucu_2008,MacLean_2018}.
Similar to the exchange-phase-cycling protocols, it relies on distinguishability
to sort photonic degrees of freedom at the detection process. In .
$\text{\ref{Fig 5}}$ we demonstrate this principle using polarization
sorting of unidirectional entangled photon-pair. Controlled distinguishability
can be employed by $\theta$--symmetrization, or by applying more
sophisticated single photon phase shaping techniques using electrooptic
modulation \citep{Specht_2009}.

All possible pathways contribute to the quantum state of the light-matter
system. Elimination of multiple pathways is possible by applying a
Fock state with fixed number of photons in conjunction with ultrafast
time-domain coincidence detection. To demonstrate this, we consider
two photons $N=2$ as shown in Fig. $\text{\ref{Fig 5}}$ that undergo
a coupling to matter and then detected in coincidence. Time domain
scanning technique is employed based on up-conversion process \citep{Kuzucu_2008,MacLean_2018}.
By fixing the detection time, only events in which both photons arrive
simultaneously are counted. We define the temporally gated coincidence
count \citep{Dorfman_2012}

\begin{subequations}
{\small{}{} 
\begin{align}
{\cal C}\left(\theta,\bar{t}_{a},\bar{t}_{b}\right) & =\int dt_{a}dt_{b}{\cal D}\left(\bar{t}_{a},t_{a}\right){\cal D}\left(\bar{t}_{b},t_{b}\right)W_{B}^{\left(2\right)}\left(t_{a},t_{b}\right),\label{TD Coincidence}\\
W_{B}^{\left(2\right)} & =\left\langle {\cal T}{\cal E}_{a,R}^{\dagger}\left(t_{a}\right){\cal E}_{b,R}^{\dagger}\left(t_{b}\right){\cal E}_{b,L}\left(t_{b}\right){\cal E}_{a,L}\left(t_{a}\right)\right.\label{Bare TD coincidence}\\
\times & \left.\exp\left\{ -\frac{i}{\hbar}\underset{-\infty}{\overset{t^{*}}{\int}}du\,H{}_{int,-}\left(u\right)\right\} \right\rangle .\nonumber 
\end{align}
}
\end{subequations}

\noindent Here ${\cal E}_{R}\left({\cal E}_{L}\right)$ denotes the
electromagnetic field superoperator that acts on a Hilbert space operator
$A$ from the right (left), ${\cal E}_{R}A\equiv A{\cal E}$ $\left({\cal E}_{L}A\equiv{\cal E}A\right)$.
Eq. $\text{\ref{TD Coincidence}}$ describes the the time-domain gated
coincidence, given by integration over the bare signal given in Eq.
$\text{ \ref{Bare TD coincidence}}$ weighted by the temporal gating
functions ${\cal D}_{i}\left(\bar{t}_{i},t_{i}\right)=\left|F_{t}\left(t_{i},\bar{t}_{i}\right)\right|^{2}$
which are determined by the pump temporal envelope. Since both photon
emission events are spontaneous, no special meaning is attributed
to the arrival time of a single photon, the time difference between
detection events reveals the time ordering of emission events. This
suggests an additional signal, that can be obtained experimentally
by scanning and summing all coincidence events in which the relative
pump delay $\tau=\tau_{b}-\tau_{a}$ is fixed. This corresponds to
integration over the last interaction time of a temporally gated coincidence
count, defining the signal

\begin{equation}
{\cal S}\left(\tau\right)=\int d\bar{t}_{a}{\cal C}\left(\theta,\bar{t}_{a},\bar{t}_{a}+\tau\right).
\end{equation}

When the pump is shorter than the measured photons wavepackets as
depicted in Fig. $\text{\ref{Fig 5}c},$each emission event is associated
with a single detection, eliminating reversed order processes contributions
in the coincidence count. Strikingly, when applied together with the
$\theta$ -- symmetrized amplitude as in Fig. $\text{\ref{Fig 5}}$b,
the signal is sensitive to the exchange in the order of the interactions
(exchanging intermediate blue and yellow arrows). Thus, potentially
sorting between pathways in addition to reversed emission discrimination.



\begin{figure}[th]
\begin{centering}
\includegraphics[bb=20bp 0bp 675bp 405bp,clip,scale=0.4]{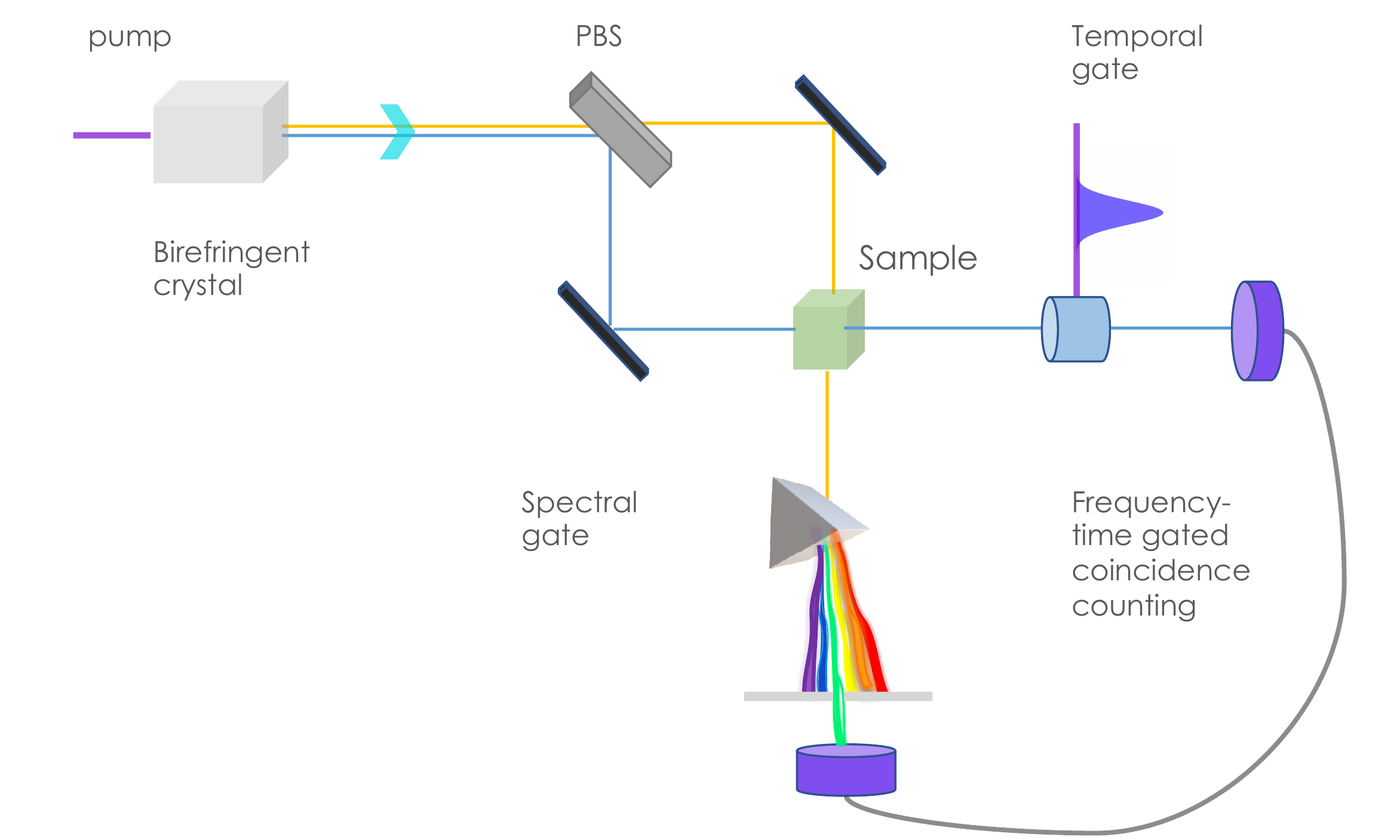}
\par\end{centering}
\caption{Frequency-time gated coincidence counting setup. Entangled photon
pair separated using a polarization beam splitter (PBS). Then coupled
to a sample and detected individually. One photon is characterized
spectrally while the other in time domain. The photons are ultimately
measured in coincidence, producing two dimensional spectra-temporal
information with superior joint resolution compared with classical
sources. \label{Fig 6}}
\end{figure}



\subsection{Time-frequency coincidence of entangled photons \label{Joint resolutions}}

Joint properties of systems described by an entangled state are not
necessarily constrained by uncertainty restrictions that apply to
single systems. By exploiting some distinguishability handles (polarization,
color, etc.), it is possible measure simultaneously conjugate properties
in an apparent violation of uncertainty relations, e.g., joint position-momentum
detection of EPR states (See Sec. 4 of the SM). Such phenomenon is
easily demystified using the appropriate definitions for conjugate
quantities in terms of the many-body wavefunction \citep{Howell_2004}.
These type of nonlocal effects can be further employed to perform
quantum microscopy and spectroscopy with unprecedented joint resolutions.

In the setup described in Fig. $\text{\ref{Fig 6}}$, time-frequency
entangled photon pair is coupled to matter. One photon is detected
in the time domain using ultrafast up-conversion detection technique,
revealing its temporal profile. Its entangled counterpart is spectrally
gated, thus recovering its frequency profile. The photons are measured
in coincidence,

{\footnotesize{}{} 
\begin{align}
{\cal C} & \left(\bar{t}_{s},\bar{\omega}_{i}\right)=\left\langle E_{s,R}^{\dagger}\left(\bar{t}_{s}\right)E_{i,R}^{\dagger}\left(\bar{\omega}_{i}\right)E_{i,L}\left(\bar{\omega}_{i}\right)E_{s,L}\left(\bar{t}_{s}\right)\right.\\
 & \left.\exp\left\{ -\frac{i}{\hbar}\underset{-\infty}{\overset{t^{*}}{\int}}duH{}_{\text{int},-}\left(u\right)\right\} \right\rangle .\nonumber 
\end{align}
}{\footnotesize\par}

\noindent Here $\bar{t}_{s}$ and $\bar{\omega}_{i}$ are the scanned
time and frequency covered by the respective gates. From the initial
frequency correlations of the photons, some knowledge can be obtained
regarding the spectral spread of the temporally gated photon, beyond
the minimal time-frequency uncertainty. This is possible thanks to
the frequency gating of its counterpart, combined with the initial
nonlocal correlations. Complementary information is obtained for the
frequency gated photon. The two dimensional time-frequency map, potentially
exhibits superior joint resolution. The main control parameters are
then the gating functions and the initial state of the light (pump
bandwidth and crystal length).


\section{Summary \label{Sec: Summary}}

This perspective article surveys several schemes in which interferometry
can be combined with quantum states of light in spectroscopy applications.
We distinguish between linear and nonlinear interferometric techniques,
and describe them using the formalism of linear transformations; relating
the input to the output ports. Inclusion of matter degrees of freedom
along the electromagnetic flux lines breaks the symmetries of these
linear transformation and induces photocurrent that carries matter
information. For simplicity, we have considered a colinear propagation
direction after the light matter interaction. Phase matching conditions,
can generate additional radiation directions induced by nonlinear
spontaneous processes, that can be associated with altered transformations.
Our algebraic-geometric description of interferometric building blocks,
offers a simple explanation for the nonlocal light-matter state after
the coupling in terms of vector rotations. These transformations can
be employed either before, or after the coupling to matter. These
affect the resulting observables, and may give access to different
processes as shown above.

Quantum interferometry combined with sequences of light-matter interactions
can single out noncausal contributions to nonlinear response functions.
Usually such contributions are uniformly summed when classical light
probes are involved. Quantum states of light offer sensitivity to
the order of events and thus give weights to different pathways. We
identify in this case the excess (quantum) information with pathway
selectivity. Distinguishability between pathways can be used to separate
decay channels in many-body systems, sorting them separately in experiments.
OTOC contribute naturally to the interferometric-spectroscopy signals.
Distinct OTOC can be extracted in full and measured by ultrafast pulse
sequences. Ranging from a single molecule to many-body systems, OTOC
tells a story regarding the quantum information scrambling; how a
single perturbation that propagates through a quantum system affects
different degrees of freedom. Characterizing such behavior becomes
increasingly important for materials designed for novel quantum technologies.
Also for detecting quantum coherent pathways in systems in which it
is not clear whether there are any. From a theoretical point of view,
such analysis poses an interesting inference-challenge for few-photon
detection of distant atomic processes occurring in curved space-time
(which are outside the scope of this perspective).

\begin{acknowledgments}
The support of the National Science Foundation Grant CHE-1953045 is
gratefully acknowledged 
\end{acknowledgments}

\bibliographystyle{apsrev4-2}
\bibliography{Perspective}

\begin{thebibliography}{89}%
\makeatletter
\providecommand \@ifxundefined [1]{%
 \@ifx{#1\undefined}
}%
\providecommand \@ifnum [1]{%
 \ifnum #1\expandafter \@firstoftwo
 \else \expandafter \@secondoftwo
 \fi
}%
\providecommand \@ifx [1]{%
 \ifx #1\expandafter \@firstoftwo
 \else \expandafter \@secondoftwo
 \fi
}%
\providecommand \natexlab [1]{#1}%
\providecommand \enquote  [1]{``#1''}%
\providecommand \bibnamefont  [1]{#1}%
\providecommand \bibfnamefont [1]{#1}%
\providecommand \citenamefont [1]{#1}%
\providecommand \href@noop [0]{\@secondoftwo}%
\providecommand \href [0]{\begingroup \@sanitize@url \@href}%
\providecommand \@href[1]{\@@startlink{#1}\@@href}%
\providecommand \@@href[1]{\endgroup#1\@@endlink}%
\providecommand \@sanitize@url [0]{\catcode `\\12\catcode `\$12\catcode
  `\&12\catcode `\#12\catcode `\^12\catcode `\_12\catcode `\%12\relax}%
\providecommand \@@startlink[1]{}%
\providecommand \@@endlink[0]{}%
\providecommand \url  [0]{\begingroup\@sanitize@url \@url }%
\providecommand \@url [1]{\endgroup\@href {#1}{\urlprefix }}%
\providecommand \urlprefix  [0]{URL }%
\providecommand \Eprint [0]{\href }%
\providecommand \doibase [0]{https://doi.org/}%
\providecommand \selectlanguage [0]{\@gobble}%
\providecommand \bibinfo  [0]{\@secondoftwo}%
\providecommand \bibfield  [0]{\@secondoftwo}%
\providecommand \translation [1]{[#1]}%
\providecommand \BibitemOpen [0]{}%
\providecommand \bibitemStop [0]{}%
\providecommand \bibitemNoStop [0]{.\EOS\space}%
\providecommand \EOS [0]{\spacefactor3000\relax}%
\providecommand \BibitemShut  [1]{\csname bibitem#1\endcsname}%
\let\auto@bib@innerbib\@empty
\bibitem [{\citenamefont {Mukamel}\ \emph {et~al.}(2020)\citenamefont
  {Mukamel}, \citenamefont {Freyberger}, \citenamefont {Schleich},
  \citenamefont {Bellini}, \citenamefont {Zavatta}, \citenamefont {Leuchs},
  \citenamefont {Silberhorn}, \citenamefont {Boyd}, \citenamefont
  {S{\'a}nchez-Soto}, \citenamefont {Stefanov} \emph {et~al.}}]{muk20}%
  \BibitemOpen
  \bibfield  {author} {\bibinfo {author} {\bibfnamefont {S.}~\bibnamefont
  {Mukamel}}, \bibinfo {author} {\bibfnamefont {M.}~\bibnamefont {Freyberger}},
  \bibinfo {author} {\bibfnamefont {W.}~\bibnamefont {Schleich}}, \bibinfo
  {author} {\bibfnamefont {M.}~\bibnamefont {Bellini}}, \bibinfo {author}
  {\bibfnamefont {A.}~\bibnamefont {Zavatta}}, \bibinfo {author} {\bibfnamefont
  {G.}~\bibnamefont {Leuchs}}, \bibinfo {author} {\bibfnamefont
  {C.}~\bibnamefont {Silberhorn}}, \bibinfo {author} {\bibfnamefont {R.~W.}\
  \bibnamefont {Boyd}}, \bibinfo {author} {\bibfnamefont {L.~L.}\ \bibnamefont
  {S{\'a}nchez-Soto}}, \bibinfo {author} {\bibfnamefont {A.}~\bibnamefont
  {Stefanov}}, \emph {et~al.},\ }\href@noop {} {\bibfield  {journal} {\bibinfo
  {journal} {Journal of Physics B: Atomic, Molecular and Optical Physics}\
  }\textbf {\bibinfo {volume} {53}},\ \bibinfo {pages} {072002} (\bibinfo
  {year} {2020})}\BibitemShut {NoStop}%
\bibitem [{\citenamefont {Dorfman}\ \emph {et~al.}(2014)\citenamefont
  {Dorfman}, \citenamefont {Schlawin},\ and\ \citenamefont {Mukamel}}]{dor14}%
  \BibitemOpen
  \bibfield  {author} {\bibinfo {author} {\bibfnamefont {K.~E.}\ \bibnamefont
  {Dorfman}}, \bibinfo {author} {\bibfnamefont {F.}~\bibnamefont {Schlawin}},\
  and\ \bibinfo {author} {\bibfnamefont {S.}~\bibnamefont {Mukamel}},\
  }\href@noop {} {\bibfield  {journal} {\bibinfo  {journal} {The journal of
  physical chemistry letters}\ }\textbf {\bibinfo {volume} {5}},\ \bibinfo
  {pages} {2843} (\bibinfo {year} {2014})}\BibitemShut {NoStop}%
\bibitem [{\citenamefont {Kalashnikov}\ \emph {et~al.}(2016)\citenamefont
  {Kalashnikov}, \citenamefont {Paterova}, \citenamefont {Kulik},\ and\
  \citenamefont {Krivitsky}}]{kal16}%
  \BibitemOpen
  \bibfield  {author} {\bibinfo {author} {\bibfnamefont {D.~A.}\ \bibnamefont
  {Kalashnikov}}, \bibinfo {author} {\bibfnamefont {A.~V.}\ \bibnamefont
  {Paterova}}, \bibinfo {author} {\bibfnamefont {S.~P.}\ \bibnamefont
  {Kulik}},\ and\ \bibinfo {author} {\bibfnamefont {L.~A.}\ \bibnamefont
  {Krivitsky}},\ }\href@noop {} {\bibfield  {journal} {\bibinfo  {journal}
  {Nature Photonics}\ }\textbf {\bibinfo {volume} {10}},\ \bibinfo {pages} {98}
  (\bibinfo {year} {2016})}\BibitemShut {NoStop}%
\bibitem [{\citenamefont {Humphreys}\ \emph {et~al.}(2013)\citenamefont
  {Humphreys}, \citenamefont {Barbieri}, \citenamefont {Datta},\ and\
  \citenamefont {Walmsley}}]{hum13}%
  \BibitemOpen
  \bibfield  {author} {\bibinfo {author} {\bibfnamefont {P.~C.}\ \bibnamefont
  {Humphreys}}, \bibinfo {author} {\bibfnamefont {M.}~\bibnamefont {Barbieri}},
  \bibinfo {author} {\bibfnamefont {A.}~\bibnamefont {Datta}},\ and\ \bibinfo
  {author} {\bibfnamefont {I.~A.}\ \bibnamefont {Walmsley}},\ }\href@noop {}
  {\bibfield  {journal} {\bibinfo  {journal} {Physical review letters}\
  }\textbf {\bibinfo {volume} {111}},\ \bibinfo {pages} {070403} (\bibinfo
  {year} {2013})}\BibitemShut {NoStop}%
\bibitem [{\citenamefont {Giovannetti}\ \emph {et~al.}(2011)\citenamefont
  {Giovannetti}, \citenamefont {Lloyd},\ and\ \citenamefont {Maccone}}]{gio11}%
  \BibitemOpen
  \bibfield  {author} {\bibinfo {author} {\bibfnamefont {V.}~\bibnamefont
  {Giovannetti}}, \bibinfo {author} {\bibfnamefont {S.}~\bibnamefont {Lloyd}},\
  and\ \bibinfo {author} {\bibfnamefont {L.}~\bibnamefont {Maccone}},\
  }\href@noop {} {\bibfield  {journal} {\bibinfo  {journal} {Nature photonics}\
  }\textbf {\bibinfo {volume} {5}},\ \bibinfo {pages} {222} (\bibinfo {year}
  {2011})}\BibitemShut {NoStop}%
\bibitem [{\citenamefont {Herrera}\ \emph {et~al.}(2014)\citenamefont
  {Herrera}, \citenamefont {Peropadre}, \citenamefont {Pachon}, \citenamefont
  {Saikin},\ and\ \citenamefont {Aspuru-Guzik}}]{her14}%
  \BibitemOpen
  \bibfield  {author} {\bibinfo {author} {\bibfnamefont {F.}~\bibnamefont
  {Herrera}}, \bibinfo {author} {\bibfnamefont {B.}~\bibnamefont {Peropadre}},
  \bibinfo {author} {\bibfnamefont {L.~A.}\ \bibnamefont {Pachon}}, \bibinfo
  {author} {\bibfnamefont {S.~K.}\ \bibnamefont {Saikin}},\ and\ \bibinfo
  {author} {\bibfnamefont {A.}~\bibnamefont {Aspuru-Guzik}},\ }\href@noop {}
  {\bibfield  {journal} {\bibinfo  {journal} {The journal of physical chemistry
  letters}\ }\textbf {\bibinfo {volume} {5}},\ \bibinfo {pages} {3708}
  (\bibinfo {year} {2014})}\BibitemShut {NoStop}%
\bibitem [{\citenamefont {Silberhorn}\ \emph {et~al.}(2001)\citenamefont
  {Silberhorn}, \citenamefont {Lam}, \citenamefont {Weiss}, \citenamefont
  {K{\"o}nig}, \citenamefont {Korolkova},\ and\ \citenamefont
  {Leuchs}}]{sil01}%
  \BibitemOpen
  \bibfield  {author} {\bibinfo {author} {\bibfnamefont {C.}~\bibnamefont
  {Silberhorn}}, \bibinfo {author} {\bibfnamefont {P.~K.}\ \bibnamefont {Lam}},
  \bibinfo {author} {\bibfnamefont {O.}~\bibnamefont {Weiss}}, \bibinfo
  {author} {\bibfnamefont {F.}~\bibnamefont {K{\"o}nig}}, \bibinfo {author}
  {\bibfnamefont {N.}~\bibnamefont {Korolkova}},\ and\ \bibinfo {author}
  {\bibfnamefont {G.}~\bibnamefont {Leuchs}},\ }\href@noop {} {\bibfield
  {journal} {\bibinfo  {journal} {Physical Review Letters}\ }\textbf {\bibinfo
  {volume} {86}},\ \bibinfo {pages} {4267} (\bibinfo {year}
  {2001})}\BibitemShut {NoStop}%
\bibitem [{\citenamefont {Varnavski}\ \emph
  {et~al.}(2017{\natexlab{a}})\citenamefont {Varnavski}, \citenamefont
  {Pinsky},\ and\ \citenamefont {Goodson~III}}]{var17}%
  \BibitemOpen
  \bibfield  {author} {\bibinfo {author} {\bibfnamefont {O.}~\bibnamefont
  {Varnavski}}, \bibinfo {author} {\bibfnamefont {B.}~\bibnamefont {Pinsky}},\
  and\ \bibinfo {author} {\bibfnamefont {T.}~\bibnamefont {Goodson~III}},\
  }\href@noop {} {\bibfield  {journal} {\bibinfo  {journal} {The journal of
  physical chemistry letters}\ }\textbf {\bibinfo {volume} {8}},\ \bibinfo
  {pages} {388} (\bibinfo {year} {2017}{\natexlab{a}})}\BibitemShut {NoStop}%
\bibitem [{\citenamefont {Dorfman}\ \emph {et~al.}(2016)\citenamefont
  {Dorfman}, \citenamefont {Schlawin},\ and\ \citenamefont
  {Mukamel}}]{Dorfman_2016}%
  \BibitemOpen
  \bibfield  {author} {\bibinfo {author} {\bibfnamefont {K.~E.}\ \bibnamefont
  {Dorfman}}, \bibinfo {author} {\bibfnamefont {F.}~\bibnamefont {Schlawin}},\
  and\ \bibinfo {author} {\bibfnamefont {S.}~\bibnamefont {Mukamel}},\ }\href
  {https://doi.org/10.1103/RevModPhys.88.045008} {\bibfield  {journal}
  {\bibinfo  {journal} {Rev. Mod. Phys.}\ }\textbf {\bibinfo {volume} {88}},\
  \bibinfo {pages} {045008} (\bibinfo {year} {2016})}\BibitemShut {NoStop}%
\bibitem [{\citenamefont {Asban}\ \emph {et~al.}(2019)\citenamefont {Asban},
  \citenamefont {Dorfman},\ and\ \citenamefont {Mukamel}}]{Asban_2019}%
  \BibitemOpen
  \bibfield  {author} {\bibinfo {author} {\bibfnamefont {S.}~\bibnamefont
  {Asban}}, \bibinfo {author} {\bibfnamefont {K.~E.}\ \bibnamefont {Dorfman}},\
  and\ \bibinfo {author} {\bibfnamefont {S.}~\bibnamefont {Mukamel}},\ }\href
  {https://doi.org/10.1073/pnas.1904839116} {\bibfield  {journal} {\bibinfo
  {journal} {Proceedings of the National Academy of Sciences}\ }\textbf
  {\bibinfo {volume} {116}},\ \bibinfo {pages} {11673} (\bibinfo {year}
  {2019})}\BibitemShut {NoStop}%
\bibitem [{\citenamefont {Brida}\ \emph {et~al.}(2010)\citenamefont {Brida},
  \citenamefont {Genovese},\ and\ \citenamefont {Berchera}}]{bri10}%
  \BibitemOpen
  \bibfield  {author} {\bibinfo {author} {\bibfnamefont {G.}~\bibnamefont
  {Brida}}, \bibinfo {author} {\bibfnamefont {M.}~\bibnamefont {Genovese}},\
  and\ \bibinfo {author} {\bibfnamefont {I.~R.}\ \bibnamefont {Berchera}},\
  }\href@noop {} {\bibfield  {journal} {\bibinfo  {journal} {Nature Photonics}\
  }\textbf {\bibinfo {volume} {4}},\ \bibinfo {pages} {227} (\bibinfo {year}
  {2010})}\BibitemShut {NoStop}%
\bibitem [{\citenamefont {Nielsen}\ and\ \citenamefont
  {Chuang}(2010)}]{nielsen_chuang_2010}%
  \BibitemOpen
  \bibfield  {author} {\bibinfo {author} {\bibfnamefont {M.~A.}\ \bibnamefont
  {Nielsen}}\ and\ \bibinfo {author} {\bibfnamefont {I.~L.}\ \bibnamefont
  {Chuang}},\ }\href {https://doi.org/10.1017/CBO9780511976667} {\emph
  {\bibinfo {title} {{Quantum Computation and Quantum Information: 10th
  Anniversary Edition}}}}\ (\bibinfo  {publisher} {Cambridge University
  Press},\ \bibinfo {year} {2010})\BibitemShut {NoStop}%
\bibitem [{\citenamefont {Helstrom}(1976)}]{Helstrom_1976}%
  \BibitemOpen
  \bibfield  {author} {\bibinfo {author} {\bibfnamefont {C.~W.}\ \bibnamefont
  {Helstrom}},\ }\href {https://books.google.com/books?id=Ne3iT\_QLcsMC} {\emph
  {\bibinfo {title} {Quantum Detection and Estimation Theory}}},\ ISSN\
  (\bibinfo  {publisher} {Elsevier Science},\ \bibinfo {year}
  {1976})\BibitemShut {NoStop}%
\bibitem [{\citenamefont {Rarity}\ \emph {et~al.}(1990)\citenamefont {Rarity},
  \citenamefont {Tapster}, \citenamefont {Jakeman}, \citenamefont {Larchuk},
  \citenamefont {Campos}, \citenamefont {Teich},\ and\ \citenamefont
  {Saleh}}]{rar90}%
  \BibitemOpen
  \bibfield  {author} {\bibinfo {author} {\bibfnamefont {J.}~\bibnamefont
  {Rarity}}, \bibinfo {author} {\bibfnamefont {P.}~\bibnamefont {Tapster}},
  \bibinfo {author} {\bibfnamefont {E.}~\bibnamefont {Jakeman}}, \bibinfo
  {author} {\bibfnamefont {T.}~\bibnamefont {Larchuk}}, \bibinfo {author}
  {\bibfnamefont {R.}~\bibnamefont {Campos}}, \bibinfo {author} {\bibfnamefont
  {M.}~\bibnamefont {Teich}},\ and\ \bibinfo {author} {\bibfnamefont
  {B.}~\bibnamefont {Saleh}},\ }\href@noop {} {\bibfield  {journal} {\bibinfo
  {journal} {Physical review letters}\ }\textbf {\bibinfo {volume} {65}},\
  \bibinfo {pages} {1348} (\bibinfo {year} {1990})}\BibitemShut {NoStop}%
\bibitem [{\citenamefont {Hong}\ \emph
  {et~al.}(1987{\natexlab{a}})\citenamefont {Hong}, \citenamefont {Ou},\ and\
  \citenamefont {Mandel}}]{hom87}%
  \BibitemOpen
  \bibfield  {author} {\bibinfo {author} {\bibfnamefont {C.-K.}\ \bibnamefont
  {Hong}}, \bibinfo {author} {\bibfnamefont {Z.-Y.}\ \bibnamefont {Ou}},\ and\
  \bibinfo {author} {\bibfnamefont {L.}~\bibnamefont {Mandel}},\ }\href@noop {}
  {\bibfield  {journal} {\bibinfo  {journal} {Physical review letters}\
  }\textbf {\bibinfo {volume} {59}},\ \bibinfo {pages} {2044} (\bibinfo {year}
  {1987}{\natexlab{a}})}\BibitemShut {NoStop}%
\bibitem [{\citenamefont {Raymer}\ \emph {et~al.}(2013)\citenamefont {Raymer},
  \citenamefont {Marcus}, \citenamefont {Widom},\ and\ \citenamefont
  {Vitullo}}]{ray13}%
  \BibitemOpen
  \bibfield  {author} {\bibinfo {author} {\bibfnamefont {M.}~\bibnamefont
  {Raymer}}, \bibinfo {author} {\bibfnamefont {A.~H.}\ \bibnamefont {Marcus}},
  \bibinfo {author} {\bibfnamefont {J.~R.}\ \bibnamefont {Widom}},\ and\
  \bibinfo {author} {\bibfnamefont {D.~L.}\ \bibnamefont {Vitullo}},\
  }\href@noop {} {\bibfield  {journal} {\bibinfo  {journal} {The Journal of
  Physical Chemistry B}\ }\textbf {\bibinfo {volume} {117}},\ \bibinfo {pages}
  {15559} (\bibinfo {year} {2013})}\BibitemShut {NoStop}%
\bibitem [{\citenamefont {Kalachev}\ \emph {et~al.}(2008)\citenamefont
  {Kalachev}, \citenamefont {Kalashnikov}, \citenamefont {Kalinkin},
  \citenamefont {Mitrofanova}, \citenamefont {Shkalikov},\ and\ \citenamefont
  {Samartsev}}]{kal08}%
  \BibitemOpen
  \bibfield  {author} {\bibinfo {author} {\bibfnamefont {A.}~\bibnamefont
  {Kalachev}}, \bibinfo {author} {\bibfnamefont {D.}~\bibnamefont
  {Kalashnikov}}, \bibinfo {author} {\bibfnamefont {A.}~\bibnamefont
  {Kalinkin}}, \bibinfo {author} {\bibfnamefont {T.}~\bibnamefont
  {Mitrofanova}}, \bibinfo {author} {\bibfnamefont {A.}~\bibnamefont
  {Shkalikov}},\ and\ \bibinfo {author} {\bibfnamefont {V.}~\bibnamefont
  {Samartsev}},\ }\href@noop {} {\bibfield  {journal} {\bibinfo  {journal}
  {Laser Physics Letters}\ }\textbf {\bibinfo {volume} {5}},\ \bibinfo {pages}
  {600} (\bibinfo {year} {2008})}\BibitemShut {NoStop}%
\bibitem [{\citenamefont {Lavoie}\ \emph {et~al.}(2020)\citenamefont {Lavoie},
  \citenamefont {Landes}, \citenamefont {Tamimi}, \citenamefont {Smith},
  \citenamefont {Marcus},\ and\ \citenamefont {Raymer}}]{Lavoie_2020}%
  \BibitemOpen
  \bibfield  {author} {\bibinfo {author} {\bibfnamefont {J.}~\bibnamefont
  {Lavoie}}, \bibinfo {author} {\bibfnamefont {T.}~\bibnamefont {Landes}},
  \bibinfo {author} {\bibfnamefont {A.}~\bibnamefont {Tamimi}}, \bibinfo
  {author} {\bibfnamefont {B.~J.}\ \bibnamefont {Smith}}, \bibinfo {author}
  {\bibfnamefont {A.~H.}\ \bibnamefont {Marcus}},\ and\ \bibinfo {author}
  {\bibfnamefont {M.~G.}\ \bibnamefont {Raymer}},\ }\href
  {https://doi.org/https://doi.org/10.1002/qute.201900114} {\bibfield
  {journal} {\bibinfo  {journal} {Advanced Quantum Technologies}\ }\textbf
  {\bibinfo {volume} {3}},\ \bibinfo {pages} {1900114} (\bibinfo {year}
  {2020})}\BibitemShut {NoStop}%
\bibitem [{\citenamefont {Caves}(1981)}]{Caves_1981}%
  \BibitemOpen
  \bibfield  {author} {\bibinfo {author} {\bibfnamefont {C.~M.}\ \bibnamefont
  {Caves}},\ }\href {https://doi.org/10.1103/PhysRevD.23.1693} {\bibfield
  {journal} {\bibinfo  {journal} {Phys. Rev. D}\ }\textbf {\bibinfo {volume}
  {23}},\ \bibinfo {pages} {1693} (\bibinfo {year} {1981})}\BibitemShut
  {NoStop}%
\bibitem [{\citenamefont {Tse}\ \emph {et~al.}(2019)\citenamefont {Tse},
  \citenamefont {Yu}, \citenamefont {Kijbunchoo}, \citenamefont
  {Fernandez-Galiana}, \citenamefont {Dupej}, \citenamefont {Barsotti},
  \citenamefont {Blair}, \citenamefont {Brown}, \citenamefont {Dwyer},
  \citenamefont {Effler}, \citenamefont {Evans}, \citenamefont {Fritschel},
  \citenamefont {Frolov}, \citenamefont {Green}, \citenamefont {Mansell},
  \citenamefont {Matichard}, \citenamefont {Mavalvala}, \citenamefont
  {McClelland}, \citenamefont {McCuller}, \citenamefont {McRae}, \citenamefont
  {Miller}, \citenamefont {Mullavey}, \citenamefont {Oelker}, \citenamefont
  {Phinney}, \citenamefont {Sigg}, \citenamefont {Slagmolen}, \citenamefont
  {Vo}, \citenamefont {Ward}, \citenamefont {Whittle}, \citenamefont {Abbott},
  \citenamefont {Adams}, \citenamefont {Adhikari}, \citenamefont {Ananyeva},
  \citenamefont {Appert}, \citenamefont {Arai}, \citenamefont {Areeda},
  \citenamefont {Asali}, \citenamefont {Aston}, \citenamefont {Austin},
  \citenamefont {Baer}, \citenamefont {Ball}, \citenamefont {Ballmer},
  \citenamefont {Banagiri}, \citenamefont {Barker}, \citenamefont {Bartlett},
  \citenamefont {Berger}, \citenamefont {Betzwieser}, \citenamefont
  {Bhattacharjee}, \citenamefont {Billingsley}, \citenamefont {Biscans},
  \citenamefont {Blair}, \citenamefont {Bode}, \citenamefont {Booker},
  \citenamefont {Bork}, \citenamefont {Bramley}, \citenamefont {Brooks},
  \citenamefont {Buikema}, \citenamefont {Cahillane}, \citenamefont {Cannon},
  \citenamefont {Chen}, \citenamefont {Ciobanu}, \citenamefont {Clara},
  \citenamefont {Cooper}, \citenamefont {Corley}, \citenamefont {Countryman},
  \citenamefont {Covas}, \citenamefont {Coyne}, \citenamefont {Datrier},
  \citenamefont {Davis}, \citenamefont {Di~Fronzo}, \citenamefont {Driggers},
  \citenamefont {Etzel}, \citenamefont {Evans}, \citenamefont {Feicht},
  \citenamefont {Fulda}, \citenamefont {Fyffe}, \citenamefont {Giaime},
  \citenamefont {Giardina}, \citenamefont {Godwin}, \citenamefont {Goetz},
  \citenamefont {Gras}, \citenamefont {Gray}, \citenamefont {Gray},
  \citenamefont {Gupta}, \citenamefont {Gustafson}, \citenamefont {Gustafson},
  \citenamefont {Hanks}, \citenamefont {Hanson}, \citenamefont {Hardwick},
  \citenamefont {Hasskew}, \citenamefont {Heintze}, \citenamefont
  {Helmling-Cornell}, \citenamefont {Holland}, \citenamefont {Jones},
  \citenamefont {Kandhasamy}, \citenamefont {Karki}, \citenamefont {Kasprzack},
  \citenamefont {Kawabe}, \citenamefont {King}, \citenamefont {Kissel},
  \citenamefont {Kumar}, \citenamefont {Landry}, \citenamefont {Lane},
  \citenamefont {Lantz}, \citenamefont {Laxen}, \citenamefont {Lecoeuche},
  \citenamefont {Leviton}, \citenamefont {Liu}, \citenamefont {Lormand},
  \citenamefont {Lundgren}, \citenamefont {Macas}, \citenamefont {MacInnis},
  \citenamefont {Macleod}, \citenamefont {M\'arka}, \citenamefont {M\'arka},
  \citenamefont {Martynov}, \citenamefont {Mason}, \citenamefont {Massinger},
  \citenamefont {McCarthy}, \citenamefont {McCormick}, \citenamefont {McIver},
  \citenamefont {Mendell}, \citenamefont {Merfeld}, \citenamefont {Merilh},
  \citenamefont {Meylahn}, \citenamefont {Mistry}, \citenamefont {Mittleman},
  \citenamefont {Moreno}, \citenamefont {Mow-Lowry}, \citenamefont {Mozzon},
  \citenamefont {Nelson}, \citenamefont {Nguyen}, \citenamefont {Nuttall},
  \citenamefont {Oberling}, \citenamefont {Oram}, \citenamefont {O'Reilly},
  \citenamefont {Osthelder}, \citenamefont {Ottaway}, \citenamefont {Overmier},
  \citenamefont {Palamos}, \citenamefont {Parker}, \citenamefont {Payne},
  \citenamefont {Pele}, \citenamefont {Perez}, \citenamefont {Pirello},
  \citenamefont {Radkins}, \citenamefont {Ramirez}, \citenamefont {Richardson},
  \citenamefont {Riles}, \citenamefont {Robertson}, \citenamefont {Rollins},
  \citenamefont {Romel}, \citenamefont {Romie}, \citenamefont {Ross},
  \citenamefont {Ryan}, \citenamefont {Sadecki}, \citenamefont {Sanchez},
  \citenamefont {Sanchez}, \citenamefont {Saravanan}, \citenamefont {Savage},
  \citenamefont {Schaetzl}, \citenamefont {Schnabel}, \citenamefont
  {Schofield}, \citenamefont {Schwartz}, \citenamefont {Sellers}, \citenamefont
  {Shaffer}, \citenamefont {Smith}, \citenamefont {Soni}, \citenamefont
  {Sorazu}, \citenamefont {Spencer}, \citenamefont {Strain}, \citenamefont
  {Sun}, \citenamefont {Szczepa\ifmmode~\acute{n}\else \'{n}\fi{}czyk},
  \citenamefont {Thomas}, \citenamefont {Thomas}, \citenamefont {Thorne},
  \citenamefont {Toland}, \citenamefont {Torrie}, \citenamefont {Traylor},
  \citenamefont {Urban}, \citenamefont {Vajente}, \citenamefont {Valdes},
  \citenamefont {Vander-Hyde}, \citenamefont {Veitch}, \citenamefont
  {Venkateswara}, \citenamefont {Venugopalan}, \citenamefont {Viets},
  \citenamefont {Vorvick}, \citenamefont {Wade}, \citenamefont {Warner},
  \citenamefont {Weaver}, \citenamefont {Weiss}, \citenamefont {Willke},
  \citenamefont {Wipf}, \citenamefont {Xiao}, \citenamefont {Yamamoto},
  \citenamefont {Yap}, \citenamefont {Yu}, \citenamefont {Zhang}, \citenamefont
  {Zucker},\ and\ \citenamefont {Zweizig}}]{LIGO_2019}%
  \BibitemOpen
  \bibfield  {author} {\bibinfo {author} {\bibfnamefont {M.}~\bibnamefont
  {Tse}}, \bibinfo {author} {\bibfnamefont {H.}~\bibnamefont {Yu}}, \bibinfo
  {author} {\bibfnamefont {N.}~\bibnamefont {Kijbunchoo}}, \bibinfo {author}
  {\bibfnamefont {A.}~\bibnamefont {Fernandez-Galiana}}, \bibinfo {author}
  {\bibfnamefont {P.}~\bibnamefont {Dupej}}, \bibinfo {author} {\bibfnamefont
  {L.}~\bibnamefont {Barsotti}}, \bibinfo {author} {\bibfnamefont {C.~D.}\
  \bibnamefont {Blair}}, \bibinfo {author} {\bibfnamefont {D.~D.}\ \bibnamefont
  {Brown}}, \bibinfo {author} {\bibfnamefont {S.~E.}\ \bibnamefont {Dwyer}},
  \bibinfo {author} {\bibfnamefont {A.}~\bibnamefont {Effler}}, \bibinfo
  {author} {\bibfnamefont {M.}~\bibnamefont {Evans}}, \bibinfo {author}
  {\bibfnamefont {P.}~\bibnamefont {Fritschel}}, \bibinfo {author}
  {\bibfnamefont {V.~V.}\ \bibnamefont {Frolov}}, \bibinfo {author}
  {\bibfnamefont {A.~C.}\ \bibnamefont {Green}}, \bibinfo {author}
  {\bibfnamefont {G.~L.}\ \bibnamefont {Mansell}}, \bibinfo {author}
  {\bibfnamefont {F.}~\bibnamefont {Matichard}}, \bibinfo {author}
  {\bibfnamefont {N.}~\bibnamefont {Mavalvala}}, \bibinfo {author}
  {\bibfnamefont {D.~E.}\ \bibnamefont {McClelland}}, \bibinfo {author}
  {\bibfnamefont {L.}~\bibnamefont {McCuller}}, \bibinfo {author}
  {\bibfnamefont {T.}~\bibnamefont {McRae}}, \bibinfo {author} {\bibfnamefont
  {J.}~\bibnamefont {Miller}}, \bibinfo {author} {\bibfnamefont
  {A.}~\bibnamefont {Mullavey}}, \bibinfo {author} {\bibfnamefont
  {E.}~\bibnamefont {Oelker}}, \bibinfo {author} {\bibfnamefont {I.~Y.}\
  \bibnamefont {Phinney}}, \bibinfo {author} {\bibfnamefont {D.}~\bibnamefont
  {Sigg}}, \bibinfo {author} {\bibfnamefont {B.~J.~J.}\ \bibnamefont
  {Slagmolen}}, \bibinfo {author} {\bibfnamefont {T.}~\bibnamefont {Vo}},
  \bibinfo {author} {\bibfnamefont {R.~L.}\ \bibnamefont {Ward}}, \bibinfo
  {author} {\bibfnamefont {C.}~\bibnamefont {Whittle}}, \bibinfo {author}
  {\bibfnamefont {R.}~\bibnamefont {Abbott}}, \bibinfo {author} {\bibfnamefont
  {C.}~\bibnamefont {Adams}}, \bibinfo {author} {\bibfnamefont {R.~X.}\
  \bibnamefont {Adhikari}}, \bibinfo {author} {\bibfnamefont {A.}~\bibnamefont
  {Ananyeva}}, \bibinfo {author} {\bibfnamefont {S.}~\bibnamefont {Appert}},
  \bibinfo {author} {\bibfnamefont {K.}~\bibnamefont {Arai}}, \bibinfo {author}
  {\bibfnamefont {J.~S.}\ \bibnamefont {Areeda}}, \bibinfo {author}
  {\bibfnamefont {Y.}~\bibnamefont {Asali}}, \bibinfo {author} {\bibfnamefont
  {S.~M.}\ \bibnamefont {Aston}}, \bibinfo {author} {\bibfnamefont
  {C.}~\bibnamefont {Austin}}, \bibinfo {author} {\bibfnamefont {A.~M.}\
  \bibnamefont {Baer}}, \bibinfo {author} {\bibfnamefont {M.}~\bibnamefont
  {Ball}}, \bibinfo {author} {\bibfnamefont {S.~W.}\ \bibnamefont {Ballmer}},
  \bibinfo {author} {\bibfnamefont {S.}~\bibnamefont {Banagiri}}, \bibinfo
  {author} {\bibfnamefont {D.}~\bibnamefont {Barker}}, \bibinfo {author}
  {\bibfnamefont {J.}~\bibnamefont {Bartlett}}, \bibinfo {author}
  {\bibfnamefont {B.~K.}\ \bibnamefont {Berger}}, \bibinfo {author}
  {\bibfnamefont {J.}~\bibnamefont {Betzwieser}}, \bibinfo {author}
  {\bibfnamefont {D.}~\bibnamefont {Bhattacharjee}}, \bibinfo {author}
  {\bibfnamefont {G.}~\bibnamefont {Billingsley}}, \bibinfo {author}
  {\bibfnamefont {S.}~\bibnamefont {Biscans}}, \bibinfo {author} {\bibfnamefont
  {R.~M.}\ \bibnamefont {Blair}}, \bibinfo {author} {\bibfnamefont
  {N.}~\bibnamefont {Bode}}, \bibinfo {author} {\bibfnamefont {P.}~\bibnamefont
  {Booker}}, \bibinfo {author} {\bibfnamefont {R.}~\bibnamefont {Bork}},
  \bibinfo {author} {\bibfnamefont {A.}~\bibnamefont {Bramley}}, \bibinfo
  {author} {\bibfnamefont {A.~F.}\ \bibnamefont {Brooks}}, \bibinfo {author}
  {\bibfnamefont {A.}~\bibnamefont {Buikema}}, \bibinfo {author} {\bibfnamefont
  {C.}~\bibnamefont {Cahillane}}, \bibinfo {author} {\bibfnamefont {K.~C.}\
  \bibnamefont {Cannon}}, \bibinfo {author} {\bibfnamefont {X.}~\bibnamefont
  {Chen}}, \bibinfo {author} {\bibfnamefont {A.~A.}\ \bibnamefont {Ciobanu}},
  \bibinfo {author} {\bibfnamefont {F.}~\bibnamefont {Clara}}, \bibinfo
  {author} {\bibfnamefont {S.~J.}\ \bibnamefont {Cooper}}, \bibinfo {author}
  {\bibfnamefont {K.~R.}\ \bibnamefont {Corley}}, \bibinfo {author}
  {\bibfnamefont {S.~T.}\ \bibnamefont {Countryman}}, \bibinfo {author}
  {\bibfnamefont {P.~B.}\ \bibnamefont {Covas}}, \bibinfo {author}
  {\bibfnamefont {D.~C.}\ \bibnamefont {Coyne}}, \bibinfo {author}
  {\bibfnamefont {L.~E.~H.}\ \bibnamefont {Datrier}}, \bibinfo {author}
  {\bibfnamefont {D.}~\bibnamefont {Davis}}, \bibinfo {author} {\bibfnamefont
  {C.}~\bibnamefont {Di~Fronzo}}, \bibinfo {author} {\bibfnamefont {J.~C.}\
  \bibnamefont {Driggers}}, \bibinfo {author} {\bibfnamefont {T.}~\bibnamefont
  {Etzel}}, \bibinfo {author} {\bibfnamefont {T.~M.}\ \bibnamefont {Evans}},
  \bibinfo {author} {\bibfnamefont {J.}~\bibnamefont {Feicht}}, \bibinfo
  {author} {\bibfnamefont {P.}~\bibnamefont {Fulda}}, \bibinfo {author}
  {\bibfnamefont {M.}~\bibnamefont {Fyffe}}, \bibinfo {author} {\bibfnamefont
  {J.~A.}\ \bibnamefont {Giaime}}, \bibinfo {author} {\bibfnamefont {K.~D.}\
  \bibnamefont {Giardina}}, \bibinfo {author} {\bibfnamefont {P.}~\bibnamefont
  {Godwin}}, \bibinfo {author} {\bibfnamefont {E.}~\bibnamefont {Goetz}},
  \bibinfo {author} {\bibfnamefont {S.}~\bibnamefont {Gras}}, \bibinfo {author}
  {\bibfnamefont {C.}~\bibnamefont {Gray}}, \bibinfo {author} {\bibfnamefont
  {R.}~\bibnamefont {Gray}}, \bibinfo {author} {\bibfnamefont {A.}~\bibnamefont
  {Gupta}}, \bibinfo {author} {\bibfnamefont {E.~K.}\ \bibnamefont
  {Gustafson}}, \bibinfo {author} {\bibfnamefont {R.}~\bibnamefont
  {Gustafson}}, \bibinfo {author} {\bibfnamefont {J.}~\bibnamefont {Hanks}},
  \bibinfo {author} {\bibfnamefont {J.}~\bibnamefont {Hanson}}, \bibinfo
  {author} {\bibfnamefont {T.}~\bibnamefont {Hardwick}}, \bibinfo {author}
  {\bibfnamefont {R.~K.}\ \bibnamefont {Hasskew}}, \bibinfo {author}
  {\bibfnamefont {M.~C.}\ \bibnamefont {Heintze}}, \bibinfo {author}
  {\bibfnamefont {A.~F.}\ \bibnamefont {Helmling-Cornell}}, \bibinfo {author}
  {\bibfnamefont {N.~A.}\ \bibnamefont {Holland}}, \bibinfo {author}
  {\bibfnamefont {J.~D.}\ \bibnamefont {Jones}}, \bibinfo {author}
  {\bibfnamefont {S.}~\bibnamefont {Kandhasamy}}, \bibinfo {author}
  {\bibfnamefont {S.}~\bibnamefont {Karki}}, \bibinfo {author} {\bibfnamefont
  {M.}~\bibnamefont {Kasprzack}}, \bibinfo {author} {\bibfnamefont
  {K.}~\bibnamefont {Kawabe}}, \bibinfo {author} {\bibfnamefont {P.~J.}\
  \bibnamefont {King}}, \bibinfo {author} {\bibfnamefont {J.~S.}\ \bibnamefont
  {Kissel}}, \bibinfo {author} {\bibfnamefont {R.}~\bibnamefont {Kumar}},
  \bibinfo {author} {\bibfnamefont {M.}~\bibnamefont {Landry}}, \bibinfo
  {author} {\bibfnamefont {B.~B.}\ \bibnamefont {Lane}}, \bibinfo {author}
  {\bibfnamefont {B.}~\bibnamefont {Lantz}}, \bibinfo {author} {\bibfnamefont
  {M.}~\bibnamefont {Laxen}}, \bibinfo {author} {\bibfnamefont {Y.~K.}\
  \bibnamefont {Lecoeuche}}, \bibinfo {author} {\bibfnamefont {J.}~\bibnamefont
  {Leviton}}, \bibinfo {author} {\bibfnamefont {J.}~\bibnamefont {Liu}},
  \bibinfo {author} {\bibfnamefont {M.}~\bibnamefont {Lormand}}, \bibinfo
  {author} {\bibfnamefont {A.~P.}\ \bibnamefont {Lundgren}}, \bibinfo {author}
  {\bibfnamefont {R.}~\bibnamefont {Macas}}, \bibinfo {author} {\bibfnamefont
  {M.}~\bibnamefont {MacInnis}}, \bibinfo {author} {\bibfnamefont {D.~M.}\
  \bibnamefont {Macleod}}, \bibinfo {author} {\bibfnamefont {S.}~\bibnamefont
  {M\'arka}}, \bibinfo {author} {\bibfnamefont {Z.}~\bibnamefont {M\'arka}},
  \bibinfo {author} {\bibfnamefont {D.~V.}\ \bibnamefont {Martynov}}, \bibinfo
  {author} {\bibfnamefont {K.}~\bibnamefont {Mason}}, \bibinfo {author}
  {\bibfnamefont {T.~J.}\ \bibnamefont {Massinger}}, \bibinfo {author}
  {\bibfnamefont {R.}~\bibnamefont {McCarthy}}, \bibinfo {author}
  {\bibfnamefont {S.}~\bibnamefont {McCormick}}, \bibinfo {author}
  {\bibfnamefont {J.}~\bibnamefont {McIver}}, \bibinfo {author} {\bibfnamefont
  {G.}~\bibnamefont {Mendell}}, \bibinfo {author} {\bibfnamefont
  {K.}~\bibnamefont {Merfeld}}, \bibinfo {author} {\bibfnamefont {E.~L.}\
  \bibnamefont {Merilh}}, \bibinfo {author} {\bibfnamefont {F.}~\bibnamefont
  {Meylahn}}, \bibinfo {author} {\bibfnamefont {T.}~\bibnamefont {Mistry}},
  \bibinfo {author} {\bibfnamefont {R.}~\bibnamefont {Mittleman}}, \bibinfo
  {author} {\bibfnamefont {G.}~\bibnamefont {Moreno}}, \bibinfo {author}
  {\bibfnamefont {C.~M.}\ \bibnamefont {Mow-Lowry}}, \bibinfo {author}
  {\bibfnamefont {S.}~\bibnamefont {Mozzon}}, \bibinfo {author} {\bibfnamefont
  {T.~J.~N.}\ \bibnamefont {Nelson}}, \bibinfo {author} {\bibfnamefont
  {P.}~\bibnamefont {Nguyen}}, \bibinfo {author} {\bibfnamefont {L.~K.}\
  \bibnamefont {Nuttall}}, \bibinfo {author} {\bibfnamefont {J.}~\bibnamefont
  {Oberling}}, \bibinfo {author} {\bibfnamefont {R.~J.}\ \bibnamefont {Oram}},
  \bibinfo {author} {\bibfnamefont {B.}~\bibnamefont {O'Reilly}}, \bibinfo
  {author} {\bibfnamefont {C.}~\bibnamefont {Osthelder}}, \bibinfo {author}
  {\bibfnamefont {D.~J.}\ \bibnamefont {Ottaway}}, \bibinfo {author}
  {\bibfnamefont {H.}~\bibnamefont {Overmier}}, \bibinfo {author}
  {\bibfnamefont {J.~R.}\ \bibnamefont {Palamos}}, \bibinfo {author}
  {\bibfnamefont {W.}~\bibnamefont {Parker}}, \bibinfo {author} {\bibfnamefont
  {E.}~\bibnamefont {Payne}}, \bibinfo {author} {\bibfnamefont
  {A.}~\bibnamefont {Pele}}, \bibinfo {author} {\bibfnamefont {C.~J.}\
  \bibnamefont {Perez}}, \bibinfo {author} {\bibfnamefont {M.}~\bibnamefont
  {Pirello}}, \bibinfo {author} {\bibfnamefont {H.}~\bibnamefont {Radkins}},
  \bibinfo {author} {\bibfnamefont {K.~E.}\ \bibnamefont {Ramirez}}, \bibinfo
  {author} {\bibfnamefont {J.~W.}\ \bibnamefont {Richardson}}, \bibinfo
  {author} {\bibfnamefont {K.}~\bibnamefont {Riles}}, \bibinfo {author}
  {\bibfnamefont {N.~A.}\ \bibnamefont {Robertson}}, \bibinfo {author}
  {\bibfnamefont {J.~G.}\ \bibnamefont {Rollins}}, \bibinfo {author}
  {\bibfnamefont {C.~L.}\ \bibnamefont {Romel}}, \bibinfo {author}
  {\bibfnamefont {J.~H.}\ \bibnamefont {Romie}}, \bibinfo {author}
  {\bibfnamefont {M.~P.}\ \bibnamefont {Ross}}, \bibinfo {author}
  {\bibfnamefont {K.}~\bibnamefont {Ryan}}, \bibinfo {author} {\bibfnamefont
  {T.}~\bibnamefont {Sadecki}}, \bibinfo {author} {\bibfnamefont {E.~J.}\
  \bibnamefont {Sanchez}}, \bibinfo {author} {\bibfnamefont {L.~E.}\
  \bibnamefont {Sanchez}}, \bibinfo {author} {\bibfnamefont {T.~R.}\
  \bibnamefont {Saravanan}}, \bibinfo {author} {\bibfnamefont {R.~L.}\
  \bibnamefont {Savage}}, \bibinfo {author} {\bibfnamefont {D.}~\bibnamefont
  {Schaetzl}}, \bibinfo {author} {\bibfnamefont {R.}~\bibnamefont {Schnabel}},
  \bibinfo {author} {\bibfnamefont {R.~M.~S.}\ \bibnamefont {Schofield}},
  \bibinfo {author} {\bibfnamefont {E.}~\bibnamefont {Schwartz}}, \bibinfo
  {author} {\bibfnamefont {D.}~\bibnamefont {Sellers}}, \bibinfo {author}
  {\bibfnamefont {T.~J.}\ \bibnamefont {Shaffer}}, \bibinfo {author}
  {\bibfnamefont {J.~R.}\ \bibnamefont {Smith}}, \bibinfo {author}
  {\bibfnamefont {S.}~\bibnamefont {Soni}}, \bibinfo {author} {\bibfnamefont
  {B.}~\bibnamefont {Sorazu}}, \bibinfo {author} {\bibfnamefont {A.~P.}\
  \bibnamefont {Spencer}}, \bibinfo {author} {\bibfnamefont {K.~A.}\
  \bibnamefont {Strain}}, \bibinfo {author} {\bibfnamefont {L.}~\bibnamefont
  {Sun}}, \bibinfo {author} {\bibfnamefont {M.~J.}\ \bibnamefont
  {Szczepa\ifmmode~\acute{n}\else \'{n}\fi{}czyk}}, \bibinfo {author}
  {\bibfnamefont {M.}~\bibnamefont {Thomas}}, \bibinfo {author} {\bibfnamefont
  {P.}~\bibnamefont {Thomas}}, \bibinfo {author} {\bibfnamefont {K.~A.}\
  \bibnamefont {Thorne}}, \bibinfo {author} {\bibfnamefont {K.}~\bibnamefont
  {Toland}}, \bibinfo {author} {\bibfnamefont {C.~I.}\ \bibnamefont {Torrie}},
  \bibinfo {author} {\bibfnamefont {G.}~\bibnamefont {Traylor}}, \bibinfo
  {author} {\bibfnamefont {A.~L.}\ \bibnamefont {Urban}}, \bibinfo {author}
  {\bibfnamefont {G.}~\bibnamefont {Vajente}}, \bibinfo {author} {\bibfnamefont
  {G.}~\bibnamefont {Valdes}}, \bibinfo {author} {\bibfnamefont {D.~C.}\
  \bibnamefont {Vander-Hyde}}, \bibinfo {author} {\bibfnamefont {P.~J.}\
  \bibnamefont {Veitch}}, \bibinfo {author} {\bibfnamefont {K.}~\bibnamefont
  {Venkateswara}}, \bibinfo {author} {\bibfnamefont {G.}~\bibnamefont
  {Venugopalan}}, \bibinfo {author} {\bibfnamefont {A.~D.}\ \bibnamefont
  {Viets}}, \bibinfo {author} {\bibfnamefont {C.}~\bibnamefont {Vorvick}},
  \bibinfo {author} {\bibfnamefont {M.}~\bibnamefont {Wade}}, \bibinfo {author}
  {\bibfnamefont {J.}~\bibnamefont {Warner}}, \bibinfo {author} {\bibfnamefont
  {B.}~\bibnamefont {Weaver}}, \bibinfo {author} {\bibfnamefont
  {R.}~\bibnamefont {Weiss}}, \bibinfo {author} {\bibfnamefont
  {B.}~\bibnamefont {Willke}}, \bibinfo {author} {\bibfnamefont {C.~C.}\
  \bibnamefont {Wipf}}, \bibinfo {author} {\bibfnamefont {L.}~\bibnamefont
  {Xiao}}, \bibinfo {author} {\bibfnamefont {H.}~\bibnamefont {Yamamoto}},
  \bibinfo {author} {\bibfnamefont {M.~J.}\ \bibnamefont {Yap}}, \bibinfo
  {author} {\bibfnamefont {H.}~\bibnamefont {Yu}}, \bibinfo {author}
  {\bibfnamefont {L.}~\bibnamefont {Zhang}}, \bibinfo {author} {\bibfnamefont
  {M.~E.}\ \bibnamefont {Zucker}},\ and\ \bibinfo {author} {\bibfnamefont
  {J.}~\bibnamefont {Zweizig}},\ }\href
  {https://doi.org/10.1103/PhysRevLett.123.231107} {\bibfield  {journal}
  {\bibinfo  {journal} {Phys. Rev. Lett.}\ }\textbf {\bibinfo {volume} {123}},\
  \bibinfo {pages} {231107} (\bibinfo {year} {2019})}\BibitemShut {NoStop}%
\bibitem [{\citenamefont {Hudelist}\ \emph {et~al.}(2014)\citenamefont
  {Hudelist}, \citenamefont {Kong}, \citenamefont {Liu}, \citenamefont {Jing},
  \citenamefont {Ou},\ and\ \citenamefont {Zhang}}]{Hudelist_2014}%
  \BibitemOpen
  \bibfield  {author} {\bibinfo {author} {\bibfnamefont {F.}~\bibnamefont
  {Hudelist}}, \bibinfo {author} {\bibfnamefont {J.}~\bibnamefont {Kong}},
  \bibinfo {author} {\bibfnamefont {C.}~\bibnamefont {Liu}}, \bibinfo {author}
  {\bibfnamefont {J.}~\bibnamefont {Jing}}, \bibinfo {author} {\bibfnamefont
  {Z.~Y.}\ \bibnamefont {Ou}},\ and\ \bibinfo {author} {\bibfnamefont
  {W.}~\bibnamefont {Zhang}},\ }\href {https://doi.org/10.1038/ncomms4049}
  {\bibfield  {journal} {\bibinfo  {journal} {Nature Communications}\ }\textbf
  {\bibinfo {volume} {5}},\ \bibinfo {pages} {3049} (\bibinfo {year}
  {2014})}\BibitemShut {NoStop}%
\bibitem [{\citenamefont {Li}\ \emph {et~al.}(2014)\citenamefont {Li},
  \citenamefont {Yuan}, \citenamefont {Ou},\ and\ \citenamefont
  {Zhang}}]{Li_2014}%
  \BibitemOpen
  \bibfield  {author} {\bibinfo {author} {\bibfnamefont {D.}~\bibnamefont
  {Li}}, \bibinfo {author} {\bibfnamefont {C.-H.}\ \bibnamefont {Yuan}},
  \bibinfo {author} {\bibfnamefont {Z.~Y.}\ \bibnamefont {Ou}},\ and\ \bibinfo
  {author} {\bibfnamefont {W.}~\bibnamefont {Zhang}},\ }\href
  {https://doi.org/10.1088/1367-2630/16/7/073020} {\bibfield  {journal}
  {\bibinfo  {journal} {New Journal of Physics}\ }\textbf {\bibinfo {volume}
  {16}},\ \bibinfo {pages} {073020} (\bibinfo {year} {2014})}\BibitemShut
  {NoStop}%
\bibitem [{\citenamefont {Anderson}\ \emph {et~al.}(2017)\citenamefont
  {Anderson}, \citenamefont {Gupta}, \citenamefont {Schmittberger},
  \citenamefont {Horrom}, \citenamefont {Hermann-Avigliano}, \citenamefont
  {Jones},\ and\ \citenamefont {Lett}}]{Anderson_2017}%
  \BibitemOpen
  \bibfield  {author} {\bibinfo {author} {\bibfnamefont {B.~E.}\ \bibnamefont
  {Anderson}}, \bibinfo {author} {\bibfnamefont {P.}~\bibnamefont {Gupta}},
  \bibinfo {author} {\bibfnamefont {B.~L.}\ \bibnamefont {Schmittberger}},
  \bibinfo {author} {\bibfnamefont {T.}~\bibnamefont {Horrom}}, \bibinfo
  {author} {\bibfnamefont {C.}~\bibnamefont {Hermann-Avigliano}}, \bibinfo
  {author} {\bibfnamefont {K.~M.}\ \bibnamefont {Jones}},\ and\ \bibinfo
  {author} {\bibfnamefont {P.~D.}\ \bibnamefont {Lett}},\ }\href
  {https://doi.org/10.1364/OPTICA.4.000752} {\bibfield  {journal} {\bibinfo
  {journal} {Optica}\ }\textbf {\bibinfo {volume} {4}},\ \bibinfo {pages} {752}
  (\bibinfo {year} {2017})}\BibitemShut {NoStop}%
\bibitem [{\citenamefont {Manceau}\ \emph {et~al.}(2017)\citenamefont
  {Manceau}, \citenamefont {Leuchs}, \citenamefont {Khalili},\ and\
  \citenamefont {Chekhova}}]{Manceau_2017}%
  \BibitemOpen
  \bibfield  {author} {\bibinfo {author} {\bibfnamefont {M.}~\bibnamefont
  {Manceau}}, \bibinfo {author} {\bibfnamefont {G.}~\bibnamefont {Leuchs}},
  \bibinfo {author} {\bibfnamefont {F.}~\bibnamefont {Khalili}},\ and\ \bibinfo
  {author} {\bibfnamefont {M.}~\bibnamefont {Chekhova}},\ }\href
  {https://doi.org/10.1103/PhysRevLett.119.223604} {\bibfield  {journal}
  {\bibinfo  {journal} {Phys. Rev. Lett.}\ }\textbf {\bibinfo {volume} {119}},\
  \bibinfo {pages} {223604} (\bibinfo {year} {2017})}\BibitemShut {NoStop}%
\bibitem [{\citenamefont {Shaked}\ \emph {et~al.}(2018)\citenamefont {Shaked},
  \citenamefont {Michael}, \citenamefont {Vered}, \citenamefont {Bello},
  \citenamefont {Rosenbluh},\ and\ \citenamefont {Pe'er}}]{Shaked_2018}%
  \BibitemOpen
  \bibfield  {author} {\bibinfo {author} {\bibfnamefont {Y.}~\bibnamefont
  {Shaked}}, \bibinfo {author} {\bibfnamefont {Y.}~\bibnamefont {Michael}},
  \bibinfo {author} {\bibfnamefont {R.~Z.}\ \bibnamefont {Vered}}, \bibinfo
  {author} {\bibfnamefont {L.}~\bibnamefont {Bello}}, \bibinfo {author}
  {\bibfnamefont {M.}~\bibnamefont {Rosenbluh}},\ and\ \bibinfo {author}
  {\bibfnamefont {A.}~\bibnamefont {Pe'er}},\ }\href
  {https://doi.org/10.1038/s41467-018-03083-5} {\bibfield  {journal} {\bibinfo
  {journal} {Nature Communications}\ }\textbf {\bibinfo {volume} {9}},\
  \bibinfo {pages} {609} (\bibinfo {year} {2018})}\BibitemShut {NoStop}%
\bibitem [{\citenamefont {Frascella}\ \emph {et~al.}(2019)\citenamefont
  {Frascella}, \citenamefont {Mikhailov}, \citenamefont {Takanashi},
  \citenamefont {Zakharov}, \citenamefont {Tikhonova},\ and\ \citenamefont
  {Chekhova}}]{Frascella_2019}%
  \BibitemOpen
  \bibfield  {author} {\bibinfo {author} {\bibfnamefont {G.}~\bibnamefont
  {Frascella}}, \bibinfo {author} {\bibfnamefont {E.~E.}\ \bibnamefont
  {Mikhailov}}, \bibinfo {author} {\bibfnamefont {N.}~\bibnamefont
  {Takanashi}}, \bibinfo {author} {\bibfnamefont {R.~V.}\ \bibnamefont
  {Zakharov}}, \bibinfo {author} {\bibfnamefont {O.~V.}\ \bibnamefont
  {Tikhonova}},\ and\ \bibinfo {author} {\bibfnamefont {M.~V.}\ \bibnamefont
  {Chekhova}},\ }\href {https://doi.org/10.1364/OPTICA.6.001233} {\bibfield
  {journal} {\bibinfo  {journal} {Optica}\ }\textbf {\bibinfo {volume} {6}},\
  \bibinfo {pages} {1233} (\bibinfo {year} {2019})}\BibitemShut {NoStop}%
\bibitem [{\citenamefont {Mukamel}(1995)}]{Mukamel_1995}%
  \BibitemOpen
  \bibfield  {author} {\bibinfo {author} {\bibfnamefont {S.}~\bibnamefont
  {Mukamel}},\ }\href@noop {} {\emph {\bibinfo {title} {{Principles of
  Nonlinear Optical Spectroscopy}}}}\ (\bibinfo  {publisher} {Oxford University
  Press},\ \bibinfo {year} {1995})\BibitemShut {NoStop}%
\bibitem [{\citenamefont {Cotler}\ \emph {et~al.}(2018)\citenamefont {Cotler},
  \citenamefont {Jian}, \citenamefont {Qi},\ and\ \citenamefont
  {Wilczek}}]{Cotler_2018}%
  \BibitemOpen
  \bibfield  {author} {\bibinfo {author} {\bibfnamefont {J.}~\bibnamefont
  {Cotler}}, \bibinfo {author} {\bibfnamefont {C.-M.}\ \bibnamefont {Jian}},
  \bibinfo {author} {\bibfnamefont {X.-L.}\ \bibnamefont {Qi}},\ and\ \bibinfo
  {author} {\bibfnamefont {F.}~\bibnamefont {Wilczek}},\ }\href
  {https://doi.org/10.1007/JHEP09(2018)093} {\bibfield  {journal} {\bibinfo
  {journal} {Journal of High Energy Physics}\ }\textbf {\bibinfo {volume}
  {2018}},\ \bibinfo {pages} {93} (\bibinfo {year} {2018})}\BibitemShut
  {NoStop}%
\bibitem [{\citenamefont {Harbola}\ and\ \citenamefont
  {Mukamel}(2008)}]{Harbola_2008}%
  \BibitemOpen
  \bibfield  {author} {\bibinfo {author} {\bibfnamefont {U.}~\bibnamefont
  {Harbola}}\ and\ \bibinfo {author} {\bibfnamefont {S.}~\bibnamefont
  {Mukamel}},\ }\href
  {https://doi.org/https://doi.org/10.1016/j.physrep.2008.05.003} {\bibfield
  {journal} {\bibinfo  {journal} {Physics Reports}\ }\textbf {\bibinfo {volume}
  {465}},\ \bibinfo {pages} {191 } (\bibinfo {year} {2008})}\BibitemShut
  {NoStop}%
\bibitem [{\citenamefont {Kryvohuz}\ and\ \citenamefont
  {Mukamel}(2012)}]{Kryvohuz_2012_I}%
  \BibitemOpen
  \bibfield  {author} {\bibinfo {author} {\bibfnamefont {M.}~\bibnamefont
  {Kryvohuz}}\ and\ \bibinfo {author} {\bibfnamefont {S.}~\bibnamefont
  {Mukamel}},\ }\href {https://doi.org/10.1103/PhysRevA.86.043818} {\bibfield
  {journal} {\bibinfo  {journal} {Phys. Rev. A}\ }\textbf {\bibinfo {volume}
  {86}},\ \bibinfo {pages} {043818} (\bibinfo {year} {2012})}\BibitemShut
  {NoStop}%
\bibitem [{\citenamefont {Kryvohuz}\ and\ \citenamefont
  {Mukamel}(2014)}]{Kryvohuz_2012_II}%
  \BibitemOpen
  \bibfield  {author} {\bibinfo {author} {\bibfnamefont {M.}~\bibnamefont
  {Kryvohuz}}\ and\ \bibinfo {author} {\bibfnamefont {S.}~\bibnamefont
  {Mukamel}},\ }\href {https://doi.org/10.1063/1.4861588} {\bibfield  {journal}
  {\bibinfo  {journal} {The Journal of Chemical Physics}\ }\textbf {\bibinfo
  {volume} {140}},\ \bibinfo {pages} {034111} (\bibinfo {year}
  {2014})}\BibitemShut {NoStop}%
\bibitem [{\citenamefont {Scully}\ and\ \citenamefont
  {Dr\"uhl}(1982)}]{Scully_1982}%
  \BibitemOpen
  \bibfield  {author} {\bibinfo {author} {\bibfnamefont {M.~O.}\ \bibnamefont
  {Scully}}\ and\ \bibinfo {author} {\bibfnamefont {K.}~\bibnamefont
  {Dr\"uhl}},\ }\href {https://doi.org/10.1103/PhysRevA.25.2208} {\bibfield
  {journal} {\bibinfo  {journal} {Phys. Rev. A}\ }\textbf {\bibinfo {volume}
  {25}},\ \bibinfo {pages} {2208} (\bibinfo {year} {1982})}\BibitemShut
  {NoStop}%
\bibitem [{\citenamefont {Kim}\ \emph {et~al.}(2000)\citenamefont {Kim},
  \citenamefont {Yu}, \citenamefont {Kulik}, \citenamefont {Shih},\ and\
  \citenamefont {Scully}}]{Kim_2000}%
  \BibitemOpen
  \bibfield  {author} {\bibinfo {author} {\bibfnamefont {Y.-H.}\ \bibnamefont
  {Kim}}, \bibinfo {author} {\bibfnamefont {R.}~\bibnamefont {Yu}}, \bibinfo
  {author} {\bibfnamefont {S.~P.}\ \bibnamefont {Kulik}}, \bibinfo {author}
  {\bibfnamefont {Y.}~\bibnamefont {Shih}},\ and\ \bibinfo {author}
  {\bibfnamefont {M.~O.}\ \bibnamefont {Scully}},\ }\href
  {https://doi.org/10.1103/PhysRevLett.84.1} {\bibfield  {journal} {\bibinfo
  {journal} {Phys. Rev. Lett.}\ }\textbf {\bibinfo {volume} {84}},\ \bibinfo
  {pages} {1} (\bibinfo {year} {2000})}\BibitemShut {NoStop}%
\bibitem [{\citenamefont {A.I.~Larkin}(1969)}]{Larkin_1969}%
  \BibitemOpen
  \bibfield  {author} {\bibinfo {author} {\bibfnamefont {Y.~N.~O.}\
  \bibnamefont {A.I.~Larkin}},\ }\href@noop {} {\bibfield  {journal} {\bibinfo
  {journal} {JEPT}\ }\textbf {\bibinfo {volume} {28}},\ \bibinfo {pages} {1200}
  (\bibinfo {year} {1969})}\BibitemShut {NoStop}%
\bibitem [{\citenamefont {Kitaev}(2014)}]{Kitaev_2014}%
  \BibitemOpen
  \bibfield  {author} {\bibinfo {author} {\bibfnamefont {A.}~\bibnamefont
  {Kitaev}},\ }in\ \href@noop {} {\emph {\bibinfo {booktitle} {Proceedings of
  the Fundamental Physics Prize Symposium}}}\ (\bibinfo {year}
  {2014})\BibitemShut {NoStop}%
\bibitem [{\citenamefont {Shenker}\ and\ \citenamefont
  {Stanford}(2014)}]{Shenker_2014}%
  \BibitemOpen
  \bibfield  {author} {\bibinfo {author} {\bibfnamefont {S.~H.}\ \bibnamefont
  {Shenker}}\ and\ \bibinfo {author} {\bibfnamefont {D.}~\bibnamefont
  {Stanford}},\ }\href {https://doi.org/10.1007/JHEP03(2014)067} {\bibfield
  {journal} {\bibinfo  {journal} {Journal of High Energy Physics}\ }\textbf
  {\bibinfo {volume} {2014}},\ \bibinfo {pages} {67} (\bibinfo {year}
  {2014})}\BibitemShut {NoStop}%
\bibitem [{\citenamefont {Roberts}\ \emph {et~al.}(2015)\citenamefont
  {Roberts}, \citenamefont {Stanford},\ and\ \citenamefont
  {Susskind}}]{Roberts_2015}%
  \BibitemOpen
  \bibfield  {author} {\bibinfo {author} {\bibfnamefont {D.~A.}\ \bibnamefont
  {Roberts}}, \bibinfo {author} {\bibfnamefont {D.}~\bibnamefont {Stanford}},\
  and\ \bibinfo {author} {\bibfnamefont {L.}~\bibnamefont {Susskind}},\ }\href
  {https://doi.org/10.1007/JHEP03(2015)051} {\bibfield  {journal} {\bibinfo
  {journal} {Journal of High Energy Physics}\ }\textbf {\bibinfo {volume}
  {2015}},\ \bibinfo {pages} {51} (\bibinfo {year} {2015})}\BibitemShut
  {NoStop}%
\bibitem [{\citenamefont {Maldacena}\ \emph {et~al.}(2016)\citenamefont
  {Maldacena}, \citenamefont {Shenker},\ and\ \citenamefont
  {Stanford}}]{Maldacena_2016}%
  \BibitemOpen
  \bibfield  {author} {\bibinfo {author} {\bibfnamefont {J.}~\bibnamefont
  {Maldacena}}, \bibinfo {author} {\bibfnamefont {S.~H.}\ \bibnamefont
  {Shenker}},\ and\ \bibinfo {author} {\bibfnamefont {D.}~\bibnamefont
  {Stanford}},\ }\href {https://doi.org/10.1007/JHEP08(2016)106} {\bibfield
  {journal} {\bibinfo  {journal} {Journal of High Energy Physics}\ }\textbf
  {\bibinfo {volume} {2016}},\ \bibinfo {pages} {106} (\bibinfo {year}
  {2016})}\BibitemShut {NoStop}%
\bibitem [{\citenamefont {Aleiner}\ \emph {et~al.}(2016)\citenamefont
  {Aleiner}, \citenamefont {Faoro},\ and\ \citenamefont
  {Ioffe}}]{Aleiner_2016}%
  \BibitemOpen
  \bibfield  {author} {\bibinfo {author} {\bibfnamefont {I.~L.}\ \bibnamefont
  {Aleiner}}, \bibinfo {author} {\bibfnamefont {L.}~\bibnamefont {Faoro}},\
  and\ \bibinfo {author} {\bibfnamefont {L.~B.}\ \bibnamefont {Ioffe}},\ }\href
  {https://doi.org/https://doi.org/10.1016/j.aop.2016.09.006} {\bibfield
  {journal} {\bibinfo  {journal} {Annals of Physics}\ }\textbf {\bibinfo
  {volume} {375}},\ \bibinfo {pages} {378 } (\bibinfo {year}
  {2016})}\BibitemShut {NoStop}%
\bibitem [{\citenamefont {Yao}\ \emph {et~al.}(2016)\citenamefont {Yao},
  \citenamefont {Grusdt}, \citenamefont {Swingle}, \citenamefont {Lukin},
  \citenamefont {Stamper-Kurn}, \citenamefont {Moore},\ and\ \citenamefont
  {Demler}}]{yao_2016}%
  \BibitemOpen
  \bibfield  {author} {\bibinfo {author} {\bibfnamefont {N.~Y.}\ \bibnamefont
  {Yao}}, \bibinfo {author} {\bibfnamefont {F.}~\bibnamefont {Grusdt}},
  \bibinfo {author} {\bibfnamefont {B.}~\bibnamefont {Swingle}}, \bibinfo
  {author} {\bibfnamefont {M.~D.}\ \bibnamefont {Lukin}}, \bibinfo {author}
  {\bibfnamefont {D.~M.}\ \bibnamefont {Stamper-Kurn}}, \bibinfo {author}
  {\bibfnamefont {J.~E.}\ \bibnamefont {Moore}},\ and\ \bibinfo {author}
  {\bibfnamefont {E.~A.}\ \bibnamefont {Demler}},\ }\href@noop {} {\bibinfo
  {title} {Interferometric approach to probing fast scrambling}} (\bibinfo
  {year} {2016}),\ \Eprint {https://arxiv.org/abs/1607.01801} {arXiv:1607.01801
  [quant-ph]} \BibitemShut {NoStop}%
\bibitem [{\citenamefont {Chen}\ \emph {et~al.}(2016)\citenamefont {Chen},
  \citenamefont {Zhou}, \citenamefont {Huse},\ and\ \citenamefont
  {Fradkin}}]{Chen_2016}%
  \BibitemOpen
  \bibfield  {author} {\bibinfo {author} {\bibfnamefont {X.}~\bibnamefont
  {Chen}}, \bibinfo {author} {\bibfnamefont {T.}~\bibnamefont {Zhou}}, \bibinfo
  {author} {\bibfnamefont {D.~A.}\ \bibnamefont {Huse}},\ and\ \bibinfo
  {author} {\bibfnamefont {E.}~\bibnamefont {Fradkin}},\ }\href
  {https://doi.org/10.1002/andp.201600332} {\bibfield  {journal} {\bibinfo
  {journal} {Annalen der Physik}\ }\textbf {\bibinfo {volume} {529}},\ \bibinfo
  {pages} {1600332} (\bibinfo {year} {2016})}\BibitemShut {NoStop}%
\bibitem [{\citenamefont {Yoshida}\ and\ \citenamefont
  {Kitaev}(2017)}]{yoshida_2017}%
  \BibitemOpen
  \bibfield  {author} {\bibinfo {author} {\bibfnamefont {B.}~\bibnamefont
  {Yoshida}}\ and\ \bibinfo {author} {\bibfnamefont {A.}~\bibnamefont
  {Kitaev}},\ }\href@noop {} {\bibinfo {title} {Efficient decoding for the
  hayden-preskill protocol}} (\bibinfo {year} {2017}),\ \Eprint
  {https://arxiv.org/abs/1710.03363} {arXiv:1710.03363 [hep-th]} \BibitemShut
  {NoStop}%
\bibitem [{\citenamefont {Kukuljan}\ \emph {et~al.}(2017)\citenamefont
  {Kukuljan}, \citenamefont {Grozdanov},\ and\ \citenamefont
  {Prosen}}]{Kukuljan_2017}%
  \BibitemOpen
  \bibfield  {author} {\bibinfo {author} {\bibfnamefont {I.}~\bibnamefont
  {Kukuljan}}, \bibinfo {author} {\bibfnamefont {S.~c.~v.}\ \bibnamefont
  {Grozdanov}},\ and\ \bibinfo {author} {\bibfnamefont {T.~c.~v.}\ \bibnamefont
  {Prosen}},\ }\href {https://doi.org/10.1103/PhysRevB.96.060301} {\bibfield
  {journal} {\bibinfo  {journal} {Phys. Rev. B}\ }\textbf {\bibinfo {volume}
  {96}},\ \bibinfo {pages} {060301} (\bibinfo {year} {2017})}\BibitemShut
  {NoStop}%
\bibitem [{\citenamefont {Swingle}\ and\ \citenamefont
  {Chowdhury}(2017)}]{Swingle_2017}%
  \BibitemOpen
  \bibfield  {author} {\bibinfo {author} {\bibfnamefont {B.}~\bibnamefont
  {Swingle}}\ and\ \bibinfo {author} {\bibfnamefont {D.}~\bibnamefont
  {Chowdhury}},\ }\href {https://doi.org/10.1103/PhysRevB.95.060201} {\bibfield
   {journal} {\bibinfo  {journal} {Phys. Rev. B}\ }\textbf {\bibinfo {volume}
  {95}},\ \bibinfo {pages} {060201} (\bibinfo {year} {2017})}\BibitemShut
  {NoStop}%
\bibitem [{\citenamefont {Pappalardi}\ \emph {et~al.}(2018)\citenamefont
  {Pappalardi}, \citenamefont {Russomanno}, \citenamefont {\ifmmode
  \check{Z}\else \v{Z}\fi{}unkovi\ifmmode~\check{c}\else \v{c}\fi{}},
  \citenamefont {Iemini}, \citenamefont {Silva},\ and\ \citenamefont
  {Fazio}}]{Pappalardi_2018}%
  \BibitemOpen
  \bibfield  {author} {\bibinfo {author} {\bibfnamefont {S.}~\bibnamefont
  {Pappalardi}}, \bibinfo {author} {\bibfnamefont {A.}~\bibnamefont
  {Russomanno}}, \bibinfo {author} {\bibfnamefont {B.}~\bibnamefont {\ifmmode
  \check{Z}\else \v{Z}\fi{}unkovi\ifmmode~\check{c}\else \v{c}\fi{}}}, \bibinfo
  {author} {\bibfnamefont {F.}~\bibnamefont {Iemini}}, \bibinfo {author}
  {\bibfnamefont {A.}~\bibnamefont {Silva}},\ and\ \bibinfo {author}
  {\bibfnamefont {R.}~\bibnamefont {Fazio}},\ }\href
  {https://doi.org/10.1103/PhysRevB.98.134303} {\bibfield  {journal} {\bibinfo
  {journal} {Phys. Rev. B}\ }\textbf {\bibinfo {volume} {98}},\ \bibinfo
  {pages} {134303} (\bibinfo {year} {2018})}\BibitemShut {NoStop}%
\bibitem [{\citenamefont {Yunger~Halpern}\ \emph {et~al.}(2019)\citenamefont
  {Yunger~Halpern}, \citenamefont {Bartolotta},\ and\ \citenamefont
  {Pollack}}]{YungerHalpern_2019}%
  \BibitemOpen
  \bibfield  {author} {\bibinfo {author} {\bibfnamefont {N.}~\bibnamefont
  {Yunger~Halpern}}, \bibinfo {author} {\bibfnamefont {A.}~\bibnamefont
  {Bartolotta}},\ and\ \bibinfo {author} {\bibfnamefont {J.}~\bibnamefont
  {Pollack}},\ }\href {https://doi.org/10.1038/s42005-019-0179-8} {\bibfield
  {journal} {\bibinfo  {journal} {Communications Physics}\ }\textbf {\bibinfo
  {volume} {2}},\ \bibinfo {pages} {92} (\bibinfo {year} {2019})}\BibitemShut
  {NoStop}%
\bibitem [{\citenamefont {Roberts}\ and\ \citenamefont
  {Stanford}(2015)}]{Roberts_2019}%
  \BibitemOpen
  \bibfield  {author} {\bibinfo {author} {\bibfnamefont {D.~A.}\ \bibnamefont
  {Roberts}}\ and\ \bibinfo {author} {\bibfnamefont {D.}~\bibnamefont
  {Stanford}},\ }\href {https://doi.org/10.1103/PhysRevLett.115.131603}
  {\bibfield  {journal} {\bibinfo  {journal} {Phys. Rev. Lett.}\ }\textbf
  {\bibinfo {volume} {115}},\ \bibinfo {pages} {131603} (\bibinfo {year}
  {2015})}\BibitemShut {NoStop}%
\bibitem [{\citenamefont {Gonz\'alez~Alonso}\ \emph {et~al.}(2019)\citenamefont
  {Gonz\'alez~Alonso}, \citenamefont {Yunger~Halpern},\ and\ \citenamefont
  {Dressel}}]{Gonzales_2019}%
  \BibitemOpen
  \bibfield  {author} {\bibinfo {author} {\bibfnamefont {J.~R.}\ \bibnamefont
  {Gonz\'alez~Alonso}}, \bibinfo {author} {\bibfnamefont {N.}~\bibnamefont
  {Yunger~Halpern}},\ and\ \bibinfo {author} {\bibfnamefont {J.}~\bibnamefont
  {Dressel}},\ }\href {https://doi.org/10.1103/PhysRevLett.122.040404}
  {\bibfield  {journal} {\bibinfo  {journal} {Phys. Rev. Lett.}\ }\textbf
  {\bibinfo {volume} {122}},\ \bibinfo {pages} {040404} (\bibinfo {year}
  {2019})}\BibitemShut {NoStop}%
\bibitem [{\citenamefont {Landsman}\ \emph {et~al.}(2019)\citenamefont
  {Landsman}, \citenamefont {Figgatt}, \citenamefont {Schuster}, \citenamefont
  {Linke}, \citenamefont {Yoshida}, \citenamefont {Yao},\ and\ \citenamefont
  {Monroe}}]{Landsman_2019}%
  \BibitemOpen
  \bibfield  {author} {\bibinfo {author} {\bibfnamefont {K.~A.}\ \bibnamefont
  {Landsman}}, \bibinfo {author} {\bibfnamefont {C.}~\bibnamefont {Figgatt}},
  \bibinfo {author} {\bibfnamefont {T.}~\bibnamefont {Schuster}}, \bibinfo
  {author} {\bibfnamefont {N.~M.}\ \bibnamefont {Linke}}, \bibinfo {author}
  {\bibfnamefont {B.}~\bibnamefont {Yoshida}}, \bibinfo {author} {\bibfnamefont
  {N.~Y.}\ \bibnamefont {Yao}},\ and\ \bibinfo {author} {\bibfnamefont
  {C.}~\bibnamefont {Monroe}},\ }\href
  {https://doi.org/10.1038/s41586-019-0952-6} {\bibfield  {journal} {\bibinfo
  {journal} {Nature}\ }\textbf {\bibinfo {volume} {567}},\ \bibinfo {pages}
  {61} (\bibinfo {year} {2019})}\BibitemShut {NoStop}%
\bibitem [{\citenamefont {Yan}\ \emph {et~al.}(2020)\citenamefont {Yan},
  \citenamefont {Cincio},\ and\ \citenamefont {Zurek}}]{Yan_2020}%
  \BibitemOpen
  \bibfield  {author} {\bibinfo {author} {\bibfnamefont {B.}~\bibnamefont
  {Yan}}, \bibinfo {author} {\bibfnamefont {L.}~\bibnamefont {Cincio}},\ and\
  \bibinfo {author} {\bibfnamefont {W.~H.}\ \bibnamefont {Zurek}},\ }\href
  {https://doi.org/10.1103/PhysRevLett.124.160603} {\bibfield  {journal}
  {\bibinfo  {journal} {Phys. Rev. Lett.}\ }\textbf {\bibinfo {volume} {124}},\
  \bibinfo {pages} {160603} (\bibinfo {year} {2020})}\BibitemShut {NoStop}%
\bibitem [{\citenamefont {Yan}\ and\ \citenamefont
  {Sinitsyn}(2020)}]{Yan_2020b}%
  \BibitemOpen
  \bibfield  {author} {\bibinfo {author} {\bibfnamefont {B.}~\bibnamefont
  {Yan}}\ and\ \bibinfo {author} {\bibfnamefont {N.~A.}\ \bibnamefont
  {Sinitsyn}},\ }\href {https://doi.org/10.1103/PhysRevLett.125.040605}
  {\bibfield  {journal} {\bibinfo  {journal} {Phys. Rev. Lett.}\ }\textbf
  {\bibinfo {volume} {125}},\ \bibinfo {pages} {040605} (\bibinfo {year}
  {2020})}\BibitemShut {NoStop}%
\bibitem [{\citenamefont {Patel}\ and\ \citenamefont
  {Sachdev}(2017)}]{Patel_2017}%
  \BibitemOpen
  \bibfield  {author} {\bibinfo {author} {\bibfnamefont {A.~A.}\ \bibnamefont
  {Patel}}\ and\ \bibinfo {author} {\bibfnamefont {S.}~\bibnamefont
  {Sachdev}},\ }\href {https://doi.org/10.1073/pnas.1618185114} {\bibfield
  {journal} {\bibinfo  {journal} {Proceedings of the National Academy of
  Sciences}\ }\textbf {\bibinfo {volume} {114}},\ \bibinfo {pages} {1844}
  (\bibinfo {year} {2017})}\BibitemShut {NoStop}%
\bibitem [{\citenamefont {Mukamel}\ \emph {et~al.}(1996)\citenamefont
  {Mukamel}, \citenamefont {Khidekel},\ and\ \citenamefont
  {Chernyak}}]{Mukamel_1996}%
  \BibitemOpen
  \bibfield  {author} {\bibinfo {author} {\bibfnamefont {S.}~\bibnamefont
  {Mukamel}}, \bibinfo {author} {\bibfnamefont {V.}~\bibnamefont {Khidekel}},\
  and\ \bibinfo {author} {\bibfnamefont {V.}~\bibnamefont {Chernyak}},\ }\href
  {https://doi.org/10.1103/PhysRevE.53.R1} {\bibfield  {journal} {\bibinfo
  {journal} {Phys. Rev. E}\ }\textbf {\bibinfo {volume} {53}},\ \bibinfo
  {pages} {R1} (\bibinfo {year} {1996})}\BibitemShut {NoStop}%
\bibitem [{\citenamefont {Aharonov}\ \emph {et~al.}(2010)\citenamefont
  {Aharonov}, \citenamefont {Popescu},\ and\ \citenamefont
  {Tollaksen}}]{Aharonov_2010}%
  \BibitemOpen
  \bibfield  {author} {\bibinfo {author} {\bibfnamefont {Y.}~\bibnamefont
  {Aharonov}}, \bibinfo {author} {\bibfnamefont {S.}~\bibnamefont {Popescu}},\
  and\ \bibinfo {author} {\bibfnamefont {J.}~\bibnamefont {Tollaksen}},\ }\href
  {https://doi.org/10.1063/1.3518209} {\bibfield  {journal} {\bibinfo
  {journal} {Physics Today}\ }\textbf {\bibinfo {volume} {63}},\ \bibinfo
  {pages} {27} (\bibinfo {year} {2010})}\BibitemShut {NoStop}%
\bibitem [{\citenamefont {Mukamel}(2011)}]{Mukamel_comments_2011}%
  \BibitemOpen
  \bibfield  {author} {\bibinfo {author} {\bibfnamefont {S.}~\bibnamefont
  {Mukamel}},\ }\href {https://doi.org/10.1063/1.3595153} {\bibfield  {journal}
  {\bibinfo  {journal} {Physics Today}\ }\textbf {\bibinfo {volume} {64}},\
  \bibinfo {pages} {9} (\bibinfo {year} {2011})}\BibitemShut {NoStop}%
\bibitem [{\citenamefont {Yurke}\ \emph {et~al.}(1986)\citenamefont {Yurke},
  \citenamefont {McCall},\ and\ \citenamefont {Klauder}}]{Yurke_1986}%
  \BibitemOpen
  \bibfield  {author} {\bibinfo {author} {\bibfnamefont {B.}~\bibnamefont
  {Yurke}}, \bibinfo {author} {\bibfnamefont {S.~L.}\ \bibnamefont {McCall}},\
  and\ \bibinfo {author} {\bibfnamefont {J.~R.}\ \bibnamefont {Klauder}},\
  }\href {https://doi.org/10.1103/PhysRevA.33.4033} {\bibfield  {journal}
  {\bibinfo  {journal} {Phys. Rev. A}\ }\textbf {\bibinfo {volume} {33}},\
  \bibinfo {pages} {4033} (\bibinfo {year} {1986})}\BibitemShut {NoStop}%
\bibitem [{\citenamefont {Reid}\ and\ \citenamefont {Walls}(1985)}]{Reid_1985}%
  \BibitemOpen
  \bibfield  {author} {\bibinfo {author} {\bibfnamefont {M.~D.}\ \bibnamefont
  {Reid}}\ and\ \bibinfo {author} {\bibfnamefont {D.~F.}\ \bibnamefont
  {Walls}},\ }\href {https://doi.org/10.1103/PhysRevA.31.1622} {\bibfield
  {journal} {\bibinfo  {journal} {Phys. Rev. A}\ }\textbf {\bibinfo {volume}
  {31}},\ \bibinfo {pages} {1622} (\bibinfo {year} {1985})}\BibitemShut
  {NoStop}%
\bibitem [{\citenamefont {Du}\ \emph {et~al.}(2018)\citenamefont {Du},
  \citenamefont {Jia}, \citenamefont {Chen}, \citenamefont {Ou},\ and\
  \citenamefont {Zhang}}]{Du_2018}%
  \BibitemOpen
  \bibfield  {author} {\bibinfo {author} {\bibfnamefont {W.}~\bibnamefont
  {Du}}, \bibinfo {author} {\bibfnamefont {J.}~\bibnamefont {Jia}}, \bibinfo
  {author} {\bibfnamefont {J.~F.}\ \bibnamefont {Chen}}, \bibinfo {author}
  {\bibfnamefont {Z.~Y.}\ \bibnamefont {Ou}},\ and\ \bibinfo {author}
  {\bibfnamefont {W.}~\bibnamefont {Zhang}},\ }\href
  {https://doi.org/10.1364/OL.43.001051} {\bibfield  {journal} {\bibinfo
  {journal} {Opt. Lett.}\ }\textbf {\bibinfo {volume} {43}},\ \bibinfo {pages}
  {1051} (\bibinfo {year} {2018})}\BibitemShut {NoStop}%
\bibitem [{\citenamefont {J.~M.~Jauch}(1976)}]{Jauche_1976}%
  \BibitemOpen
  \bibfield  {author} {\bibinfo {author} {\bibfnamefont {F.~R.}\ \bibnamefont
  {J.~M.~Jauch}},\ }\href
  {https://doi.org/https://doi.org/10.1007/978-3-642-80951-4} {\emph {\bibinfo
  {title} {{ The Theory of Photons and Electrons}}}}\ (\bibinfo  {publisher}
  {Springer, Berlin, Heidelberg},\ \bibinfo {year} {1976})\BibitemShut
  {NoStop}%
\bibitem [{\citenamefont {Mota}\ \emph {et~al.}(2016)\citenamefont {Mota},
  \citenamefont {Ojeda-Guill\'{e}n}, \citenamefont {Salazar-Ram\'{i}rez},\ and\
  \citenamefont {Granados}}]{Mota_2016}%
  \BibitemOpen
  \bibfield  {author} {\bibinfo {author} {\bibfnamefont {R.~D.}\ \bibnamefont
  {Mota}}, \bibinfo {author} {\bibfnamefont {D.}~\bibnamefont
  {Ojeda-Guill\'{e}n}}, \bibinfo {author} {\bibfnamefont {M.}~\bibnamefont
  {Salazar-Ram\'{i}rez}},\ and\ \bibinfo {author} {\bibfnamefont {V.~D.}\
  \bibnamefont {Granados}},\ }\href {https://doi.org/10.1364/JOSAB.33.001696}
  {\bibfield  {journal} {\bibinfo  {journal} {J. Opt. Soc. Am. B}\ }\textbf
  {\bibinfo {volume} {33}},\ \bibinfo {pages} {1696} (\bibinfo {year}
  {2016})}\BibitemShut {NoStop}%
\bibitem [{\citenamefont {Mota}\ \emph {et~al.}(2004)\citenamefont {Mota},
  \citenamefont {Xicoténcatl},\ and\ \citenamefont {Granados}}]{Mota_2004_2}%
  \BibitemOpen
  \bibfield  {author} {\bibinfo {author} {\bibfnamefont {R.~D.}\ \bibnamefont
  {Mota}}, \bibinfo {author} {\bibfnamefont {M.~A.}\ \bibnamefont
  {Xicoténcatl}},\ and\ \bibinfo {author} {\bibfnamefont {V.~D.}\ \bibnamefont
  {Granados}},\ }\href {https://doi.org/10.1139/p04-051} {\bibfield  {journal}
  {\bibinfo  {journal} {Canadian Journal of Physics}\ }\textbf {\bibinfo
  {volume} {82}},\ \bibinfo {pages} {767} (\bibinfo {year} {2004})}\BibitemShut
  {NoStop}%
\bibitem [{\citenamefont {Hong}\ \emph
  {et~al.}(1987{\natexlab{b}})\citenamefont {Hong}, \citenamefont {Ou},\ and\
  \citenamefont {Mandel}}]{Hong_1987}%
  \BibitemOpen
  \bibfield  {author} {\bibinfo {author} {\bibfnamefont {C.~K.}\ \bibnamefont
  {Hong}}, \bibinfo {author} {\bibfnamefont {Z.~Y.}\ \bibnamefont {Ou}},\ and\
  \bibinfo {author} {\bibfnamefont {L.}~\bibnamefont {Mandel}},\ }\href
  {https://doi.org/10.1103/PhysRevLett.59.2044} {\bibfield  {journal} {\bibinfo
   {journal} {Phys. Rev. Lett.}\ }\textbf {\bibinfo {volume} {59}},\ \bibinfo
  {pages} {2044} (\bibinfo {year} {1987}{\natexlab{b}})}\BibitemShut {NoStop}%
\bibitem [{\citenamefont {Glauber}(1963)}]{Glauber_1963}%
  \BibitemOpen
  \bibfield  {author} {\bibinfo {author} {\bibfnamefont {R.~J.}\ \bibnamefont
  {Glauber}},\ }\href {https://doi.org/10.1103/PhysRev.130.2529} {\bibfield
  {journal} {\bibinfo  {journal} {Phys. Rev.}\ }\textbf {\bibinfo {volume}
  {130}},\ \bibinfo {pages} {2529} (\bibinfo {year} {1963})}\BibitemShut
  {NoStop}%
\bibitem [{Note1()}]{Note1}%
  \BibitemOpen
  \bibinfo {note} {S. Asban and S. Mukamel. Out-of-Time-Ordering Matter
  Correlators in Quantum Interferometric Spectroscopy, -- \protect \emph {to be
  published \label {fn:OTOC paper}}}\BibitemShut {NoStop}%
\bibitem [{\citenamefont {Kamenev}(2011)}]{kamenev2011field}%
  \BibitemOpen
  \bibfield  {author} {\bibinfo {author} {\bibfnamefont {A.}~\bibnamefont
  {Kamenev}},\ }\href {https://books.google.es/books?id=CwlrUepnla4C} {\emph
  {\bibinfo {title} {Field Theory of Non-Equilibrium Systems}}}\ (\bibinfo
  {publisher} {Cambridge University Press},\ \bibinfo {year}
  {2011})\BibitemShut {NoStop}%
\bibitem [{\citenamefont {Hong}\ and\ \citenamefont
  {Mandel}(1985)}]{Mandel_1985}%
  \BibitemOpen
  \bibfield  {author} {\bibinfo {author} {\bibfnamefont {C.~K.}\ \bibnamefont
  {Hong}}\ and\ \bibinfo {author} {\bibfnamefont {L.}~\bibnamefont {Mandel}},\
  }\href {https://doi.org/10.1103/PhysRevA.31.2409} {\bibfield  {journal}
  {\bibinfo  {journal} {Phys. Rev. A}\ }\textbf {\bibinfo {volume} {31}},\
  \bibinfo {pages} {2409} (\bibinfo {year} {1985})}\BibitemShut {NoStop}%
\bibitem [{\citenamefont {Bocquillon}\ \emph {et~al.}(2013)\citenamefont
  {Bocquillon}, \citenamefont {Freulon}, \citenamefont {Berroir}, \citenamefont
  {Degiovanni}, \citenamefont {Pla{\c c}ais}, \citenamefont {Cavanna},
  \citenamefont {Jin},\ and\ \citenamefont {F{\`e}ve}}]{Bocquillon_2013}%
  \BibitemOpen
  \bibfield  {author} {\bibinfo {author} {\bibfnamefont {E.}~\bibnamefont
  {Bocquillon}}, \bibinfo {author} {\bibfnamefont {V.}~\bibnamefont {Freulon}},
  \bibinfo {author} {\bibfnamefont {J.-M.}\ \bibnamefont {Berroir}}, \bibinfo
  {author} {\bibfnamefont {P.}~\bibnamefont {Degiovanni}}, \bibinfo {author}
  {\bibfnamefont {B.}~\bibnamefont {Pla{\c c}ais}}, \bibinfo {author}
  {\bibfnamefont {A.}~\bibnamefont {Cavanna}}, \bibinfo {author} {\bibfnamefont
  {Y.}~\bibnamefont {Jin}},\ and\ \bibinfo {author} {\bibfnamefont
  {G.}~\bibnamefont {F{\`e}ve}},\ }\href
  {https://doi.org/10.1126/science.1232572} {\bibfield  {journal} {\bibinfo
  {journal} {Science}\ }\textbf {\bibinfo {volume} {339}},\ \bibinfo {pages}
  {1054} (\bibinfo {year} {2013})}\BibitemShut {NoStop}%
\bibitem [{\citenamefont {de~C.~Chamon}\ \emph {et~al.}(1997)\citenamefont
  {de~C.~Chamon}, \citenamefont {Freed}, \citenamefont {Kivelson},
  \citenamefont {Sondhi},\ and\ \citenamefont {Wen}}]{Chamon_1997}%
  \BibitemOpen
  \bibfield  {author} {\bibinfo {author} {\bibfnamefont {C.}~\bibnamefont
  {de~C.~Chamon}}, \bibinfo {author} {\bibfnamefont {D.~E.}\ \bibnamefont
  {Freed}}, \bibinfo {author} {\bibfnamefont {S.~A.}\ \bibnamefont {Kivelson}},
  \bibinfo {author} {\bibfnamefont {S.~L.}\ \bibnamefont {Sondhi}},\ and\
  \bibinfo {author} {\bibfnamefont {X.~G.}\ \bibnamefont {Wen}},\ }\href
  {https://doi.org/10.1103/PhysRevB.55.2331} {\bibfield  {journal} {\bibinfo
  {journal} {Phys. Rev. B}\ }\textbf {\bibinfo {volume} {55}},\ \bibinfo
  {pages} {2331} (\bibinfo {year} {1997})}\BibitemShut {NoStop}%
\bibitem [{\citenamefont {Branning}\ \emph {et~al.}(1999)\citenamefont
  {Branning}, \citenamefont {Grice}, \citenamefont {Erdmann},\ and\
  \citenamefont {Walmsley}}]{Branning_1999}%
  \BibitemOpen
  \bibfield  {author} {\bibinfo {author} {\bibfnamefont {D.}~\bibnamefont
  {Branning}}, \bibinfo {author} {\bibfnamefont {W.~P.}\ \bibnamefont {Grice}},
  \bibinfo {author} {\bibfnamefont {R.}~\bibnamefont {Erdmann}},\ and\ \bibinfo
  {author} {\bibfnamefont {I.~A.}\ \bibnamefont {Walmsley}},\ }\href
  {https://doi.org/10.1103/PhysRevLett.83.955} {\bibfield  {journal} {\bibinfo
  {journal} {Phys. Rev. Lett.}\ }\textbf {\bibinfo {volume} {83}},\ \bibinfo
  {pages} {955} (\bibinfo {year} {1999})}\BibitemShut {NoStop}%
\bibitem [{Note2()}]{Note2}%
  \BibitemOpen
  \bibinfo {note} {S.Asban and S. Mukamel, Exchange Phase Cycling -- pathway
  selection in Hong-Ou-Mandel Interferometric Spectroscopy. \protect \emph {In
  preparation}}\BibitemShut {NoStop}%
\bibitem [{\citenamefont {Riek}\ \emph {et~al.}(2015)\citenamefont {Riek},
  \citenamefont {Seletskiy}, \citenamefont {Moskalenko}, \citenamefont
  {Schmidt}, \citenamefont {Krauspe}, \citenamefont {Eckart}, \citenamefont
  {Eggert}, \citenamefont {Burkard},\ and\ \citenamefont
  {Leitenstorfer}}]{Riek_2015}%
  \BibitemOpen
  \bibfield  {author} {\bibinfo {author} {\bibfnamefont {C.}~\bibnamefont
  {Riek}}, \bibinfo {author} {\bibfnamefont {D.~V.}\ \bibnamefont {Seletskiy}},
  \bibinfo {author} {\bibfnamefont {A.~S.}\ \bibnamefont {Moskalenko}},
  \bibinfo {author} {\bibfnamefont {J.~F.}\ \bibnamefont {Schmidt}}, \bibinfo
  {author} {\bibfnamefont {P.}~\bibnamefont {Krauspe}}, \bibinfo {author}
  {\bibfnamefont {S.}~\bibnamefont {Eckart}}, \bibinfo {author} {\bibfnamefont
  {S.}~\bibnamefont {Eggert}}, \bibinfo {author} {\bibfnamefont
  {G.}~\bibnamefont {Burkard}},\ and\ \bibinfo {author} {\bibfnamefont
  {A.}~\bibnamefont {Leitenstorfer}},\ }\href
  {https://doi.org/10.1126/science.aac9788} {\bibfield  {journal} {\bibinfo
  {journal} {Science}\ }\textbf {\bibinfo {volume} {350}},\ \bibinfo {pages}
  {420} (\bibinfo {year} {2015})}\BibitemShut {NoStop}%
\bibitem [{\citenamefont {Riek}\ \emph {et~al.}(2017)\citenamefont {Riek},
  \citenamefont {Sulzer}, \citenamefont {Seeger}, \citenamefont {Moskalenko},
  \citenamefont {Burkard}, \citenamefont {Seletskiy},\ and\ \citenamefont
  {Leitenstorfer}}]{Riek_2017}%
  \BibitemOpen
  \bibfield  {author} {\bibinfo {author} {\bibfnamefont {C.}~\bibnamefont
  {Riek}}, \bibinfo {author} {\bibfnamefont {P.}~\bibnamefont {Sulzer}},
  \bibinfo {author} {\bibfnamefont {M.}~\bibnamefont {Seeger}}, \bibinfo
  {author} {\bibfnamefont {A.~S.}\ \bibnamefont {Moskalenko}}, \bibinfo
  {author} {\bibfnamefont {G.}~\bibnamefont {Burkard}}, \bibinfo {author}
  {\bibfnamefont {D.~V.}\ \bibnamefont {Seletskiy}},\ and\ \bibinfo {author}
  {\bibfnamefont {A.}~\bibnamefont {Leitenstorfer}},\ }\href
  {https://doi.org/10.1038/nature21024} {\bibfield  {journal} {\bibinfo
  {journal} {Nature}\ }\textbf {\bibinfo {volume} {541}},\ \bibinfo {pages}
  {376} (\bibinfo {year} {2017})}\BibitemShut {NoStop}%
\bibitem [{\citenamefont {Boitier}\ \emph {et~al.}(2009)\citenamefont
  {Boitier}, \citenamefont {Godard}, \citenamefont {Rosencher},\ and\
  \citenamefont {Fabre}}]{Boitier_2009}%
  \BibitemOpen
  \bibfield  {author} {\bibinfo {author} {\bibfnamefont {F.}~\bibnamefont
  {Boitier}}, \bibinfo {author} {\bibfnamefont {A.}~\bibnamefont {Godard}},
  \bibinfo {author} {\bibfnamefont {E.}~\bibnamefont {Rosencher}},\ and\
  \bibinfo {author} {\bibfnamefont {C.}~\bibnamefont {Fabre}},\ }\href
  {https://doi.org/10.1038/nphys1218} {\bibfield  {journal} {\bibinfo
  {journal} {Nature Physics}\ }\textbf {\bibinfo {volume} {5}},\ \bibinfo
  {pages} {267} (\bibinfo {year} {2009})}\BibitemShut {NoStop}%
\bibitem [{\citenamefont {Kuzucu}\ \emph {et~al.}(2008)\citenamefont {Kuzucu},
  \citenamefont {Wong}, \citenamefont {Kurimura},\ and\ \citenamefont
  {Tovstonog}}]{Kuzucu_2008}%
  \BibitemOpen
  \bibfield  {author} {\bibinfo {author} {\bibfnamefont {O.}~\bibnamefont
  {Kuzucu}}, \bibinfo {author} {\bibfnamefont {F.~N.~C.}\ \bibnamefont {Wong}},
  \bibinfo {author} {\bibfnamefont {S.}~\bibnamefont {Kurimura}},\ and\
  \bibinfo {author} {\bibfnamefont {S.}~\bibnamefont {Tovstonog}},\ }\href
  {https://doi.org/10.1103/PhysRevLett.101.153602} {\bibfield  {journal}
  {\bibinfo  {journal} {Phys. Rev. Lett.}\ }\textbf {\bibinfo {volume} {101}},\
  \bibinfo {pages} {153602} (\bibinfo {year} {2008})}\BibitemShut {NoStop}%
\bibitem [{\citenamefont {MacLean}\ \emph {et~al.}(2018)\citenamefont
  {MacLean}, \citenamefont {Donohue},\ and\ \citenamefont
  {Resch}}]{MacLean_2018}%
  \BibitemOpen
  \bibfield  {author} {\bibinfo {author} {\bibfnamefont {J.-P.~W.}\
  \bibnamefont {MacLean}}, \bibinfo {author} {\bibfnamefont {J.~M.}\
  \bibnamefont {Donohue}},\ and\ \bibinfo {author} {\bibfnamefont {K.~J.}\
  \bibnamefont {Resch}},\ }\href
  {https://doi.org/10.1103/PhysRevLett.120.053601} {\bibfield  {journal}
  {\bibinfo  {journal} {Phys. Rev. Lett.}\ }\textbf {\bibinfo {volume} {120}},\
  \bibinfo {pages} {053601} (\bibinfo {year} {2018})}\BibitemShut {NoStop}%
\bibitem [{\citenamefont {Specht}\ \emph {et~al.}(2009)\citenamefont {Specht},
  \citenamefont {Bochmann}, \citenamefont {M\"ucke}, \citenamefont {Weber},
  \citenamefont {Figueroa}, \citenamefont {Moehring},\ and\ \citenamefont
  {Rempe}}]{Specht_2009}%
  \BibitemOpen
  \bibfield  {author} {\bibinfo {author} {\bibfnamefont {H.~P.}\ \bibnamefont
  {Specht}}, \bibinfo {author} {\bibfnamefont {J.}~\bibnamefont {Bochmann}},
  \bibinfo {author} {\bibfnamefont {M.}~\bibnamefont {M\"ucke}}, \bibinfo
  {author} {\bibfnamefont {B.}~\bibnamefont {Weber}}, \bibinfo {author}
  {\bibfnamefont {E.}~\bibnamefont {Figueroa}}, \bibinfo {author}
  {\bibfnamefont {D.~L.}\ \bibnamefont {Moehring}},\ and\ \bibinfo {author}
  {\bibfnamefont {G.}~\bibnamefont {Rempe}},\ }\href
  {https://doi.org/10.1038/nphoton.2009.115} {\bibfield  {journal} {\bibinfo
  {journal} {Nature Photonics}\ }\textbf {\bibinfo {volume} {3}},\ \bibinfo
  {pages} {469} (\bibinfo {year} {2009})}\BibitemShut {NoStop}%
\bibitem [{\citenamefont {Dorfman}\ and\ \citenamefont
  {Mukamel}(2012)}]{Dorfman_2012}%
  \BibitemOpen
  \bibfield  {author} {\bibinfo {author} {\bibfnamefont {K.~E.}\ \bibnamefont
  {Dorfman}}\ and\ \bibinfo {author} {\bibfnamefont {S.}~\bibnamefont
  {Mukamel}},\ }\href {https://doi.org/10.1103/PhysRevA.86.013810} {\bibfield
  {journal} {\bibinfo  {journal} {Phys. Rev. A}\ }\textbf {\bibinfo {volume}
  {86}},\ \bibinfo {pages} {013810} (\bibinfo {year} {2012})}\BibitemShut
  {NoStop}%
\bibitem [{\citenamefont {Howell}\ \emph {et~al.}(2004)\citenamefont {Howell},
  \citenamefont {Bennink}, \citenamefont {Bentley},\ and\ \citenamefont
  {Boyd}}]{Howell_2004}%
  \BibitemOpen
  \bibfield  {author} {\bibinfo {author} {\bibfnamefont {J.~C.}\ \bibnamefont
  {Howell}}, \bibinfo {author} {\bibfnamefont {R.~S.}\ \bibnamefont {Bennink}},
  \bibinfo {author} {\bibfnamefont {S.~J.}\ \bibnamefont {Bentley}},\ and\
  \bibinfo {author} {\bibfnamefont {R.~W.}\ \bibnamefont {Boyd}},\ }\href
  {https://doi.org/10.1103/PhysRevLett.92.210403} {\bibfield  {journal}
  {\bibinfo  {journal} {Phys. Rev. Lett.}\ }\textbf {\bibinfo {volume} {92}},\
  \bibinfo {pages} {210403} (\bibinfo {year} {2004})}\BibitemShut {NoStop}%
\bibitem [{\citenamefont {Cresser}(1984)}]{Cresser_1984}%
  \BibitemOpen
  \bibfield  {author} {\bibinfo {author} {\bibfnamefont {J.~D.}\ \bibnamefont
  {Cresser}},\ }\href {https://doi.org/10.1103/PhysRevA.29.1984} {\bibfield
  {journal} {\bibinfo  {journal} {Phys. Rev. A}\ }\textbf {\bibinfo {volume}
  {29}},\ \bibinfo {pages} {1984} (\bibinfo {year} {1984})}\BibitemShut
  {NoStop}%
\bibitem [{Note3()}]{Note3}%
  \BibitemOpen
  \bibinfo {note} {Detection operators are defined at the detection plane, and
  thus given in their natural basis.}\BibitemShut {Stop}%
\bibitem [{Note4()}]{Note4}%
  \BibitemOpen
  \bibinfo {note} {\noindent one can also derive this by first expanding the
  interaction propagator, then transforming the fields. This will be the
  formally correct approach. Here we have transformed the field prior to the
  expansion to express the relative shift between field and matter operators
  already at the interaction Hamiltonian level in the detection picture. It is
  confusing to think of the time ordering in terms of the shifted fields and
  thus here only serves for demonstration purposes}\BibitemShut {NoStop}%
\bibitem [{Note5()}]{Note5}%
  \BibitemOpen
  \bibinfo {note} {The fields are described at the same position. The
  untranslated and dipoles are transforms to the far-field basis, also denoted
  as detection basis.}\BibitemShut {Stop}%
\bibitem [{\citenamefont {Mukamel}(2008)}]{Mukamel_2008}%
  \BibitemOpen
  \bibfield  {author} {\bibinfo {author} {\bibfnamefont {S.}~\bibnamefont
  {Mukamel}},\ }\href {https://doi.org/10.1103/PhysRevA.77.023801} {\bibfield
  {journal} {\bibinfo  {journal} {Phys. Rev. A}\ }\textbf {\bibinfo {volume}
  {77}},\ \bibinfo {pages} {023801} (\bibinfo {year} {2008})}\BibitemShut
  {NoStop}%
\bibitem [{\citenamefont {Dorfman}\ \emph {et~al.}(2019)\citenamefont
  {Dorfman}, \citenamefont {Asban}, \citenamefont {Ye}, \citenamefont {Rouxel},
  \citenamefont {Cho},\ and\ \citenamefont {Mukamel}}]{Dorfman_2019}%
  \BibitemOpen
  \bibfield  {author} {\bibinfo {author} {\bibfnamefont {K.~E.}\ \bibnamefont
  {Dorfman}}, \bibinfo {author} {\bibfnamefont {S.}~\bibnamefont {Asban}},
  \bibinfo {author} {\bibfnamefont {L.}~\bibnamefont {Ye}}, \bibinfo {author}
  {\bibfnamefont {J.~R.}\ \bibnamefont {Rouxel}}, \bibinfo {author}
  {\bibfnamefont {D.}~\bibnamefont {Cho}},\ and\ \bibinfo {author}
  {\bibfnamefont {S.}~\bibnamefont {Mukamel}},\ }\href
  {https://doi.org/10.1021/acs.jpclett.9b00071} {\bibfield  {journal} {\bibinfo
   {journal} {The Journal of Physical Chemistry Letters}\ }\textbf {\bibinfo
  {volume} {10}},\ \bibinfo {pages} {768} (\bibinfo {year} {2019})}\BibitemShut
  {NoStop}%
\bibitem [{\citenamefont {Lee}\ and\ \citenamefont {Goodson}(2006)}]{Lee_2006}%
  \BibitemOpen
  \bibfield  {author} {\bibinfo {author} {\bibfnamefont {D.-I.}\ \bibnamefont
  {Lee}}\ and\ \bibinfo {author} {\bibfnamefont {T.}~\bibnamefont {Goodson}},\
  }\href {https://doi.org/10.1021/jp066767g} {\bibfield  {journal} {\bibinfo
  {journal} {The Journal of Physical Chemistry B}\ }\textbf {\bibinfo {volume}
  {110}},\ \bibinfo {pages} {25582} (\bibinfo {year} {2006})}\BibitemShut
  {NoStop}%
\bibitem [{\citenamefont {Upton}\ \emph {et~al.}(2013)\citenamefont {Upton},
  \citenamefont {Harpham}, \citenamefont {Suzer}, \citenamefont {Richter},
  \citenamefont {Mukamel},\ and\ \citenamefont {Goodson}}]{Upton_2013}%
  \BibitemOpen
  \bibfield  {author} {\bibinfo {author} {\bibfnamefont {L.}~\bibnamefont
  {Upton}}, \bibinfo {author} {\bibfnamefont {M.}~\bibnamefont {Harpham}},
  \bibinfo {author} {\bibfnamefont {O.}~\bibnamefont {Suzer}}, \bibinfo
  {author} {\bibfnamefont {M.}~\bibnamefont {Richter}}, \bibinfo {author}
  {\bibfnamefont {S.}~\bibnamefont {Mukamel}},\ and\ \bibinfo {author}
  {\bibfnamefont {T.}~\bibnamefont {Goodson}},\ }\href
  {https://doi.org/10.1021/jz400851d} {\bibfield  {journal} {\bibinfo
  {journal} {The Journal of Physical Chemistry Letters}\ }\textbf {\bibinfo
  {volume} {4}},\ \bibinfo {pages} {2046} (\bibinfo {year} {2013})}\BibitemShut
  {NoStop}%
\bibitem [{\citenamefont {Varnavski}\ \emph
  {et~al.}(2017{\natexlab{b}})\citenamefont {Varnavski}, \citenamefont
  {Pinsky},\ and\ \citenamefont {Goodson}}]{Varnavski_2017}%
  \BibitemOpen
  \bibfield  {author} {\bibinfo {author} {\bibfnamefont {O.}~\bibnamefont
  {Varnavski}}, \bibinfo {author} {\bibfnamefont {B.}~\bibnamefont {Pinsky}},\
  and\ \bibinfo {author} {\bibfnamefont {T.}~\bibnamefont {Goodson}},\ }\href
  {https://doi.org/10.1021/acs.jpclett.6b02378} {\bibfield  {journal} {\bibinfo
   {journal} {The Journal of Physical Chemistry Letters}\ }\textbf {\bibinfo
  {volume} {8}},\ \bibinfo {pages} {388} (\bibinfo {year}
  {2017}{\natexlab{b}})}\BibitemShut {NoStop}%
\bibitem [{\citenamefont {Schlawin}\ \emph {et~al.}(2018)\citenamefont
  {Schlawin}, \citenamefont {Dorfman},\ and\ \citenamefont
  {Mukamel}}]{Schlawin_2018}%
  \BibitemOpen
  \bibfield  {author} {\bibinfo {author} {\bibfnamefont {F.}~\bibnamefont
  {Schlawin}}, \bibinfo {author} {\bibfnamefont {K.~E.}\ \bibnamefont
  {Dorfman}},\ and\ \bibinfo {author} {\bibfnamefont {S.}~\bibnamefont
  {Mukamel}},\ }\href {https://doi.org/10.1021/acs.accounts.8b00173} {\bibfield
   {journal} {\bibinfo  {journal} {Accounts of Chemical Research}\ }\textbf
  {\bibinfo {volume} {51}},\ \bibinfo {pages} {2207} (\bibinfo {year}
  {2018})}\BibitemShut {NoStop}%
\bibitem [{\citenamefont {Yang}\ \emph {et~al.}(2020)\citenamefont {Yang},
  \citenamefont {Saurabh}, \citenamefont {Schlawin}, \citenamefont {Mukamel},\
  and\ \citenamefont {Dorfman}}]{Yang_2020}%
  \BibitemOpen
  \bibfield  {author} {\bibinfo {author} {\bibfnamefont {Z.}~\bibnamefont
  {Yang}}, \bibinfo {author} {\bibfnamefont {P.}~\bibnamefont {Saurabh}},
  \bibinfo {author} {\bibfnamefont {F.}~\bibnamefont {Schlawin}}, \bibinfo
  {author} {\bibfnamefont {S.}~\bibnamefont {Mukamel}},\ and\ \bibinfo {author}
  {\bibfnamefont {K.~E.}\ \bibnamefont {Dorfman}},\ }\href
  {https://doi.org/10.1063/5.0009575} {\bibfield  {journal} {\bibinfo
  {journal} {Applied Physics Letters}\ }\textbf {\bibinfo {volume} {116}},\
  \bibinfo {pages} {244001} (\bibinfo {year} {2020})}\BibitemShut {NoStop}%
\end{thebibliography}%


%

\newpage{}

\appendix
\begin{widetext}

 \renewcommand{\thefigure}{S\arabic{figure}} 
 \renewcommand{\theequation}{S\arabic{equation}} 
 \renewcommand{\thetable}{S\arabic{table}} 
 \renewcommand{\thesection}{S\arabic{section}} 
 \renewcommand{\thepage}{S\arabic{page}} 
\noindent \begin{center}
\textbf{\textsc{\Large{}Interferometric-Spectroscopy With Quantum-Light;
Revealing Out-of-Time-Ordering Correlators}}\\
\textbf{\textsc{\textcolor{gray}{\Large{}--- Supplementary Information
---}}}{\Large\par}
\par\end{center}

$ $

$ $

$ $

\tableofcontents{}

\newpage{}

\section{Heisenberg picture -- field of a dipole \label{Field of a dipole}}

In this section we solve the Heisenberg equation and obtain the displacement
operator space-time representation. This provides physical intuition
for the seeming irregularity of the in the time ordering of the nonlinear
response when quantum interferometers are involved. We begin by closely
following the derivation done in \citep{Cresser_1984} for single
particle, and then expand it for  multiple scatterers. We calculate
the displacement operator far from the location of the sample cavity.
The sample is composed of multiple scatterers, and is much smaller
than the wavelength of the applied field. In this case, the displacement
operator coincides with the electric operator within the multipolar
expansion. This is a direct result of the localized scatterers model
(for more details see Eq. 17 and 18 in \citep{Cresser_1984}). \\
\\

The field operator far from the sample cavity is given by

\noindent 
\begin{equation}
\boldsymbol{E}\left(\boldsymbol{r},t\right)=\sum_{\boldsymbol{k},s}\sqrt{\frac{2\pi k}{\Omega}}\hat{\epsilon}_{s}\left(\boldsymbol{k}\right)a_{\boldsymbol{k},s}\left(t\right)e^{i\boldsymbol{k}\cdot\boldsymbol{r}}+H.c.,\label{Field operator}
\end{equation}

\noindent where $\hat{\epsilon}_{s}\left(\boldsymbol{k}\right)$ is
the polarization, the speed of light $c=1$, $k=\omega$ and $\Omega$
is the quantization volume. We shall solve Heisenberg's equation of
motion for the annihilation operator 

\noindent 
\begin{align}
\frac{d}{dt}a_{\boldsymbol{k},s} & =i\left[H,a_{\boldsymbol{k},s}\right],
\end{align}

\noindent where the coupling is given by ${\cal H}_{\mu\phi}=\int d^{3}r\boldsymbol{\,E}\left(\boldsymbol{r},t\right)\cdot\boldsymbol{V}\left(\boldsymbol{r},t\right)$
and the dipole operator is $\boldsymbol{V}\left(\boldsymbol{r}\right)=-e\sum_{\alpha}\left(\boldsymbol{r}_{e,\alpha}-\boldsymbol{r}_{\alpha}\right)\delta\left(\boldsymbol{r}-\boldsymbol{r}_{\alpha}\right)$.
$\boldsymbol{r}_{e,\alpha}$ is the position operator of the electron
relative to the nucleus positioned in $\boldsymbol{r}_{\alpha}$.
This equation admits the formal solution

\noindent 
\begin{equation}
a_{\boldsymbol{k},s}\left(t\right)=a_{\boldsymbol{k},s}e^{-ikt}+\sqrt{\frac{2\pi k}{\Omega}}\underset{0}{\overset{t}{\int}}d\tau\underset{{\cal V}_{\mu}}{\int}d^{3}r^{\prime}\,e^{-i\boldsymbol{k}\cdot\boldsymbol{r}^{\prime}-ik\left(t-\tau\right)}\boldsymbol{V}\left(\boldsymbol{r}^{\prime},\tau\right)\cdot\hat{\epsilon}_{s}^{*}\left(\boldsymbol{k}\right).\label{a_k solution}
\end{equation}

\noindent Here ${\cal V}_{\mu}$ is the sample cavity volume. Plugging
in Eq. $\text{\ref{a_k solution}}$ into Eq. $\text{\ref{Field operator}}$
we obtain two contributions, one from the free evolution and the other
due to the dipole acting as a source term,

\noindent 
\begin{equation}
\boldsymbol{E}\left(\boldsymbol{r},t\right)=\boldsymbol{E}_{0}^{\left(+\right)}\left(\boldsymbol{r},t\right)+\frac{1}{\left(2\pi\right)^{2}}\sum_{s}\underset{0}{\overset{t}{\int}}d\tau\,\underset{{\cal V}_{\mu}}{\int}d^{3}r^{\prime}\underset{{\cal V}_{k}}{\int}d^{3}\boldsymbol{k}k\hat{\epsilon}_{s}\left(\boldsymbol{k}\right)e^{-ik\left(t-\tau\right)}\boldsymbol{V}\left(\boldsymbol{r}^{\prime},\tau\right)\cdot\hat{\epsilon}_{s}^{*}\left(\boldsymbol{k}\right)e^{i\boldsymbol{k}\cdot\left(\boldsymbol{r}-\boldsymbol{r}^{\prime}\right)}+H.c.,
\end{equation}

\noindent where ${\cal V}_{k}$ is the relevant integration domain
of the field momentum, and

\noindent 
\begin{equation}
\boldsymbol{E}_{0}\left(\boldsymbol{r},t\right)=\sum_{\boldsymbol{k},s}\sqrt{\frac{2\pi k}{\Omega}}\hat{\epsilon}_{s}\left(\boldsymbol{k}\right)a_{\boldsymbol{k},s}e^{i\boldsymbol{k}\cdot\boldsymbol{r}-ikt}+H.c.
\end{equation}

\noindent Summing over the polarizations yields 

\noindent 
\begin{align*}
\Delta E_{i}\left(\boldsymbol{r},t\right)\equiv E_{j}\left(\boldsymbol{r},t\right)-E_{j,0}\left(\boldsymbol{r},t\right) & =\frac{1}{\left(2\pi\right)^{2}}\underset{0}{\overset{t}{\int}}d\tau\,\underset{{\cal V}_{\mu}}{\int}d^{3}r^{\prime}\int d^{3}\boldsymbol{k}k\left[\sum_{s}\hat{\epsilon}_{s}^{\left(j\right)}\left(\boldsymbol{k}\right)\hat{\epsilon}_{s}^{*\left(i\right)}\left(\boldsymbol{k}\right)\right]V_{i}\left(\boldsymbol{r}^{\prime},\tau\right)e^{i\boldsymbol{k}\cdot\left(\boldsymbol{r}-\boldsymbol{r}^{\prime}\right)-ik\left(t-\tau\right)}+H.c.\\
= & \frac{1}{\left(2\pi\right)^{2}}\underset{0}{\overset{t}{\int}}d\tau\,\underset{{\cal V}_{\mu}}{\int}d^{3}r^{\prime}\int d^{3}\boldsymbol{k}k\left[\delta_{ij}-\frac{k_{i}k_{j}}{k^{2}}\right]V_{i}\left(\boldsymbol{r}^{\prime},\tau\right)e^{i\boldsymbol{k}\cdot\left(\boldsymbol{r}-\boldsymbol{r}^{\prime}\right)-ik\left(t-\tau\right)}+H.c.
\end{align*}

\noindent The $n^{th}$ order polarization is obtained from the $n$
interactions with the same scatterer, we average over the position
of the scatterer (in the cavity volume), and express the $n$-wave-mixing
phase factors explicitly, resulting in the familiar phase-matching
factor

\noindent 
\begin{align}
\Delta E_{i}\left(\boldsymbol{r},t\right)= & \frac{1}{\left(2\pi\right)^{2}}\underset{0}{\overset{t}{\int}}d\tau\,\int d^{3}\boldsymbol{k}\,k\,\left[\delta_{ij}-\frac{k_{i}k_{j}}{k^{2}}\right]\underset{{\cal V}_{\mu}}{\int}d^{3}r^{\prime}e^{i\left(\boldsymbol{k}\pm\sum_{m}^{n}\boldsymbol{k}_{m}\right)\cdot\boldsymbol{r}^{\prime}}\left\langle V_{j}\left(\boldsymbol{r}_{\alpha},\tau\right)\right\rangle _{\left\{ n,\boldsymbol{r}_{\alpha}\right\} }e^{i\boldsymbol{k}\cdot\boldsymbol{r}-ik\left(t-\tau\right)}+H.c.
\end{align}

\noindent We denote $\Delta\boldsymbol{k}=\boldsymbol{k}\pm\sum_{j}^{n}\boldsymbol{k}_{j}$
and carry the spatial integration -- which is possible due to the
assumed uniform distribution of scatterers -- introducing the phase-matching
factor

\noindent 
\begin{equation}
\varphi_{V}\left(\Delta\boldsymbol{k}\right)\equiv2^{d}\underset{{\cal V}_{\mu}}{\int}d^{3}r^{\prime}e^{i\left(\boldsymbol{k}\pm\sum_{j}^{n}\boldsymbol{k}_{j}\right)\cdot\boldsymbol{r}^{\prime}},
\end{equation}

\noindent where $d$ is the dimension of the cavity (normalization),
and obtain

\noindent 
\begin{equation}
\Delta E_{i}\left(\boldsymbol{r},t\right)=\frac{2^{d}}{\left(2\pi\right)^{2}}\underset{0}{\overset{t}{\int}}d\tau\,\int d^{3}\boldsymbol{k}\,k\,\left[\delta_{ij}-\frac{k_{i}k_{j}}{k^{2}}\right]\varphi_{V}\left(\Delta\boldsymbol{k}\right)\left\langle V_{j}\left(\boldsymbol{r}_{\alpha},\tau\right)\right\rangle _{\left\{ n,\boldsymbol{r}_{\alpha}\right\} }e^{i\boldsymbol{k}\cdot\boldsymbol{r}-ik\left(t-\tau\right)}+H.c.
\end{equation}

\noindent Computing the angular integration, using $\int d\Omega\left(\delta_{ij}-\hat{\boldsymbol{k}}_{i}\hat{\boldsymbol{k}}_{j}\right)e^{i\boldsymbol{k}\cdot\boldsymbol{r}}=$$-\left(\partial_{ii}^{2}-\partial_{i}\partial_{j}\right)\frac{\sin\left(kr\right)}{k^{3}r}$
then yields

\noindent 
\begin{equation}
\Delta E_{i}\left(\boldsymbol{r},t\right)=\frac{2^{d}}{\left(2\pi\right)^{2}}\underset{0}{\overset{t}{\int}}d\tau\,\int dk\,k^{3}\left[-\left(\partial_{ii}^{2}-\partial_{i}\partial_{j}\right)\frac{\sin\left(kr\right)}{k^{3}r}\right]\varphi_{V}\left(\Delta\boldsymbol{k}\right)\left\langle V_{j}\left(\boldsymbol{r}_{\alpha},\tau\right)\right\rangle _{\left\{ n,\boldsymbol{r}_{\alpha}\right\} }e^{-ik\left(t-\tau\right)}+H.c.
\end{equation}

\noindent Next, we average over the rotations $\left\langle \left(-\partial_{ii}^{2}+\partial_{i}\partial_{j}\right)\frac{e^{ikr}}{r}\right\rangle _{\text{rotation}}\rightarrow\frac{2}{3}k^{2}\delta_{ij}\frac{e^{ikr}}{r}$
and obtain

\noindent 
\begin{equation}
\Delta E_{i}\left(\boldsymbol{r},t\right)=\frac{2}{3}\frac{2^{d}}{\left(2\pi\right)^{2}}\underset{0}{\overset{t}{\int}}d\tau\,\int dk\left[k^{2}\frac{e^{ikr-ik\left(t-\tau\right)}}{2ir}\right]\varphi_{V}\left(\Delta\boldsymbol{k}\right)\left\langle V_{i}\left(\boldsymbol{r}_{\alpha},\tau\right)\right\rangle _{\left\{ n,\boldsymbol{r}_{\alpha}\right\} }+H.c.
\end{equation}

\noindent Here, the backward propagation was eliminated due to causality.
We assumed the slowly varying amplitude approximation for the field,
extracting the central $k_{0}$ from the integration. The phase matching
approaches unity for $\Delta\boldsymbol{k}\rightarrow0$. We assume
small $\Delta\boldsymbol{k}L$ and carry the integration (considering
the phase-matching a multiplicative prefactor)

\noindent 
\begin{equation}
\Delta E_{i}\left(\boldsymbol{r},t\right)=-\frac{ik_{0}^{2}2^{d}}{12\pi^{2}r}\varphi_{V}\left(\Delta\boldsymbol{k}\right)\underset{0}{\overset{t}{\int}}d\tau\,\int dke^{ikr-ik\left(t-\tau\right)}\left\langle V_{i}\left(\boldsymbol{r}_{\alpha},\tau\right)\right\rangle _{\left\{ n,\boldsymbol{r}_{\alpha}\right\} }+H.c.
\end{equation}

\noindent resulting in

\noindent 
\begin{equation}
\Delta E_{i}\left(\boldsymbol{r},t\right)=-i\frac{2^{d}\pi}{3\lambda_{0}^{2}r}\left\langle V_{i}\left(\boldsymbol{r}_{\alpha},t-r\right)\right\rangle _{\left\{ n,\boldsymbol{r}_{\alpha}\right\} }\varphi_{V}\left(\Delta\boldsymbol{k}\right)\Theta\left(t-r\right)+H.c.\label{eq: n^th order polarization radiation}
\end{equation}

\noindent where we have used the step function $\Theta\left(t\right)$,
and defined the central wavelength $\lambda_{0}=\nicefrac{2\pi}{k_{0}}$.
Equation $\text{\ref{eq: n^th order polarization radiation}}$ is
constitutes a microscopic derivation for dipole radiation due to nonlinear
field interactions -- corresponding to Eq. $\text{4.75}$ in \citep{Mukamel_1995}.
Averaging over nuclei coordinates gives rise to the structure factor
in addition. 

Note that the $\nicefrac{\pi}{2}$ factor of the dipole contribution
to the radiation is a result of point-like source radiation density
of the scatterer while the incoming fields are understood as directional.
This phase is closely related to the Gouy phase and require integration
of phase-matching condition as well. For a narrow cylinder shaped
cavity of length $L$, the phase matching takes the familiar expression
(for which $d=1$)

\noindent 
\begin{equation}
\varphi_{V}\left(\Delta\boldsymbol{k}\right)\equiv\varphi_{L}\left(\Delta\boldsymbol{k}\right)=\frac{L}{2}\text{sinc}\left(\Delta\boldsymbol{k}\frac{L}{2}\right)
\end{equation}

\noindent resulting in 

\begin{equation}
\Delta E_{i}\left(\boldsymbol{r},t\right)=-\frac{2\pi i}{3\lambda_{0}^{2}r}\left\langle V_{i}\left(\boldsymbol{r}_{\alpha},t-r\right)\right\rangle _{\left\{ n,\boldsymbol{r}_{\alpha}\right\} }\left[\frac{L}{2}\text{sinc}\left(\Delta\boldsymbol{k}\frac{L}{2}\right)\right]\Theta\left(t-r\right)+H.c.
\end{equation}

Generally, when a single wave is considered, the electric field in
the detection plane is given using similar calculation by 

\begin{equation}
\boldsymbol{E}\left(\boldsymbol{r},t\right)=\boldsymbol{E}_{0}\left(\boldsymbol{r},t\right)-i\boldsymbol{V}\left(t-\frac{r}{c}\right)\Theta\left(t-\frac{r}{c}\right),\label{Field after interaction}
\end{equation}

\noindent where we have restored the speed of light to highlight the
retardation in the observed time of the dipole oscillation. Here we
also assume spatial averaging over scatterer positions, yet express
the fact that $\boldsymbol{V}$ maintain operator form. Eq. $\text{\ref{Field after interaction}}$
implies that path transformation of the field is associated with a
similar transformation to the respective dipole vector (in the transverse
plane). Explicitly, the translation $r\rightarrow r+\Delta r$ implies
the temporal shift $t\rightarrow t-T$ where $T=\nicefrac{\Delta r}{c}$.
This means that the dipole radiation transforms as a vector as one
would naturally expect. 

similar results are obtained in nonlinear optics macroscopically by
using the Maxwell equations \citet{Mukamel_1995}. The present derivation
is fully microscopic.

\section{Out-of-Time-Ordering Matter Correlators}

\subsection{Definitions}

\noindent We consider the setup depicted in Fig. $\text{\ref{Fig 1-1}a}$,
as described in the main text. Two modes $\left\{ E_{a},E_{b}\right\} $
are prepared in a superposition using BS1, then interact with a sample.
Finally the modes are counter rotated by BS2 and measured in coincidence
respectively at spacetime coordinates $\left\{ \boldsymbol{r}_{a}t_{a},\boldsymbol{r}_{b}t_{b}\right\} $.
The recorded signal is given by simultaneous annihilation of two photons
from both sides of the density operator \footnote{Detection operators are defined at the detection plane, and thus given
in their natural basis.}\\

\begin{align}
{\cal C}_{ab}\left(T,\tau\right)= & \left\langle {\cal T}{\cal O}_{t_{a},t_{b}}\left(\boldsymbol{{\cal E}},\boldsymbol{{\cal E}}^{\dagger}\right)\exp\left\{ -\frac{i}{\hbar}\underset{t_{0}}{\overset{t}{\int}}du{\cal H}_{\text{\ensuremath{\mu\phi}},-}\left(u\right)\right\} \right\rangle \label{Main signal}\\
{\cal O}_{t_{a},t_{b}} & \left(\boldsymbol{{\cal E}},\boldsymbol{{\cal E}}^{\dagger}\right)={\cal E}_{R,a}^{\dagger}\left(\boldsymbol{r}_{a}t_{a}\right){\cal E}_{R,b}^{\dagger}\left(\boldsymbol{r}_{b}t_{b}\right){\cal E}_{L,b}\left(\boldsymbol{r}_{b}t_{b}\right){\cal E}_{L,a}\left(\boldsymbol{r}_{a}t_{a}\right).\nonumber 
\end{align}
\\

\noindent Here, the field is broken down into its positive and negative
frequency components according to $\boldsymbol{E}=\boldsymbol{{\cal E}}+\boldsymbol{{\cal E}}^{\dagger}$,
and ${\cal E}_{L,i},{\cal E}_{R,i}\left({\cal E}_{L,i}^{\dagger},{\cal E}_{R,i}^{\dagger}\right)$
describe electric field annihilation (creation) operators of mode
$i\in a,b$, whereby the subscripts $\left(L,R\right)$ correspond
to left and right operation on the density operator. The signal is
obtained by a coincidence measurement of temporally resolved individual
photons. We are interested in calculating the contribution due to
the diagram given in Fig. $\text{\ref{Fig 1-1}b}$, according to the
\emph{order of arrival }(detection plane). The field $\boldsymbol{E}\left(t\right)=\left[E_{a}\left(t\right),E_{b}\left(t\right)\right]^{T}$
undergoes transformation with respect to the initial state using the
rotation matrix (in frequency domain),\\

\noindent 
\begin{equation}
\hat{{\cal R}}_{T}=\frac{1}{\sqrt{2}}\left(\begin{array}{cc}
1 & ie^{-i\omega T}\\
ie^{i\omega T} & 1
\end{array}\right).
\end{equation}
\\

\noindent Here $T$ is a the relative delay. The initial state of
the field is given by, $\vert0\rangle$\\

\noindent 
\begin{equation}
\vert\Psi_{0}\rangle=\int d\omega_{a}d\omega_{b}\Phi\left(\omega_{a},\omega_{b}\right)a^{\dagger}\left(\omega_{a}\right)b^{\dagger}\left(\omega_{b}\right)\vert0\rangle,
\end{equation}
\\
the initial state can be represented in the detection basis using
the inverse rotation

\noindent 
\[
\left(\begin{array}{c}
a\left(\omega\right)\\
b\left(\omega\right)
\end{array}\right)_{\text{input}}=\frac{1}{\sqrt{2}}\left(\begin{array}{cc}
1 & -ie^{-i\omega T}\\
-ie^{i\omega T} & 1
\end{array}\right)\left(\begin{array}{c}
a\left(\omega\right)\\
b\left(\omega\right)
\end{array}\right)_{\text{detection}}.
\]

\noindent This gives rise to the following transformation 

\noindent {\small{}
\begin{equation}
a^{\dagger}\left(\omega_{a}\right)b^{\dagger}\left(\omega_{b}\right)\rightarrow\frac{1}{2}\left[a^{\dagger}\left(\omega_{a}\right)b^{\dagger}\left(\omega_{b}\right)-e^{i\left(\omega_{a}-\omega_{b}\right)T}a^{\dagger}\left(\omega_{b}\right)b^{\dagger}\left(\omega_{a}\right)+ie^{-i\omega_{b}T}a^{\dagger}\left(\omega_{a}\right)a^{\dagger}\left(\omega_{b}\right)+ie^{i\omega_{a}T}b^{\dagger}\left(\omega_{b}\right)b^{\dagger}\left(\omega_{a}\right)\right].
\end{equation}
}{\small\par}

\noindent In the derivation below we only keep the terms that contain
two modes.

\noindent 
\begin{figure}[h]
\begin{centering}
\includegraphics[scale=0.35]{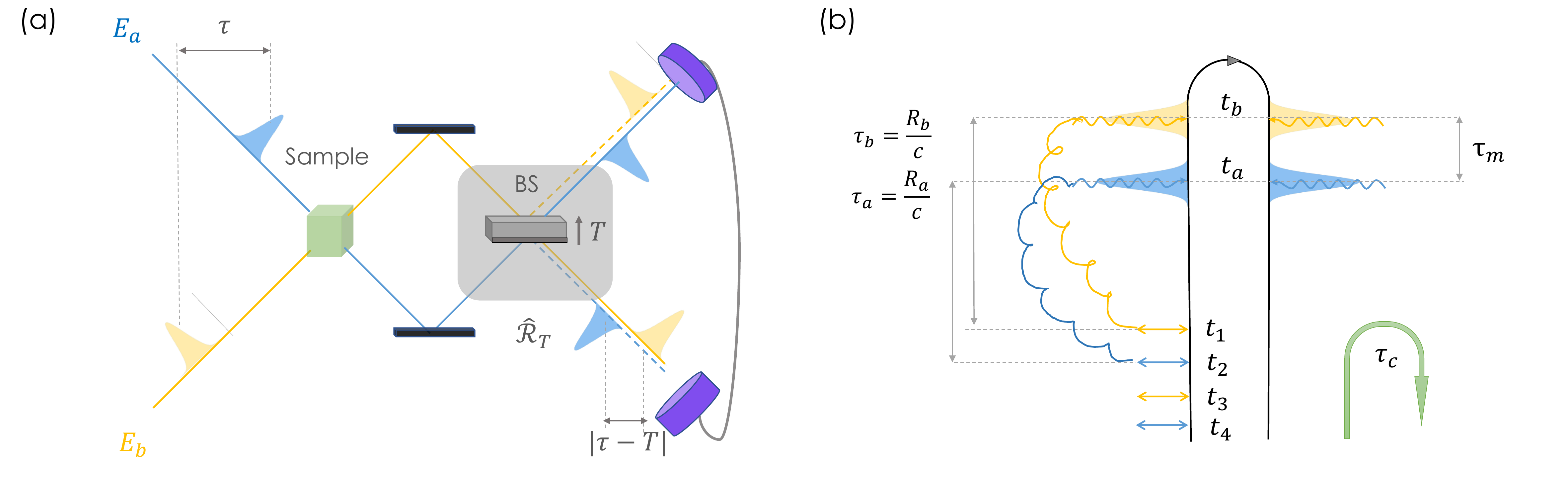}
\par\end{centering}
\caption{\textbf{Demonstration of time-ordering wiggles}. a.) Two photons with
$\tau$ relative delay time, interact with a sample, then combined
a beam-splitter BS. Each photon is scanned in time $\left(t_{a},t_{b}\right)$,
forming time resolved coincidence detection at two detectors b.) The
diagrammatic representation of the considered in the derivation below
that contributes to the entire signal, and results in OTOC in some
parameter regime. \label{Fig 1-1}}
\end{figure}

The Interaction Hamiltonian is initially given by ${\cal H}_{\mu\phi}=\boldsymbol{E}\left(t\right)\cdot\boldsymbol{V}\left(t\right)$
is transformed according to $\boldsymbol{E}\vert_{\text{detected}}=\hat{{\cal R}}^{\dagger}\boldsymbol{E}_{\text{interaction}}$,
and thus given in the detection basis by \footnote{\noindent one can also derive this by first expanding the interaction
propagator, then transforming the fields. This will be the formally
correct approach. Here we have transformed the field prior to the
expansion to express the relative shift between field and matter operators
already at the interaction Hamiltonian level in the detection picture.
It is confusing to think of the time ordering in terms of the shifted
fields and thus here only serves for demonstration purposes}. Similarly, the dipole radiation is transformed as a vector -- see
Sec. $\text{\ref{Field of a dipole}}$ and the discussion following
Eq. $\text{\ref{Field after interaction}}$. Each interaction in the
far-field is given by a superposition of terms 

\noindent 
\begin{align}
{\cal H}_{\mu\phi}\left(t\right)\vert_{\text{detected}} & =\left[\hat{{\cal R}}^{\dagger}\boldsymbol{V}\left(t\right)\right]\cdot\left[\hat{{\cal R}}^{\dagger}\boldsymbol{E}_{\text{detected}}\left(t\right)\right]\nonumber \\
= & \frac{1}{2}\left[\boldsymbol{E}\left(t\right)\cdot\boldsymbol{V}\left(t\right)-E_{a}\left(t+T\right)V_{a}\left(t+T\right)-E_{b}\left(t-T\right)V_{b}\left(t-T\right)\right.\label{eq: Interaction in detection picture}\\
 & \left.-iE_{b}\left(t+T\right)V_{a}\left(t-T\right)-iE_{a}\left(t-T\right)V_{b}\left(t+T\right)\right]
\end{align}
{\small{}}\\

\noindent where we have omitted the spatial coordinate for brevity,
defined $\boldsymbol{E}_{\text{detected}}\equiv\boldsymbol{E}$, and
wrote explicitly the polarization operator with respect to its two
components $\boldsymbol{V}=\left(V_{a},V_{b}\right)^{T}$. The first
two terms constrain the light-matter interaction of individual mode
as a single time event. The last two terms account for a mode generated
by oscillation of $V_{a}\left(V_{b}\right)$, appearing at the $b\left(a\right)$
detector -- hence the apparent $2T$ relative shift. Having all the
ingredients defined, we proceed to the derivation of the contribution
of the process in Fig. $\text{\ref{Fig 1-1}}$b. \\

\subsection{Derivation}

\noindent The coincidence counting of the two fields in the detection
picture is given by the time-ordered product (near field),\\

\noindent 
\begin{align}
{\cal C}_{ab}\left(T\right) & ={\cal T}\underset{-\infty}{\overset{t_{a}}{\int}}dt_{1}\underset{-\infty}{\overset{t_{1}}{\int}}dt_{2}\underset{-\infty}{\overset{t_{2}}{\int}}dt_{3}\underset{-\infty}{\overset{t_{3}}{\int}}dt_{4}\left\langle {\cal E}_{a}^{\dagger}\left(t_{a}-\frac{R_{a}}{c}\right){\cal E}_{b}^{\dagger}\left(t_{b}-\frac{R_{b}}{c}\right){\cal E}_{b}\left(t_{b}-\frac{R_{b}}{c}\right){\cal E}_{a}\left(t_{a}-\frac{R_{a}}{c}\right)\right.\nonumber \\
\times & \left.E_{b}\left(t_{1}\right)E_{a}\left(t_{2}\right)E_{b}\left(t_{3}\right)E_{a}\left(t_{4}\right)\right\rangle \left\langle V_{a}\left(t_{1}\right)V_{b}\left(t_{2}\right)V_{a}\left(t_{3}\right)V_{b}\left(t_{4}\right)\right\rangle ,
\end{align}

\noindent In the far-field (detection plane), multiple combinations
of Eq. $\text{\ref{eq: Interaction in detection picture}}$ contribute
\footnote{The fields are described at the same position. The untranslated and
dipoles are transforms to the far-field basis, also denoted as detection
basis.}. We focus on one contribution 

\noindent 
\begin{align}
{\cal C}_{ab}^{\left(I\right)}\left(T\right) & =\underset{-\infty}{\overset{t_{a}}{\int}}dt_{1}\underset{-\infty}{\overset{t_{1}}{\int}}dt_{2}\underset{-\infty}{\overset{t_{2}}{\int}}dt_{3}\underset{-\infty}{\overset{t_{3}}{\int}}dt_{4}\left\langle {\cal E}_{a}^{\dagger}\left(t_{a}-\frac{R_{a}}{c}\right){\cal E}_{b}^{\dagger}\left(t_{b}-\frac{R_{b}}{c}\right){\cal E}_{b}\left(t_{b}-\frac{R_{b}}{c}\right){\cal E}_{a}\left(t_{a}-\frac{R_{a}}{c}\right)\right.\nonumber \\
\times & \left.E_{a}\left(t_{1}+T\right)E_{b}\left(t_{2}-T\right)E_{a}\left(t_{3}+T\right)E_{b}\left(t_{4}-T\right)\right\rangle \left\langle V_{a}\left(t_{1}+T\right)V_{b}\left(t_{2}-T\right)V_{a}\left(t_{3}+T\right)V_{b}\left(t_{4}-T\right)\right\rangle ,
\end{align}

\noindent Note that time ordering is applied by the integration boundaries.
The fields expectation value we are interested in is given by\\

\noindent {\small{}
\[
\left\langle \boldsymbol{1}_{a'},\boldsymbol{1}_{b'}\vert{\cal E}_{a}^{\dagger}\left(t_{a}-\frac{R_{a}}{c}\right){\cal E}_{b}^{\dagger}\left(t_{b}-\frac{R_{b}}{c}\right){\cal E}_{b}\left(t_{b}-\frac{R_{b}}{c}\right){\cal E}_{a}\left(t_{a}-\frac{R_{a}}{c}\right)E_{a}\left(t_{1}+T\right)E_{b}\left(t_{2}-T\right)E_{a}\left(t_{3}+T\right)E_{b}\left(t_{4}-T\right)\vert\boldsymbol{1}_{a},\boldsymbol{1}_{b}\right\rangle .
\]
}{\small\par}

\noindent We are interested in the following contraction (according
diagram in Fig. $\text{\ref{Fig 1-1}}$b)\\

\noindent {\small{}$$
\langle  
\wick{         
	\c3 {\boldsymbol{1}_{a'}}, \c4 {\boldsymbol{1}_{b'}}\vert		      	         
	\c3 {{\cal E}_{a}^{\dagger}\left(t_{a}-\frac{R_{a}}{c}\right)} 
	\c4 {{\cal E}_{b}^{\dagger}\left(t_{b}-\frac{R_{b}}{c}\right)}
	\c4 {\cal E}_{b}\left(t_{b}-\frac{R_{b}}{c}\right)  
	\c3 {\cal E}_{a}\left(t_{a}-\frac{R_{a}}{c}\right)%
	\c3 {E_{a}\left(t_{1}+T\right)} 
	\c4 {E_{b}\left(t_{2}-T\right)} 
	\c2 {E_{a}\left(t_{3}+T\right)} 
	\c1 {E_{b}\left(t_{4}-T\right)}\vert         
	\c2 {\boldsymbol{1}_{a}}, \c1 {\boldsymbol{1}_{b}} %
} 
\rangle %
$$

}{\small\par}

\noindent resulting in \\

\noindent {\footnotesize{}
\[
\langle\boldsymbol{1}_{a'},\boldsymbol{1}_{b'}\vert{\cal E}_{a}^{\dagger}\left(t_{a}-\frac{R_{a}}{c}\right){\cal E}_{b}^{\dagger}\left(t_{b}-\frac{R_{b}}{c}\right)\rvac\lvac{\cal E}_{a}\left(t_{a}-\frac{R_{a}}{c}\right)E_{a}\left(t_{1}+T\right)\rvac\lvac{\cal E}_{b}\left(t_{b}-\frac{R_{b}}{c}\right)E_{b}\left(t_{2}-T\right)\rvac\lvac E_{a}\left(t_{3}+T\right)E_{b}\left(t_{4}-T\right)\vert\boldsymbol{1}_{a},\boldsymbol{1}_{b}\rangle.
\]
}\\

\noindent This process corresponds to two interactions with both photons,
resulting in two photon populated state which freely propagates from
the sample to the detector. We remember that we focus on a single
term to express the OTOC contribution and take the limit of large
$t_{a}$ to obtain\\

\noindent {\small{}
\begin{align}
{\cal C}_{ab}^{\left(I\right)}\left(T\right)= & \underset{-\infty}{\overset{\infty}{\int}}dt_{1}\underset{-\infty}{\overset{t_{1}}{\int}}dt_{2}\underset{-\infty}{\overset{t_{2}}{\int}}dt_{3}\underset{-\infty}{\overset{t_{3}}{\int}}dt_{4}\int d\omega_{a}^{'}d\omega_{b}^{'}d\omega_{a}d\omega_{b}\Phi^{*}\left(\omega_{a}^{'},\omega_{b}^{'}\right)\Phi\left(\omega_{a},\omega_{b}\right)e^{i\omega_{a}^{'}\left(t_{a}-\frac{R_{a}}{c}\right)}e^{i\omega_{b}^{'}\left(t_{b}-\frac{R_{b}}{c}\right)}e^{-i\omega_{a}\left(t_{3}+T\right)}e^{-i\omega_{b}\left(t_{4}-T\right)}\nonumber \\
\times & \delta\left(t_{a}-\frac{R_{a}}{c}-t_{1}-T\right)\delta\left(t_{b}-\frac{R_{b}}{c}-t_{2}+T\right)\left\langle V_{a}\left(t_{a}-\frac{R_{a}}{c}\right)V_{b}\left(t_{b}-\frac{R_{b}}{c}\right)V_{a}\left(t_{3}+T\right)V_{b}\left(t_{4}-T\right)\right\rangle ,
\end{align}
}\\

\noindent The delta distributions arise due to the curved lines depicted
in Fig. $\text{\ref{Fig 1-1}}$a. They reflect the free photon propagation
from the sample to the detector at the speed of light. Their computation
follows similar steps to the ones taken in the derivation of Eq. $\text{\ref{Field after interaction}}$.
We assume the detection occurs much later than the interaction process
by taking $\left\{ t_{a/b},\frac{R_{a/b}}{c}\right\} \rightarrow\infty$
such that their difference is finite. Usually we are interested in
this point in expressing the signal in frequency domain. Here, we
will express the signal in time domain to highlight the time ordering
of the correlation function by taking the frequency integrals first,\\

\noindent 
\begin{align}
{\cal C}_{ab}^{\left(I\right)}\left(T\right)\propto & \underset{-\infty}{\overset{t_{b}-\frac{R_{b}}{c}+T}{\int}}dt_{3}\underset{-\infty}{\overset{t_{3}}{\int}}dt_{4}\Phi^{*}\left(t_{a}-\frac{R_{a}}{c},t_{b}-\frac{R_{b}}{c}\right)\Phi\left(t_{3}+T,t_{4}-T\right)\nonumber \\
\times & \left\langle V_{a}\left(t_{a}-\frac{R_{a}}{c}\right)V_{b}\left(t_{b}-\frac{R_{b}}{c}\right)V_{a}\left(t_{3}+T\right)V_{b}\left(t_{4}-T\right)\right\rangle .
\end{align}
\\
Separating the initial state amplitudes reads\\

\noindent 
\begin{align}
{\cal C}_{ab}^{\left(I\right)}\left(T\right)\propto & \underset{-\infty}{\overset{t_{b}-\frac{R_{b}}{c}+T}{\int}}dt_{3}\underset{-\infty}{\overset{t_{3}}{\int}}dt_{4}\phi_{a}^{*}\left(t_{a}-\frac{R_{a}}{c}\right)\phi_{b}^{*}\left(t_{b}-\frac{R_{b}}{c}\right)\phi_{a}\left(t_{3}+T\right)\phi_{b}\left(t_{4}-T\right)\nonumber \\
\times & \left\langle V_{a}\left(t_{a}-\frac{R_{a}}{c}\right)V_{b}\left(t_{b}-\frac{R_{b}}{c}\right)V_{a}\left(t_{3}+T\right)V_{b}\left(t_{4}-T\right)\right\rangle ,
\end{align}
\\

\noindent To get some physical intuition on the measured dynamics,
we consider $\phi_{a}\rightarrow\delta_{\epsilon}\left(t-\tau\right)$
and $\phi_{b}\rightarrow\delta_{\epsilon}\left(t\right)$. Plugging
in the coincidence count we obtain\\

\begin{align}
{\cal C}_{ab}^{\left(I\right)}\left(T,\tau\right) & \propto\left\langle V_{a}\left(\tau\right)V_{b}\left(0\right)V_{a}\left(\tau\right)V_{b}\left(0\right)\right\rangle ,\label{OTOC expression}
\end{align}
\\

\noindent for $\tau=2T$. Eq. $\text{\ref{OTOC expression}}$ reflects
the out-of-time-ordering correlator (OTOC) -- reading from right
to left -- the time evolution is positive then negative, and finally
positive again. Such contributions can be calculated according to
the time-contour depicted in Fig. $\text{\ref{Fig 2-1}}$. Crucially,
for wider photon distributions the OTOC contributes in some intervals.

\begin{figure}[h]
\begin{centering}
\includegraphics[scale=0.37]{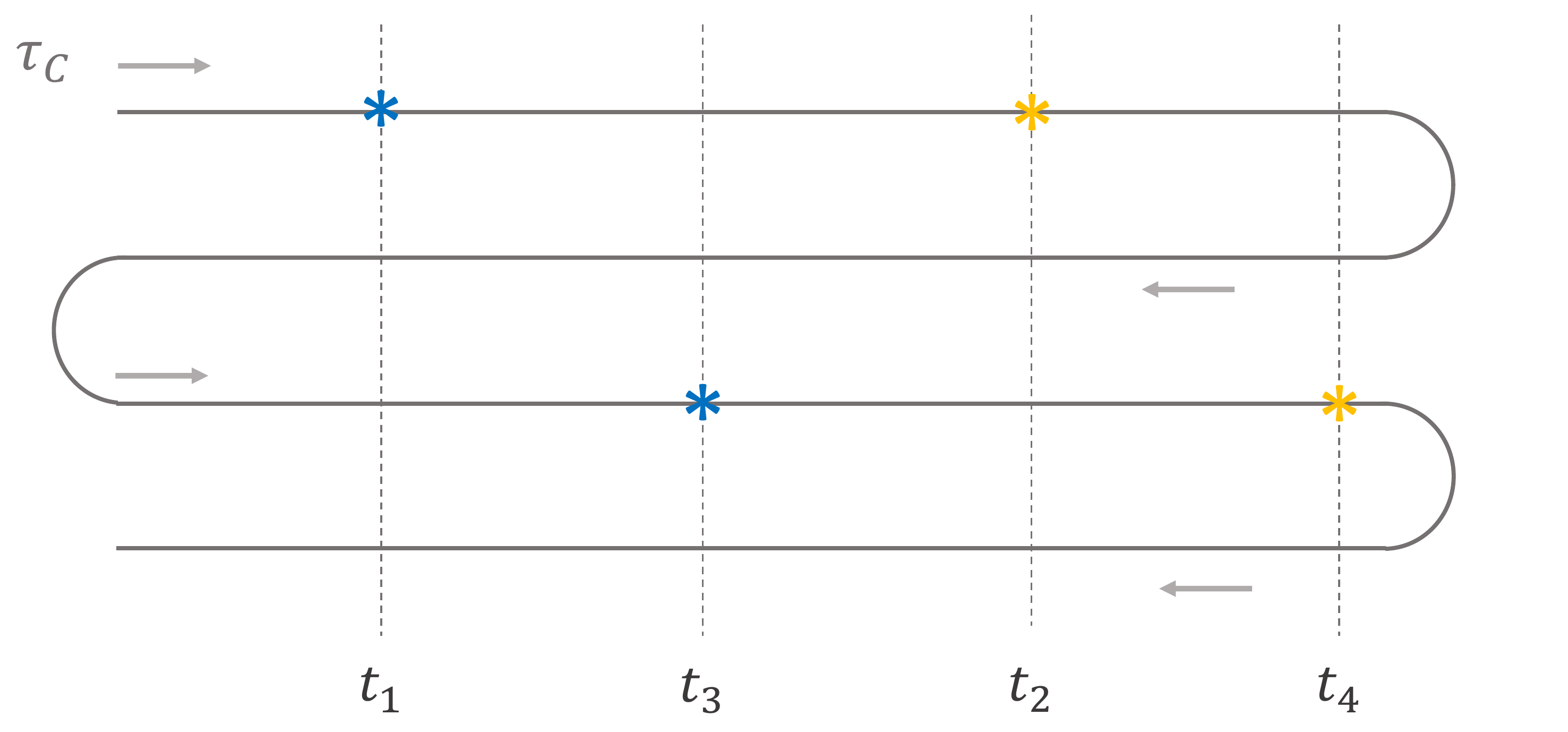}
\par\end{centering}
\caption{The wiggling-time contour $\tau_{C}$, along which the matter OTOC
demonstrated by Eq. $\text{\ref{OTOC expression}}$ can be computed,
as proposed in \citep{Aleiner_2016}. \label{Fig 2-1}}
\end{figure}

\subsubsection*{The rotating wave approximation}

In the above derivation, the field operators took specific form due
to the free propagators and inhomogenity (initial photons), while
the polarization assumed both components. This means that matter vacuum
fluctuations contribute to the over all signal. Now, we invoke the
rotating wave approximation in which absorption and emission are associated
with decrease and increase of photons respectively. This yields the
following modification to the OTOC\\

\begin{align}
{\cal C}_{ab}^{\left(I\right)}\left(T,\tau\right) & \rightarrow\left\langle \mu_{a}\left(\tau\right)\mu_{b}\left(0\right)\mu_{a}^{\dagger}\left(\tau\right)\mu_{b}^{\dagger}\left(0\right)\right\rangle .
\end{align}

\subsubsection*{Implementation using a narrowband entangled photon pair}

OTOC contributions appear also when an entangled photon state is considered.
We consider a narrowband SPDC entangled photon pair, and compute the
coincidence signal. When the entangled pair coherence time $T_{\text{e}}$
is much shorter than the characteristic timescale of the sample (dipole
temporal variation), the two photon can be approximated to arrive
within a narrow time window \citep{Dorfman_2016}

\[
\lvac E_{a}\left(t_{1}\right)E_{b}\left(t_{2}\right)\vert\Phi\rangle\propto T_{\text{e}}\delta\left(t_{1}-t_{2}\right).
\]

\noindent Plugging this into the coincidence signal, we obtain integrated
correlation function that can be divided into OTOC contribution and
a time ordered correlator (TOC) 

\noindent 
\begin{equation}
{\cal C}_{ab}^{\left(I\right)}\left(T,\tau\right)\propto\Pi\left(\frac{t_{a}-t_{b}+2T-\tau}{T_{\text{e}}}\right)\underset{0}{\overset{2T}{\int}}dt\left\langle V_{a}G\left(\tau\right)V_{b}G\left(-t\right)V_{a}G\left(\tau\right)V_{b}\right\rangle +\text{TOC}.
\end{equation}

\noindent where we have used the superoperator Green's function $G\left(t\right)=\left(-\frac{i}{\hbar}\right)\theta\left(t\right)\exp\left(-\frac{i}{\hbar}H_{-}t\right)$
\citep{Mukamel_2008}. Here $\Pi\left(t\right)=1$ $\forall t\in\left(-\nicefrac{1}{2},\nicefrac{1}{2}\right)$
and vanishes out of this interval (approached $\delta\left(t\right)$
as $T_{\text{e}}\rightarrow0$) . Clearly the time flow of the first
contribution wiggles from positive to negative and back to positive
( for $\tau>0$). The narrow arrival time window eliminates some of
the temporal integration. The remaining integration is a result of
the spontaneous generation time of the SPDC entangled pair.

\section{Homodyne vs heterodyne detection}

Homodyne detection constitutes a measurement of the signal photons
generated from a vacuum (e.g. spontaneous emission). Such signals
are relatively weak. The other, heterodyne detection, is based on
interference of the weak signal photons with a much stronger local
oscillator field. The resulting interferometric pattern has improved
signal-to-noise ratio and further provides the phase of the electric
field. In classical spectroscopy, i.e. using classical light the signal
and local oscillator fields are independently generated and thus,
uncorrelated. Combining interferometric tools typically used in quantum
optics such as: Mach-Zehnder, Hong-Ou-Mandel and Franson interferometers
with quantum light allows to utilize correlations between the signal
and local oscillator fields. Rather than measuring a single intensity
(photon number $G^{(1)}$), photon statistics measurements (e.g. photon
coincidence or intensity variance $G^{(2)}$) can exploit the quantum
nature of light. Examples involving two or more intensities include
biphoton spectroscopy, ghost imaging, and photon counting spectroscopy
all relying on interferometric setups. Photon counting signals may
be expressed in terms of multipoint correlation functions of the incoming
fields. Spectroscopy is classical if all fields are in a coherent
state and the observables are given by normally ordered products of
field amplitudes. Field correlation functions may reflect genuine
quantum field effects but may also arise from stochastic classical
fields. The two should be sorted out. Glauber's celebrated hierarchical
correlation formulation of the radiation field aims at field characterization.
For spectroscopy applications it must be extended to explicitly include
the interaction with matter. Nonlinear optical signals induced by
quantized light fields are expressed using \textit{time-ordered} multipoint
correlation functions of \textit{superoperators} in the joint field
plus matter phase space. These are distinct from Glauber's photon
counting formalism which employs \textit{normally ordered} products
of \textit{ordinary operators} in the field phase space. Glauber's
$G^{(2)}$ function of the incoming light is directly related to its
ability to induce correlations in matter. The exploitation of strong
correlations with quantum light in nonlinear spectroscopy offers new
means for probing complex quantum systems.

Note, that the same interferometer can be used for both homodyne and
heterodyne detection of optical signals. For instance, MZ interferometer
can detect linear $\chi^{(3)}$ if all field-matter interactions including
incoming and detected fields are treated on equal footing. In this
case the coincidence signal can be written as a four-point correlation
function of the dipole operators. On the other hand, if detected modes
are treated separately and traced over the vacuum states, while the
incoming modes are traced over the incoming state. In this case the
four-point correlation function factorizes into a product of $\chi^{(3)*}\chi^{(3)}$
which yields a homodyne signal.

\section{Light sources \label{Sec: Light-sources}}

Quantum light sources are typically classified as such, if they have
statistics different from that of a coherent state (e.g., laser).
Coherent sources have a Poissonian distribution of photon number.
Therefore, the average photon number of a coherent state $|\alpha\rangle$
coincides with its variance $N\equiv\left\langle N\right\rangle =\left\langle \Delta N^{2}\right\rangle =\left|\alpha\right|^{2}$.
This yields the known average-standard deviation ratio $\nicefrac{N}{\sqrt{\left\langle \Delta N^{2}\right\rangle }}=N^{-\nicefrac{1}{2}}$.
The coherent state can be represented as a superposition of Fock states
$|n\rangle$. A single photon Fock state is the most basic quantum
state which can rely on the benefits of the interferometric setup.
Unlike classical states single photon state can interfere with itself
giving rise to bunching and antibunching effect, depending on whether
interference is constructive or destructive.

Multiphoton Fock states can be generated with a fixed photon number
in each spatial mode using the nonlinear interferometric setups discussed
above. Such states have very different properties from the single
photon Fock state due to their many-body characteristics such as exchange
statistics and entanglement. This gives rise to a different spectroscopic
signals and yield different phase matching \citep{Dorfman_2019,Asban_2019}.

Entangled states of light are commonly generated via a nonlinear process
in which one pump input is converted into two photons, or $\chi^{\left(2\right)}$
parametric down conversion (PDC). The most striking properties of
entangled light is its vanishing two-photon correlation function $G^{(2)}$
due to strong antibunching, and sub-Poissonian counting statistics.
Entangled photons have several important properties stemming from
their quantum many-body (more than one) characteristics. While individual
photons are constrained to uncertainty relations in the form $\Delta t\Delta\omega\geq1$
(Fourier uncertainty), the joint probability of the entangled photons
is not, e.g. $\Delta(\omega_{1}+\omega_{s})\Delta(t_{1}-t_{2})<1$.
This results in EPR characteristics, which make them perfect candidate
for two-photon absorption measurements \citep{Lee_2006,Upton_2013,Varnavski_2017,Schlawin_2018}.
Second, the parametric process output is sensitive to the pump properties,
this yields novel control parameters absent in classical light. For
instance, the pump bandwidth $\sigma_{p}$ controls the degree of
the frequency anti correlations in PDC, and also determines energy
conservation. The group velocity in type II PDC of each photon is
different, this results in a delay between photon propagating times
through the nonlinear crystal and captured by the entanglement time
$T$. EPR correlations are observed when $\sigma_{p}T<1$. In two-photon
absorption measurements $T$ characterizes an upper bound for the
duration in which the system can spend in the intermediate single
photon state, before being prompt to double excited state. Complementarily,
$\sigma_{p}$ gives spectral bandwidth of the excited two electron/exciton
states. 

Squeezed light is a multiphoton quantum state composed of photon pairs.
Quantum squeezing manifests as a below shot-noise counting error.
Such states are typically generated via nonlinear parametric processes
such as four-wave mixing. The latter is an interesting scheme, which
on its own accord, provides a spectroscopic information about the
media where the squeezing is generated via its third order nonlinear
response $\chi^{(3)}$. Thus, for the studies of $\chi^{(3)}$ it
is not necessary to first generate the squeezed state and then use
it to probe the system response, generation itself serves as a spectroscopic
probe \citep{Yang_2020}. The control parameters for the squeezed
state include pump and probe phase-matching geometry, intensity of
the incoming probe field, four-wave mixing gain and squeezing phase,
which is related to the microscopic details of the $\chi^{(3)}$.
Cascading interferometric elements allow to improve the degree of
squeezing and provides an additional control over the signals.

\end{widetext}
\end{document}